\documentclass[twocolumn]{aastex63}
\usepackage{natbib}
\usepackage{color}
\usepackage{mathtools}
\usepackage{booktabs}
\usepackage{hyperref} 

\def    \apjl  		{\rm {ApJL}}

\def    \apjl  		{\rm {ApJL}}

\def	\cm		{\,{\rm {cm}}}
\def	\K		{\,{\rm K}}
\def	\g		{\,{\rm {g}}}
\def	\mum	{\,{\mu \rm{m}}}

\def \bea {\begin{eqnarray}}
\def \ena {\end{eqnarray}}

\def	\bB	{\boldsymbol{B}}

\def 	\bE	{\boldsymbol{E}}

\def	\C	{{\rm C}}

\def	\cm	{\,{\rm cm}}

\def	\d	{{\rm d}}

\def	\D	{{\rm D}}

\def	\erg	{\,{\rm erg}}

\def	\g	{\,{\rm g}}
\def	\gas	{\,{\rm gas}}
\def	\G	{{\rm G}}

\def	\H	{{\rm H}}

\def	\N	{{\rm N}}

\def	\s	{\,{\rm s}}

\def	\V	{{\rm V}}

\def	\xhat		{\hat{\bf x}}
\def	\yhat		{\hat{\bf y}}
\def	\zhat		{\hat{\bf z}}

\def    \bB     	{\boldsymbol{B}} 
\def    \bE     	{\boldsymbol{E}}

\newcommand{\alfvenic}{Alfv$\acute{\text{e}}$nic~}

\begin{document}
\shorttitle{3D B-fields using starlight polarization}
\shortauthors{Truong and Hoang}
\title{Probing 3D magnetic fields using starlight polarization and grain alignment theory}

\author{Bao Truong}

\affiliation{Korea Astronomy and Space Science Institute, Daejeon 34055, Republic of Korea} 
\email{baotruong@kasi.re.kr}
\affiliation{Department of Astronomy and Space Science, University of Science and Technology, 217 Gajeong-ro, Yuseong-gu, Daejeon, 34113, Republic of Korea}

\author{Thiem Hoang}
\affiliation{Korea Astronomy and Space Science Institute, Daejeon 34055, Republic of Korea} 
\email{thiemhoang@kasi.re.kr}
\affiliation{Department of Astronomy and Space Science, University of Science and Technology, 217 Gajeong-ro, Yuseong-gu, Daejeon, 34113, Republic of Korea}


\begin{abstract}
Polarization of starlight induced by dust grains aligned with the magnetic field (hereafter B-field) is widely used to measure the two-dimensional B-fields projected onto the plane-of-sky. Here, we introduce a new method to infer three-dimensional B-fields using starlight polarization. We show that the inclination angle or line-of-sight (LOS) component of B-fields can be constrained by the starlight polarization efficiency from observations, the alignment degree provided by the magnetically enhanced radiative torque (MRAT) alignment theory, and the effect of B-field tangling. We first perform synthetic observations of starlight polarization of magnetohydrodynamic (MHD) simulations of a filamentary cloud with our updated POLARIS code incorporating the modern MRAT theory. We test the new technique with synthetic observations and find that the B-field inclination angles can be accurately determined by the synthetic starlight polarization efficiency once the effects of grain alignment, dust properties, and B-field fluctuations are well characterized. The technique can provide an accurate constraint on B-field inclination angles using optical polarization in low-density regions $A_{\rm V}< 3$ with efficient MRAT alignment, whereas the technique can infer further to high-density regions with significant alignment loss at $A_{\rm V} \sim 8 - 30$ by using near-infrared polarization. Our new technique unlocks the full potential of tracing 3D B-fields and constraining dust properties and grain alignment physics on multiple scales of the diffuse interstellar medium and star-forming regions using multi-wavelength starlight polarization observations. 
\end{abstract}

\section{Introduction}
Magnetic fields (B-fields) are thought to play an important role in the evolution of the interstellar medium (ISM) and the formation of young stars and planets \citep{Crutcher:2010p318}. Interstellar magnetic field is first recognized through the discovery of the polarization of distant starlight \citep{Hall.1949,Hiltner.1949}. Subsequently, starlight polarization is widely used to trace the large-scale B-fields in the diffuse ISM \citep{Serkowski:1975p6681,Heiles.2000} and molecular clouds \citep{Goodman.1990,Goodman.1995a} and dense cores \citep{Kandori.2020}. Starlight polarization angles could be used to measure the plane-of-sky (POS) B-fields $B_{\rm POS}$ \citep{Chapman.2011} using the Davis-Chandrasekhar-Fermi (DCF, \citealt{Davis.1951,ChandraFermi.1953}) technique.

The degree of starlight polarization encodes unique information on fundamental dust properties (grain size, shape, and composition), the B-field properties (inclination angle and tangling), and grain alignment efficiency with B-fields. \cite{Lee.1985} demonstrated that, in the observer's frame, the absorbed polarization degree of aligned grains is reduced by a factor of $f_{\rm align}\sin^2{\gamma}F_{\rm turb}$, where $f_{\rm align}$ is the degree of grain alignment, $\gamma$ is the angle between the magnetic field $\bB$ and the line-of-sight (LOS), and $F_{\rm turb}$ is the factor describing the effect of B-field tangling. Using the observed polarization degree of silicate feature at 9.7$\mum$, \cite{Lee.1985} first suggested that the B-fields along the LOS toward the Backlin-Neugebauer (BN) object are inclined by an angle $\sin^{2}\gamma_{\rm max}<0.5$ and is highly ordered ($F_{\rm turb} =1$) once the grain alignment is perfect ($f_{\rm align} = 1$). Therefore, the starlight polarization degree can provide information about the B-field inclination angle and the tangling (randomness) when the grain alignment and dust properties are well constrained \citep{Jones.1989,Jones.1992}.

The comprehensive starlight polarization survey for 364 stars by \cite{Serkowski:1975p6681} showed the maximum ratio of polarization to reddening of $(P_{V}/E(B-V))_{\rm max}\approx 9\%/\rm mag$, where $P_{\rm V}$ is the starlight polarization degree at the visible wavelength $\lambda_{\rm V} = 0.55\,\rm\mu m$ and $E(B - V) = A_B - A_V$ is the color excess describing the differential extinction at B-band ($\lambda_{\rm B} = 0.45\,\rm\mu m$) and V-band. The maximum $(P_{V}/E(B-V))_{\rm max}$ demonstrates the highest starlight polarization degree in the ideal conditions: (1) full perfect grain alignment and (2) uniform B-fields parallel to the plane-of-sky (POS). However, the latest studies by \cite{Panopoulou.2019a} using R-band RoboPol toward a sample of 22 stars that have the highest polarization degree of thermal dust emission observed by Planck of $P_{\rm 353GHz}\approx 20\%$ found a higher $P_{\rm V}/E(\rm B-V) \geq 13\%/\rm mag$ (see \citealt{Hensley.2021} for more details). Recently, several surveys of interstellar starlight polarization are available \citep{Clemens.2016,Clemens.2020,Versteeg.2023}, which provide crucial constraints on interstellar magnetic field and dust. \cite{Panopoulou.2019a} suggested the combination of starlight polarization with star distance from Gaia to reconstruct the variation of the POS component of magnetic fields along the LOS (2D B-field tomography). The degree of starlight polarization per reddening, $P_{V}/E(B-V)$, is also used to constrain the magnetic field inclination angle (\citealt{Angarita.2023}). The practice is done as follows. By observing many stars from a sky region, one can measure the $P_{V}/E(B-V)$, for all these stars. The upper bound $(P_{V}/E(B-V))_{\rm max}$ corresponds to the optimal conditions of the starlight polarization, including perfect grain alignment and a well-ordered $B$ lying completely in the POS. The sightlines with lower than $(P_{V}/E(B-V))_{\rm max}$ are interpreted as having an inclined magnetic field. However, this approach assumes that grain alignment is perfect, the effect of magnetic turbulence is minimal, and the depolarization is entirely attributed to the B-field inclination angle (\citealt{Angarita.2023,Doi.2024}). To unleash the full potential of starlight polarization in probing dust properties and B-fields, a detailed study on the effect of grain alignment and B-fields on the starlight polarization must be quantified to accurately infer the inclination angle using the starlight polarization.


Grain alignment physics is a long-standing problem in astrophysics \citep{DavisGreenstein.1951,JonesSpitzer.1967,Purcell.1979}. The leading theory of grain alignment is now based on the Radiative Torques (RATs), initially studied by \cite{Dolginov:1976p2480} and numerically quantified by \cite{DraineWein.1996,DraineWein.1997}. The quantitative theory of grain alignment based on RATs is developed in \cite{LazHoang.2007,HoangLaz.2008,Hoang.2014}. The successful development of a unified theory of grain alignment by the joint effect of magnetic relaxation and RATs by \cite{LazarianHoang.2008,HoangLaz.2016} allows us to predict the alignment degree as a function of the local properties and dust properties. \cite{HoangBao.2024} (Paper I) first incorporated grain alignment physics into the thermal dust polarization model and introduced an improved technique to constrain 3D B-fields using the observed polarization data. The testing using synthetic polarization of magnetohydrodynamic (MHD) simulations revealed a successful recovery of the B-fields. In this follow-up paper, for the first time, we aim to extend the technique to starlight polarization using a similar approach in Paper I (\citealt{HoangBao.2024}) by using synthetic observations of starlight polarization using MHD simulations for the turbulent magnetized filament using the modern MRAT alignment theory. 

The structure of this paper is as follows. In Section \ref{sec:method} we introduce a physical model of starlight polarization based on the MRAT mechanism and present a new method for constraining 3D B-fields using starlight polarization. In Section \ref{sec:simul}, we will test the starlight polarization model using synthetic polarization observations of MHD simulations. In Section \ref{sec:results}, we will present our numerical results, compare them with the polarization from the analytical model, and apply our method for inferring the inclination angles using the synthetic starlight polarization. An extended discussion of our results and implications is presented in Section \ref{sec:discuss}, and a summary is shown in Section \ref{sec:summ}.

\section{Methods}\label{sec:method}
\subsection{Starlight Polarization Model}
\label{sec:method_starlight_model}
In this section, we first present the detailed physical model of starlight polarization incorporated with grain alignment theory and the calculations of starlight polarization efficiency $P/N_{\rm H}$ and the starlight polarization integral $\Pi$ (see \citealt{Draine.2021no}; \citealt{Hensley.2023}) in the case of uniform B-fields. The effect of magnetic turbulence on the calculation of starlight polarization will be discussed in Section \ref{sec:method_star_fluc}.
\subsubsection{Uniform magnetic field}
\label{sec:method_star_uniform}
\begin{figure}
    \includegraphics[width=0.45\textwidth]{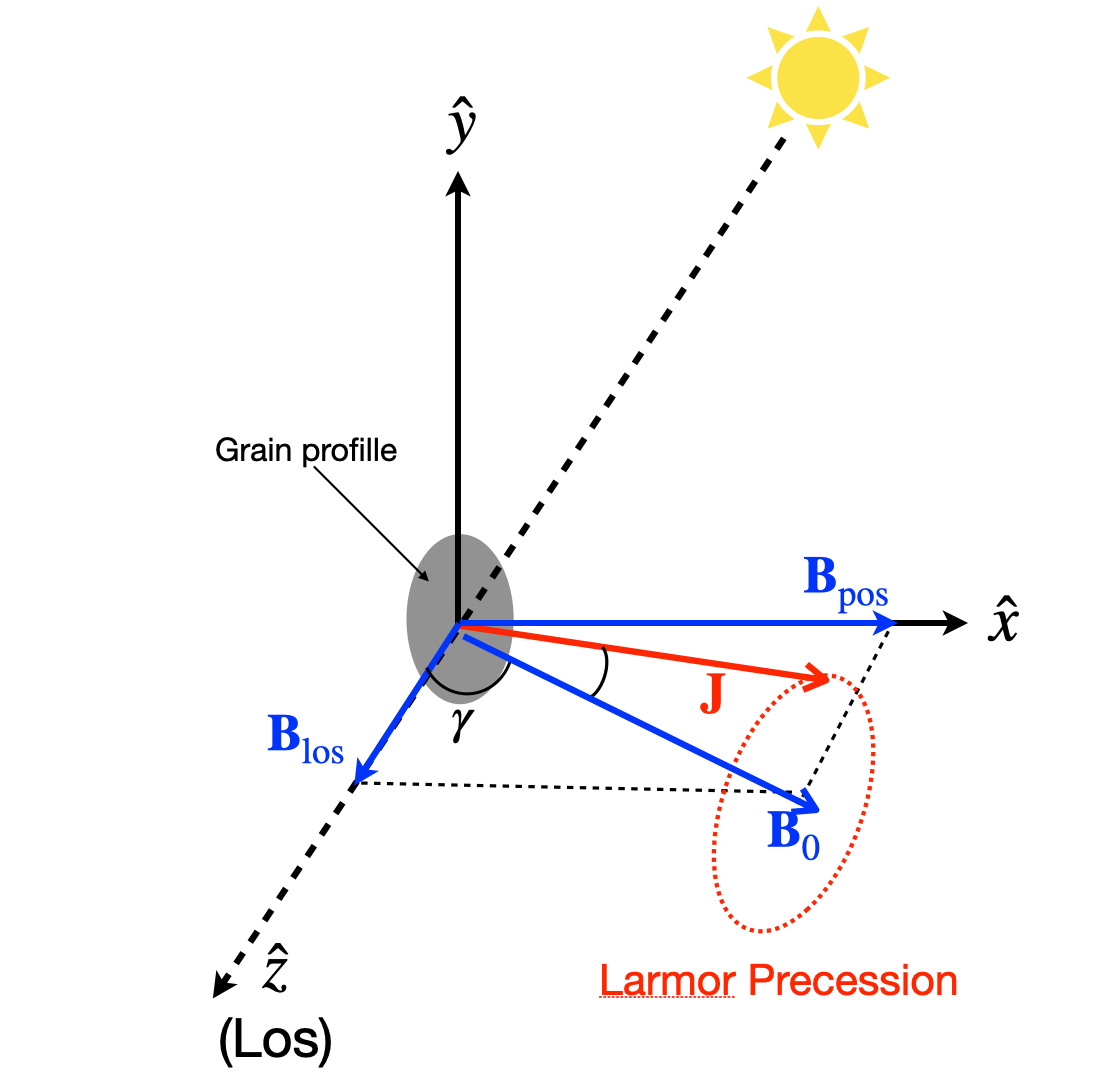}
    \caption{Illustration of the effect of grain alignment, the mean magnetic field $\bB$, and the line of sight $\zhat$ from Paper I (\citealt{HoangBao.2024}). The mean B-field lies in the plane $\xhat \zhat$ and makes an angle $\gamma$ with the line of sight. The grain axis of maximum inertia is assumed to be precessed along $\bB$ by the Larmor precession and aligned with $\bB$ by RATs. The difference in the cross-sections along $\xhat$ and $\yhat$ induces the polarization of transmission starlight.}
    \label{fig:JB_Cxy}
\end{figure}

Let $\zhat$ be the LOS, and the plane of the sky is described by $\xhat\yhat$ (see Figure \ref{fig:JB_Cxy}). The regular magnetic field $\bB_{0}$ is assumed to make an angle $\gamma$ with the LOS, and the POS component of the B-field is lying along $\xhat$. The fluctuations of the local magnetic field along the LOS are first ignored for simplicity. Let $C_{x},C_{y}$ be the extinction cross-section of grain for the electric field of the incident radiation along $\xhat$ and $\yhat$, respectively. The difference in $C_{x}$ and $C_{y}$ induced by partially aligned grains is given by (\citealt{Lee.1985}; \citealt{HoangBao.2024})
\bea
C_{y}-C_{x}=C_{\rm pol}f_{\rm align}\sin^{2}\gamma,\label{eq:CxCy}
\ena
where $C_{\rm pol}$ is the polarization cross-section of the grain, and $f_{\rm align}$ is the degree of grain alignment with the magnetic field, which will be determined in Section \ref{sec:method_align_degree}.


The intensity of transmission light from stars with the incident electric field along $\xhat$ and $\yhat$ is given by, respectively
\bea
I_{x}(\bE\|\xhat)=I_{0x}e^{-\tau_{x}},~~I_{y}(\bE\|\yhat)=I_{0y}e^{-\tau_{y}},
\ena
where $I_{0x}=I_{0y}=I_{0}/2$ with $I_{0}$ is the total intensity of unpolarized incident starlight, and $\tau_{x},\tau_{y}$ are the optical depth for the incident electric field along the $\xhat$ and $\yhat$ axes.

The polarization degree of transmission starlight is defined by
\bea
P_{\rm ext}=\frac{I_{x}({\bE}\|\xhat)-I_{y}({\bE}\|\yhat)}{I_{x}({\bE}\|\xhat)+I_{y}({\bE}\|\yhat)}=\frac{e^{-\tau_{x}}-e^{-\tau_{y}}}{e^{-\tau_{x}}+e^{-\tau_{y}}}=\tanh\left(\frac{\tau_{y}-\tau_{x}}{2}\right),\label{eq:pol_ext}
\ena
where $\tau_{\rm pol}=\tau_{y}-\tau_{x}$ is the polarized optical depth.

For the case of small polarized extinction with $\tau_{\rm pol}\ll 1$, one obtains
\bea
P_{\rm ext}\approx \frac{\tau_{y}-\tau_{x}}{2+\tau_{y}+\tau_{x}}\approx \left(\frac{\tau_{y}-\tau_{x}}{2}\right).\label{eq:pext1}
\ena

For the case of large polarized extinction with $\tau_{\rm  pol}\gg 1$, one has $\tanh(\tau_{\rm pol})\rightarrow 1$, so that the polarization degree by dichroic extinction achieves the maximum level given by $P_{\rm ext} = 1.$

In this subsection, we first consider the case of a uniform environment where dust properties and grain alignment do not change significantly along the LOS, and the magnetic fields are uniform and lying in the POS. Using the relationship between $\tau_{y}-\tau_{x}$ and the grain alignment model in Paper I (\citealt{HoangBao.2024}), one obtains the polarization degree of starlight by extinction for the case of small polarized extinction (Equation \ref{eq:pext1})
\bea
P_{\rm ext}(\lambda) =N_{\rm H}\sin^{2}\gamma\int_{a_{\rm min}}^{a_{\rm max}} f_{\rm align}(a)C_{\rm pol}(a,\lambda)n_{d}(a) da,
\label{eq:pext2}
\ena
where $N_{\H}$ is the column density of hydrogen nucleon along the LOS, $n_{d}(a)$ is the distribution of the effective grain size $a$ with the lower and upper cutoff of $a_{\rm min}$ to $a_{\rm max}$.

\subsubsection{Polarization Efficiency}
\label{sec:method_pol_eff}
The degree of starlight polarization per hydrogen atom (i.e., polarization efficiency) can also be written as
\bea
\frac{P_{\rm ext}(\lambda)}{ N_{\rm H}}=\sin^{2}\gamma \times \sigma_{\rm pol, align}(\lambda),\label{eq:p_NH}
\ena
where 
\bea
\sigma_{\rm pol, align}(\lambda) = \int_{a_{\rm min}}^{a_{\rm max}}f_{\rm align}(a)C_{\rm pol}(a,\lambda) n_{d}(a)da,
\ena
is the polarization coefficient of aligned grains at the wavelength $\lambda$ (in units of $\cm^{2}$ per H atoms). 



\subsubsection{Starlight Polarization Integral}
\label{sec:method_pol_int}
Following \cite{Draine.2021no}, we define the starlight polarization integral as
\bea
{\Pi} = \frac{\int_{\lambda} P_{\rm ext}(\lambda) d\lambda}{N_{\rm H}},
\ena
which integrates over the starlight polarization spectrum.

Using the polarization extinction from Equation \ref{eq:p_NH} for the above equation, one can write
\bea
\frac{\Pi}{ \sin^{2}\gamma} &=& \frac{\int_{\lambda} P_{\rm ext}(\lambda) d\lambda}{N_{\rm H } \sin^{2}\gamma}\nonumber\\
&=&  \int_{a_{\rm amin}}^{a_{\rm max}} \left[\int_{\lambda} C_{\rm pol}(a,\lambda)d\lambda\right]f_{\rm align}(a)n_{d}(a)da\nonumber\\
&=&  \int_{a_{\rm amin}}^{a_{\rm max}} \left[\frac{\int C_{\rm pol}(a,\lambda)d\lambda}{V_{\rm gr}(a)}\right]V_{gr}(a)f_{\rm align}(a)n_{d}(a)da\nonumber\\
&=&  \int_{a_{\rm amin}}^{a_{\rm max}} f_{\rm align}(a)\Phi(a) V_{\rm gr}(a)n_{d}(a)da,\label{eq:Pi_obs}
\ena
where 

\bea
\Phi(a)=\frac{\int_{\lambda} C_{\rm pol}(a,\lambda)d\lambda}{V_{gr}(a)}=\frac{3}{4\pi}\int \frac{Q_{\rm pol}(a,\lambda)}{a}d\lambda,
\ena
is the dimensionless starlight polarization efficiency integral and $V_{\rm gr}(a) = (4\pi/3)a^{3}$ is the grain volume. Here, $Q_{\rm pol}(a, \lambda)$ is the polarization cross-section efficiency taken from the Astrodust model (\citealt{Draine.2021no}; \citealt{Hensley.2023})\footnote{http://arks.princeton.edu/ark:/88435/dsp01qb98mj541}. The starlight polarization efficiency integral depends on both grain sizes and grain elongation.

\subsubsection{Effect of magnetic field fluctuations}
\label{sec:method_star_fluc}
Magnetic fluctuations with respect to the mean B-field have a significant influence on the observed starlight polarization degree (\citealt{Lee.1985}). In this subsection, we discuss the effect of magnetic fluctuations on starlight polarization. Assuming that the turbulent field component is perpendicular to the mean field (i.e., \alfvenic turbulence), $\bB_{0}$, the local magnetic field reads 
\bea
\bB =\bB_{0}+\delta \bB_{\perp}.
\ena
Note that B-fields in the ISM are highly anisotropic and compressible (see, e.g., \citealt{Skalidis.Tassis2021}; \citealt{Skalidis.etal.2021}). \cite{Cho.Lazarian.2003} numerically studied compressible MHD turbulence for different MHD modes as \alfvenic, compressible fast, and slow modes. The authors suggested that the ISM could be dominated by \alfvenic and compressible slow modes, with the Kolmonogrov $k^{-5/3}$ spectrum and the preferential dominance of perpendicular fluctuations to the mean B-fields (see also \citealt{Kandel.2017}; \citealt{HuLaz.2023}). Thus, our assumption of perpendicular magnetic fluctuations is valid in the compressible ISM. 




Let $\Delta \theta$ be the deviation angle between the local magnetic field $\bB$ and the mean B-field $\bB_0$ in the three-dimensional space. From the MHD simulation, the deviation angle is calculated from the local and mean B-fields in each cell of the simulation box as
\bea
\Delta\theta = \cos^{-1}\frac{(\bB \cdot \bB_0)}{\|\bB\| \|\bB_0\|}.\label{eq:delta_theta}
\ena

Then, the polarization cross-section (Equation \ref{eq:CxCy}) becomes (\citealt{Lee.1985})
\bea
C_{y}-C_{x}=C_{\rm pol}f_{\rm align}\sin^{2}\gamma F_{\rm turb},\label{eq:Cy-Cx_Fturb}
\ena
where 
\bea
F_{\rm turb}=\frac{1}{2} \left[(3\langle \cos^{2}(\Delta \theta)\rangle -1\right] = 1 - \frac{3}{2}\langle \sin^{2}(\Delta \theta)\rangle,\label{eq:F}
\ena
which describes the effect of magnetic fluctuations along the LOS. In this study, we integrate the local $\cos^2\Delta\theta$ over the LOS, and calculate the factor $F_{\rm turb}$ as shown in Equation \ref{eq:F}.

For small perpendicular magnetic fluctuations of $\Delta \theta \ll 1$ (i.e., sub-\alfvenic turbulence and $\sin\Delta \theta \approx \Delta\theta$), one can make the approximation \citep{HoangBao.2024}
\bea
F_{\rm turb} \approx 1-\frac{3}{2}\langle (\Delta \theta)^{2}\rangle \approx 1-1.5(\delta \theta)^{2}\label{eq:F_weak},
\ena
where $\delta \theta = \langle (\Delta \theta)^{2}\rangle^{1/2}$ is the angle dispersion, which is derived from the integrated $\langle (\Delta \theta)^{2}\rangle$ along the sightline.

\begin{figure}
     \centering
     \includegraphics[width = 0.48\textwidth]{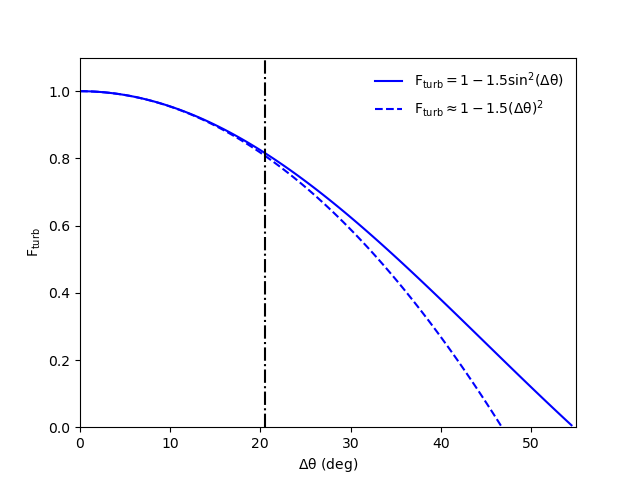}
     \caption{The magnetic turbulence factor $F_{\rm turb}$ versus the deviation angle $\Delta \theta$ between local and mean B-fields from Equation \ref{eq:F} (solid blue line). The factor $F_{\rm turb}$ decreases with increasing $\Delta \theta$. It can be approximated by $1 - 1.5(\Delta \theta)^2$ (dashed blue line) when the deviation is less than $20.5^{\circ}$ (i.e., the relative error is less than $1\%$, dashed-dotted black vertical line).}
     \label{fig:Fturb_approx}
 \end{figure}

Figure \ref{fig:Fturb_approx} shows the analytical calculation of magnetic turbulence factor $F_{\rm turb}$ varying with the deviation angle $\Delta \theta$. One obtains that the increase in the deviation angle $\Delta \theta$ reduces the value of $F_{\rm turb}$, which then decreases the starlight polarization efficiency (\citealt{Lee.1985}). Therefore, when taking into account the magnetic fluctuations, one needs to incorporate the factor $F_{\rm turb}$. The factor $F_{\rm turb}$ can be approximated by $1 - 1.5(\Delta\theta)^2$ when the deviation is smaller than $20.5^{\circ}$ (or the dispersion $\delta \theta < 20.5^{\circ}$) with the relative error less than $1\%$. In addition to the dominant perpendicular fluctuations, there is the presence of parallel fluctuations to the mean fields $\delta \bB_{\|}$ (i.e., compressive turbulence, see \citealt{Skalidis.Tassis2021}; \citealt{Skalidis.etal.2021}). In that case, the local B-fields become parallel to $\bB_0$ with $\Delta\theta \approx 0$ and $F_{\rm turb} \approx 1$; thus, the effect of compressive turbulence is negligible to the absorbed depolarization in the ISM.

The polarization efficiency with magnetic turbulence is now given by
\bea
\frac{P_{\rm ext}(\lambda)}{ N_{\rm H}}=\sin^{2}\gamma \times F_{\rm turb} \times \sigma_{\rm pol, align}(\lambda),\label{eq:p_NH_Fturb}
\ena
and the polarization starlight integral becomes
\bea
\frac{\Pi}{\sin^{2}\gamma F_{\rm turb}} =  \int_{a_{\rm amin}}^{a_{\rm max}} \Phi(a) V_{\rm gr}(a)f_{\rm align}(a)n_{d}(a)da.\label{eq:Pi_obs_Fturb}
\ena

\subsection{Grain alignment from the MRAT theory}
\label{sec:method_grain_align}
From the environmental conditions and magnetic properties of grains (see, e.g., \citealt{HoangLaz.2016}), we can determine the local alignment properties produced by RATs. In this section, we present the calculation of grain alignment properties in local environments from the modern MRAT alignment theory.

\subsubsection{Minimum size for grain alignment}
\label{sec:method_align}
The key parameter of the RAT alignment theory is the minimum grain size required for grain alignment, denoted by $a_{\rm align}$, which is given by \citep{Hoang.2021}:
\bea
a_{\rm align}&=&\left(\frac{1.2n_{\rm H}T_{\rm gas}}{\gamma_{\rm rad} u_{\rm rad}\bar{\lambda}^{-2}} \right)^{2/7}\left(\frac{15m_{\rm H}k^{2}}{4\rho}\right)^{1/7}(1+F_{\rm IR})^{2/7}\nonumber\\
&\simeq &0.055\hat{\rho}^{-1/7} \left(\frac{\gamma_{-1}U}{n_{3}T_{\rm gas,1}}\right)^{-2/7} \nonumber\\
    &&\times \left(\frac{\bar{\lambda}}{1.2\mum}\right)^{4/7} (1+F_{\rm IR})^{2/7} ~\mum,\label{eq:align_ana}
\ena 
where $m_{\rm H} = 1.00784\,\rm u$ is the hydrogen atom mass, $\hat{\rho}=\rho/(3\g\cm^{-3})$ is the normalized mass density of grain material, $T_{\rm gas,1}=T_{\rm gas}/10\K$ is the normalized gas temperature, $n_{3}=n_{\H}/(10^{3}\cm^{-3})$ is the normalized gas density, $\gamma_{-1}=\gamma_{\rm rad}/0.1$ is the normalized anisotropy degree of radiation, $\bar{\lambda}$ is the mean wavelength of the radiation field and $U=u_{\rm rad}/u_{\rm ISRF}$ is the strength of radiation field with the energy density $u_{\rm rad}$ over the energy density of the interstellar radiation field (ISRF) in the solar neighborhood with $u_{\rm ISRF}=8.6\times 10^{-13}\erg\cm^{-3}$ \citep{Mathis.1983}. $F_{\rm IR}$ is a dimensionless parameter that describes the grain rotational damping by infrared emission, which can be negligible in the ISM and dense clouds as $F_{\rm IR}\ll 1$. For typical values of $n_{\rm H} = 10^3\,\rm cm^{-3}, T_{\rm gas} = T_{\rm dust} = 10\,\rm K$, $\gamma_{\rm rad} = 0.1$ and $U = 1$, $a_{\rm align} \sim 0.055\,\rm\mu m$.


\subsubsection{Grain Alignment Degree and Alignment Function}
\label{sec:method_align_degree}
The alignment efficiency for an ensemble of grains having a given size $a$ is described by the Rayleigh reduction factor, $R$ (see \citealt{Greenberg.1968}), which characterizes the average alignment degree of the grain axis of major inertia with the ambient magnetic field. From the modern theory of grain alignment based on RATs, the Rayleigh reduction factor is parameterized by the fractions of grains aligned at low-J and high-J attractors (see in Paper I, \citealt{HoangBao.2024}),
\bea
R=f_{\rm high-J}Q_{X}^{\rm high-J}+ (1-f_{\rm high-J})Q_{X}^{\rm low-J}.\label{eq:Ralign}
\ena
Numerical simulations in \cite{HoangLaz.2016} show that if the RAT alignment has a high-J attractor point, then, grains can be perfectly aligned when they are spun up to suprathermal rotation with $Q_{X}^{\rm high-J} = 1$. Meanwhile, grains at low-J attractors rotating subthermally by gas collision have smaller alignment efficiency with $Q_{X}^{\rm low-J} \sim 0.3 $ \citep{HoangLaz.2016,HoangLaz.2016b}

Within the MRAT mechanism, the fraction $f_{\rm high-J}$ is determined by both RATs, magnetic relaxation, and gas randomization, which is approximately given by the function 
\bea
f_{\rm high-J}=f(\delta_{\rm mag}),\label{eq:fhighJ}
\ena
and $\delta_{\rm mag}$ is the magnetic relaxation strength which measures the relative strength of the magnetic relaxation to gas randomization. For grains without iron inclusions (i.e., ordinary paramagnetic material, a.k.a, PM), high-J attractors are only present for a limited range of the radiation direction that depends on the grain shape (\citealt{HoangLaz.2016}).

For superparamagnetic (SPM) grains with embedded iron inclusions, the magnetic relaxation strength is enhanced and calculated as
\bea
\delta_{\rm mag}&=&\frac{\tau_{\rm gas}}{\tau_{\rm mag,sp}}\nonumber\\
&= &5.6a_{-5}^{-1}\frac{N_{\rm cl}\phi_{\rm sp,-2}\hat{p}^{2}B_{2}^{2}}{\hat{\rho} n_{3}T_{\gas,1}^{1/2}}\frac{k_{\rm sp}(\Omega)}{T_{\rm d,1}},\label{eq:delta_m}~~~~
\ena
where $N_{\rm cl}$ and $\phi_{\rm sp,-2} = \phi_{\rm sp}/10^{-2}$ are the number of iron inclusions and the volume filling factor, $a_{-5} = a /10^{-5}\,\rm cm$, $\hat{p} = p/5.5$ ($p \simeq 5.5$ for silicate dust, see \citealt{Draine.1996}), $B_{2}=B/10^{2}\mu G$ is the normalized magnetic field strength and $T_{\rm d,1}=T_{\rm d}/10\K$ is the normalized dust temperature. The relaxation strength is higher if SPM grains have an enhanced magnetic susceptibility by including higher levels of iron clusters (\citealt{LazarianHoang.2008}; \citealt{HoangLaz.2016}).

Detailed calculations of \cite{HoangLaz.2016} shows the increase of $f_{\rm high-J}$ with $\delta_{\rm mag}$, which can be parameterized by \citep{Giang.2023}:
\bea 
f_{\rm high-J}(\delta_{\rm mag}) = \left\{
\begin{array}{l l}    
    0.25 ~ ~  {\rm ~ for~ } \delta_{\rm mag} < 1   \\
    0.5 ~ ~  {\rm ~ for~ } 1 \leq \delta_{\rm mag} \leq 10 \\
    1    ~ ~ ~ ~  {\rm ~ for~}  \delta_{\rm mag} > 10 \\
\end{array}\right..
\label{eq:fhiJ_deltam}
\ena

From numerical calculations in \cite{HoangLaz.2016} using the MRAT alignment theory, grains smaller than the critical size $a_{\rm align}$ are weakly aligned only by paramagnetic relaxation (also \citealt{HoangLaz.2016b}), and grains much larger than $a_{\rm align}$ have an alignment degree described by $R$. Moreover, the alignment degree increases smoothly with the grain size across $a=a_{\rm align}$. Therefore, we can describe the degree of grain alignment as a function of the grain size (i.e., alignment function) by multiplying $R$ with an exponential function \citep{LeeHoang.2020}:
\bea 
f_{\rm align}(a) = R\times \left[1 - \exp\left(-\left(\frac{0.5a}{a_{\rm align}}\right)^{3}\right)\right],
\label{eq:falign}
\ena
which implies that $f_{\rm align}$ approaches $R$ for $a\gg a_{\rm align}$ and is negligible for $a\ll a_{\rm align}$.

\begin{figure}
     \centering
     \includegraphics[width = 0.48\textwidth]{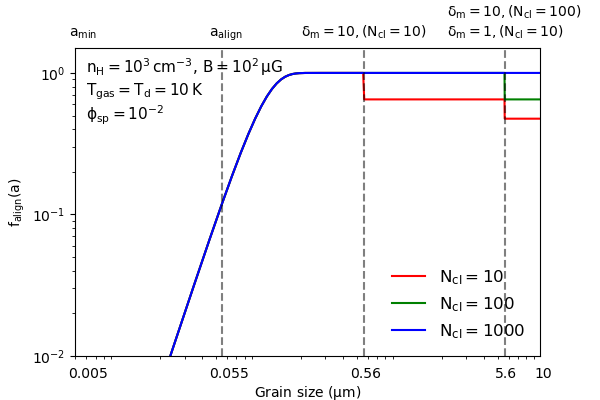}
     \caption{The analytical calculation of the alignment degree of SPM grains with $N_{\rm cl} = 10-1000$ as an exponential form in Equation \ref{eq:falign} over the grain size with $a_{\rm align} = 0.055\,\rm\mu m$. The alignment degree increases with the grain size, and large grains $a>a_{\rm align}$ can achieve perfect magnetic alignment for $N_{\rm cl}\sim 1000$ by the MRAT mechanism.}
     \label{fig:falign_a}
 \end{figure}

For a given grain size distribution and magnetic properties, using $a_{\rm align}$ from the RAT theory, $\delta_{\rm mag}$ and $R$, one can calculate the alignment function $f_{\rm align}(a)$ as shown in Equation \ref{eq:falign}. Figure \ref{fig:falign_a} shows an example of the analytical calculation of the grain alignment function $f_{\rm align}$ for SPM grains and $N_{\rm cl} = 10, 100$ and 1000, assuming $T_{\rm gas} = T_{\rm d} = 10\,\K$, $n_{\H} = 10^3\,\cm^{-3}$, $B = 10^2\,\rm \mu G$ with $a_{\rm align} = 0.055\,\rm\mu m$ (see Equation \ref{eq:align_ana}). The volume filling factor of iron clusters of $\phi_{\rm sp} = 0.01$ is chosen. Compared to the volume filling factor of entire chemical elements (including Fe, Mg, Si, and O) incorporated into dust to form the structure $\rm MgSiFeO_{4}$ as $\phi_{\rm sp} = 0.29$ (\citealt{HoangLaz.2016}), it demonstrates that about $0.01/0.29 \approx 3\%$ of Fe incorporated into dust as a form of iron inclusions. Small grains $a < a_{\rm align}$ are weakly aligned with B-fields with a very low alignment degree $f_{\rm align}(a) < 0.1$. Large grains of $a > a_{\rm align}$ are efficiently aligned by the MRAT mechanism and can achieve perfect alignment of $f_{\rm align}(a) = 1$ as $N_{\rm cl} \sim 1000$. This is important for very large grains (VLG) of $a > 10\,\rm\mu m$  formed in dense environments as protostellar cores/disks (\citealt{Hoangetal.2022,Giang.2023}). 

We have updated the POLARIS with this alignment function $f_{\rm align}(a)$ and compared the polarization results with the piece-wise function $R$ originally implemented in POLARIS \citep{Reissl.2016,Giang.2023}. We found the exponential function in $f_{\rm align}(a)$ mostly affects starlight polarization in optical-UV, which are dominated by aligned grains of $a\sim a_{\rm align}$, but not important for starlight polarization in NIR or polarized thermal dust emission at far-IR/sub-mm which are dominated by large grains of $a>a_{\rm align}$. Therefore, the previous results in Paper I (\citealt{HoangBao.2024}) on far-IR/sub-mm polarization using POLARIS remain valid.

\subsection{Inferring the inclination angle of the mean magnetic fields}
\label{sec:method_incl_cal}
Given the information on both dust and magnetic field properties, we are able to infer the inclination angles of B-fields using starlight polarization. In this section, we describe the calculation of inferred inclination angles from the starlight polarization efficiency and the starlight polarization integral.

\subsubsection{Using Polarization Efficiency}
\label{sec:method_incl_pol_eff}
To infer the inclination angle of the mean magnetic field wrt to the LOS from starlight polarization observed at a wavelength $\lambda$, we use the observed polarization efficiency described in Equation \ref{eq:p_NH_Fturb}. The maximum polarization efficiency occurs for the magnetic field perpendicular to the LOS (i.e., $\sin^{2}\gamma=1$), and perfect grain alignment of all grain sizes. This corresponds to the intrinsic polarization efficiency

\bea
\frac{P_{\rm ext,i}(\lambda)}{N_{\rm H}}=\int_{a_{\rm amin}}^{a_{\rm max}}C_{\rm pol}(a,\lambda)n_{d}(a)da = \sigma_{\rm pol, i}(\lambda),\label{eq:pext_i}
\ena
where $\sigma_{\rm pol, i}(\lambda)$ is the intrinsic polarization coefficient for all grain sizes perfectly aligned with magnetic fields with $f_{\rm align}(a) =1$. 

Here, we introduce the fraction of the polarization coefficient $f_{\rm pol} (\lambda)$ as
\bea
f_{\rm pol} (\lambda) = \frac{\sigma_{\rm pol, align}(\lambda)}{\sigma_{\rm pol, i}(\lambda)}.\label{eq:fpol}
\ena

Therefore, the observed polarization efficiency (Equation \ref{eq:p_NH_Fturb}) can be rewritten as
\bea
\frac{P_{\rm ext}(\lambda)}{N_{\rm H}} = \frac{P_{\rm ext,i}(\lambda)}{N_{\rm H}} \times f_{\rm pol} (\lambda) F_{\rm turb} \sin^{2}\gamma.\label{eq:p_NH_fpol}
\ena

Thus, one can infer the inclination angle from the observed starlight polarization at the wavelength $\lambda$ as
\bea
\sin^{2}\gamma = \frac{1}{f_{\rm pol}(\lambda)F_{\rm turb}}\frac{(P_{\rm ext}/N_{\rm H})_{\rm obs}}{(P_{\rm ext,i}/N_{\rm H})},\label{eq:chi_ext}
\ena
which gives the physical solution of the absolute $|\gamma|$ (i.e., the orientation of the mean B-field with respect to the LOS) only when $\sin^2\gamma \lesssim 1$. The direction of the inclined B-field is not given due to the unknown sign of $\gamma$.

Consider the perfect alignment case ($f_{\rm pol}(\lambda) = 1$), and the magnetic fluctuations are negligible ($F_{\rm turb} = 1$). The inferred inclination angle becomes
\bea
(\sin^{2}\gamma)_{\rm per} = \frac{(P_{\rm ext}/N_{\rm H})_{\rm obs}}{(P_{\rm ext}/N_{\rm H})_{\rm max}}.\label{eq:chi_ext_perp}
\ena
Thus,
\bea
\sin^{2}\gamma = \frac{(\sin^{2}\gamma)_{\rm per}}{f_{\rm pol}(\lambda)F_{\rm turb}}.\label{eq:chi_ext_compare}
\ena

\begin{figure*}
    \centering
    \includegraphics[width=0.48\linewidth]{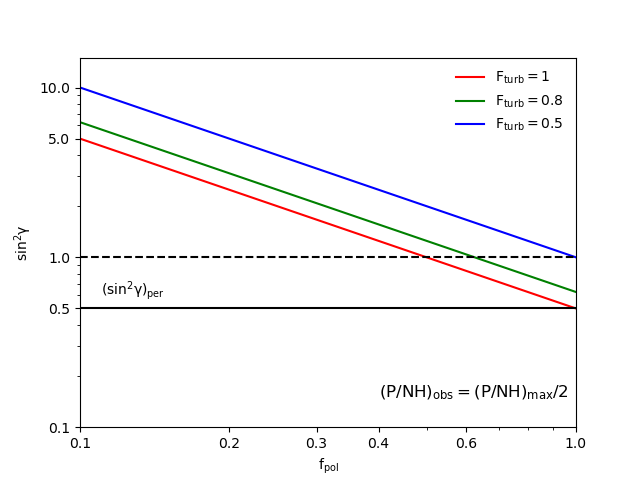}
    \includegraphics[width=0.48\linewidth]{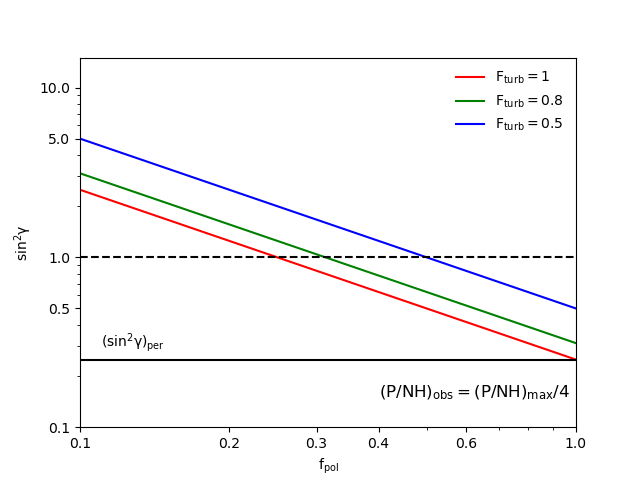}
    \caption{The analytical calculations of $\sin^2\gamma$ from the starlight polarization efficiency (Equation \ref{eq:chi_ext}) with respect to the polarization coefficient fraction $f_{\rm pol}$ and the magnetic turbulence factor $F_{\rm turb}$. The fixed observed polarization efficiencies are assumed to be lower than the maximum ones by a factor of 2 (left panel) and 4 (right panel). The presence of grain alignment and magnetic fluctuations causes the variation of inferred inclination angles, in comparison to the perfect model $(\sin^2\gamma)_{\rm per}$ (i.e., $f_{\rm pol}  = 1$ and $F_{\rm turb}$) illustrated by the solid black horizontal line. The inferred angles cannot be retrieved (i.e., $\sin^2\gamma > 1$) when $f_{\rm pol} \ll 1$ or $F_{\rm turb} \ll 1$.}
    \label{fig:sin_fpol_Fturb}
\end{figure*}

Figure \ref{fig:sin_fpol_Fturb} demonstrates the importance of grain alignment and magnetic fluctuations on the calculation of the inferred inclination angles $\sin^{2}\gamma$, in comparison to the perfect case $(\sin^{2}\gamma)_{\rm per}$ (solid black horizontal line). We assume the observed polarization efficiencies are fixed and lower than the maximum values by a factor of 2 and 4, corresponding to the absolute inclination angles of $|\gamma_{\rm per}| \sim 45^{\circ}$ (left panel) and $|\gamma_{\rm per}| \sim 30^{\circ}$ (right panel) in the perfect model, respectively. In the presence of grain alignment and magnetic turbulence, the inclination angles vary considerably. For instance, with $f_{\rm pol} \sim 0.5$ or $F_{\rm turb} \sim 0.5$, the inclination angles could increase by $60\%$. The inclination angles could not be retrieved with $\sin^2\gamma > 1$ and return to NAN when $f_{\rm pol} \ll 1$ (i.e., significant loss of grain alignment) or $F_{\rm turb} \ll 1$ (i.e., stronger magnetic turbulence). This is realistic in high-density environments, such as starless cores and protostellar cores/disks, where the loss of RAT alignment could dominate (\citealt{Hoang.2021}; \citealt{Hoangetal.2022}). Note that Equation \ref{eq:chi_ext} is valid for the case of small polarized optical depth $\tau_{\rm pol} \ll 1$, and could return NAN values when $\tau_{\rm pol} \gg 1$ (i.e., high polarized extinction, see Section \ref{sec:method_star_uniform}).


\subsubsection{Using Starlight Polarization Integral}
\label{sec:method_polint}

From the starlight polarization integral described in Equation \ref{eq:Pi_obs_Fturb}, we also obtain the inclination angles of the mean fields as
\bea
\sin^{2}\gamma=\frac{\Pi}{F_{\rm turb}\int_{a_{\rm amin}}^{a_{\rm max}} \Phi(a) V_{gr}(a)f_{\rm align}(a)n_{d}(a)da}.\label{eq:incl_pi_obs}
\ena

For grains in the diffuse ISM, \cite{Draine.2021no} evaluated the starlight polarization efficiency integral by taking
\bea
\Phi(a)\approx \Phi(a_{\rm align})=\frac{3}{4\pi}\int \frac{Q_{\rm pol}(a_{\rm align},\lambda)}{a_{\rm align}}d\lambda,
\ena
which now only depends on the grain elongation (see also \citealt{Draine2024a}; \citealt{Draine2024b}).

Consequently,
\bea
\frac{\Pi}{\sin^{2}\gamma F_{\rm turb}} &=& \Phi(a_{\rm align})\int_{a_{\rm amin}}^{a_{\rm max}}  V_{gr}(a)f_{\rm align}(a)n_{d}(a)da\nonumber\\
&=&\Phi(a_{\rm align})V_{d}\langle f_{\rm align}\rangle,
\ena
where 
\bea
\langle f_{\rm align}\rangle = \frac{\int_{a_{\rm min}}^{a_{\rm max}}(4\pi a^{3}/3)f_{\rm align}(a)n_{d}(a)da}{V_{d}},\label{eq:falign_mass}
\ena
is the mass-weighted alignment efficiency and $V_{d}=\int (4\pi a^{3}/3)n_{d}(a)da$ is the total dust volume per H atoms. In this synthetic modeling, we assume ISM grains followed by the power-law distribution $n_{\rm d}(a)da \sim C a^{-\eta}da$, where $C$ is the normalization constant derived from the dust-to-gas mass ratio $M_{\rm d/g}$ and the grain size distribution slope $\eta$ (see Equation 25 in \citealt{TramHoang.2020}). With $\eta = -3.5$, $a_{\rm min} = 0.005\,\rm\mu m$, $a_{\rm max} = 0.25\,\rm\mu m$, $M_{\rm d/g} = 0.01$ (\citealt{Mathis.1977}), $\rho_{\rm astro} = 2.74\,\rm g\,cm^{-3}$ for the Astrodust model with the porosity of 0.2 (\citealt{Hensley.2023}), $C \simeq 1.684 \times 10^{-25}\,\cm^{-2.5}$, giving $V_{\rm d} \simeq 6.057 \times 10^{-27}\,\cm^{3}\,\H^{-1}$.


The observed starlight polarization integral becomes
\bea
\Pi = \Phi(a_{\rm align})V_{d}\langle f_{\rm align}\rangle\sin^{2}\gamma F_{\rm turb},
\ena
which yields the inclination angle
\bea
\sin^{2}\gamma=\frac{\Pi}{\Phi(a_{\rm align})V_{d}\langle f_{\rm align}\rangle F_{\rm turb}}.\label{eq:incl_pi_ism}
\ena

\section{Synthetic observations of starlight polarization using MHD simulations}\label{sec:simul}
In real observations, the properties of local gas, dust, and magnetic fields change along the LOS due to the nature of turbulence and gravity in the ISM and star-forming regions and grain evolution. In addition, the polarized optical depth could be large in high-density regions with $\tau_{\rm pol} \gg 1$ and affect the derivation of inferred inclination angles based on the analytical model of starlight polarization in Equation \ref{eq:p_NH_fpol}. Consequently, we need to test whether the analytical formula given by Equation \ref{eq:p_NH_fpol} can adequately describe the observed starlight polarization to verify the capability of inferring the inclination angle of the magnetic field using starlight polarization efficiency in such an inhomogeneous environment. In this study, we will post-process MHD simulations with our POLARIS code, updated to include MRAT (\citealt{Giang.2023}), to obtain the synthetic starlight polarization. We will then apply our technique to infer B-field inclination angles using the synthetic observations, testing whether we can accurately extract the effects of RAT and MRAT included in the simulations.

\subsection{MHD simulation datacube}
Following Paper I (\citealt{HoangBao.2024}), we reuse the snapshot of the evolution of a filamentary cloud under the effects of self-gravitating and B-fields taken from the MHD simulations by \cite{Ntormousi.2019}. The filamentary cloud was set up with a length of 66 pc and an effective thickness of 33 pc onto a three-dimensional data cube of $256^{3}$ pixels. The simulation box of 66 pc was chosen to cover the entire filamentary cloud, corresponding to a pixel size of $\sim 0.25$ pc. The initial mean B-field has a strength of $B=5\mu$G and is perpendicular to the filament axis. The filamentary cloud is mostly sub-\alfvenic with the mean $\langle M_{A} \rangle \sim 0.2 - 0.8$ (see in Paper I, \citealt{HoangBao.2024}).

\begin{figure*}
    \centering
    \includegraphics[width = 0.48\textwidth]{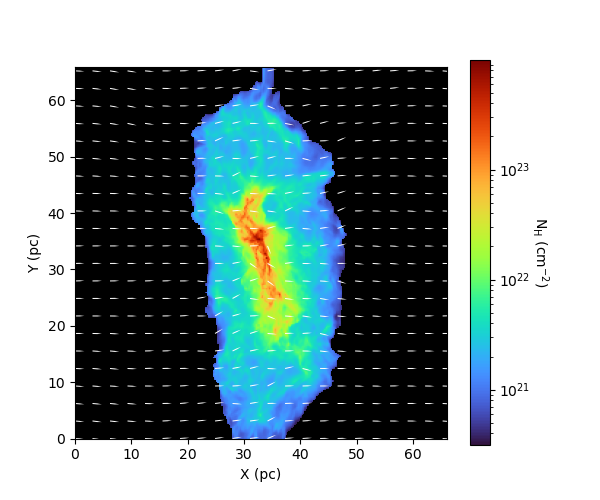}
    \includegraphics[width = 0.51\textwidth]{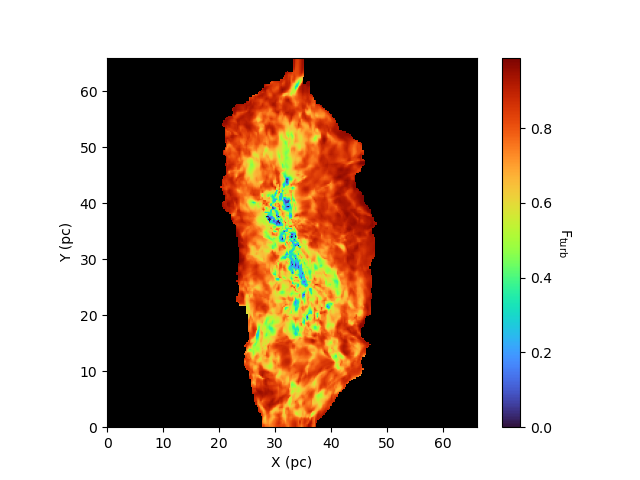}
    \caption{Left panel: The map of gas column density $N_{\rm H}$ of a filamentary cloud taken from the MHD simulations by \cite{Ntormousi.2019}. The white segments demonstrate the B-field morphology perpendicular to the filament axis. Right panel: The distribution of the magnetic turbulence factor $F_{\rm turb}$ across the filamentary cloud calculated from the MHD data by \cite{HoangBao.2024}.}
    \label{fig:NH_map}
\end{figure*}

The left panel of Figure \ref{fig:NH_map} shows the map of gas column density and the B-field morphology represented by the white segments. The gas column density is linearly correlated with the optical extinction $A_{\rm V}$ with $A_{\rm V}/N_{\rm H} \approx 6.23 \times 10^{-22}\,\rm mag\,cm^{2}$. The gas column density changes significantly within the filament, from $N_{\rm H} \sim 10^{21}-10^{22}\,\cm^{-2}$ in the outer region to $N_{\rm H} \sim 4-5\times 10^{23}\,\cm^{-2}$ in the central region, corresponding to the increase in visual extinction from $A_{\rm V} \sim 1 - 10$ to $A_{\rm V} > 100$. The variation of gas column density is expected to cause the variation of grain alignment and directly affect the observed starlight polarization and the results of inferred inclination angles of mean B-fields (see, e.g., \citealt{Whittet.2008}; \citealt{Andersson:2015bq}; \citealt{Vaillancourt:2020ch}; \citealt{Hoang.2021}).


The effect of magnetic turbulence is significant within the filamentary cloud and tends to increase toward the denser part of the cloud (see in Paper I, \citealt{HoangBao.2024}). We retake the magnetic turbulence factor $F_{\rm turb}$ derived from the deviation angle between local and mean B-fields from the MHD data by \cite{HoangBao.2024}, which is illustrated in the right panel of Figure \ref{fig:NH_map} (see Equation \ref{eq:F}). The factor $F_{\rm turb}$ is about $0.8 - 0.9$ in the outer cloud with weaker magnetic turbulence and lower dispersion angle $\delta\theta < 20^{\circ}$ (see Figure 14 in Paper I, \citealt{HoangBao.2024}), which can be approximated by $1 - 1.5(\delta\theta)^2$ (Figure \ref{fig:Fturb_approx}). The factor $F_{\rm turb}$ decreases to $< 0.5$ in high-density regions of the cloud having stronger magnetic fluctuations. The values of $F_{\rm turb}$ are included in the calculations of the starlight polarization efficiency and the inferred inclination angles of the mean B-fields in the filamentary cloud.



\subsection{Dust and Grain Alignment Models}
\label{sec:syn_dust_align}
In this work, we use the Astrodust model of interstellar dust in which silicate and carbonaceous components are well mixed together within one single dust population from \cite{Draine.2021no}. The grain size distribution is assumed to follow the Mathis-Rumpl-Nordsieck size distribution described by the power law of $dn \varpropto Ca^{-3.5}da$ in the range of grain sizes from $a_{\rm min} = 5\,\rm nm$ to $a_{\rm max} = 0.25\,\rm\mu m$ \citep{Mathis.1977}. The grain shape is considered to be oblate spheroid with an axial ratio of 1.4, which was previously constrained through the starlight polarization observation in the ISM environment (\citealt{Hensley.2023}). We take the cross-sections for this oblate shape from \cite{Draine.2021no}.\footnote{The variation of the grain elongation could directly affect both the starlight polarization efficiency $P/N_{\rm H}$ and the results of inferred inclination angles (see Appendix \ref{sec:appendix_ratio} for different grain axial ratios).}

We examine the effect of grain alignment on the synthetic starlight polarization and the inferred inclination angle by considering different grain alignment models as (1) Ideal RAT and (2) MRAT with different grain magnetic properties. For the radiation field, we consider grains to be irradiated by only the ISRF, which is described by the average radiation field in the solar neighborhood \citep{Mathis.1977}. Grains can then be aligned by RATs with the exponential alignment degree as shown in Equation \ref{eq:falign}, which is first implemented in the updated POLARIS. For grain magnetic properties, we consider both PM grains with the fraction of iron of $f_p = 0.1$, and SPM grains with increasing levels of iron inclusion $N_{\rm cl}$ from $50$ to $10^3$. The SPM grains are expected to achieve more efficient magnetic alignment than PM grains due to higher magnetic susceptibility, according to the MRAT mechanism (\citealt{HoangLaz.2016}, see also in Equation \ref{eq:fhighJ} - \ref{eq:fhiJ_deltam}). The assumption of alignment models is summarized in Table \ref{tab:align_model}.

Note that the presence of grain growth can significantly affect the multi-band starlight polarization, especially toward dense regions (e.g., \citealt{Whittet:2003p4164}; \citealt{Vaillancourt:2020ch}; \citealt{Hoang.2021}). We consider the variation of the maximum grain size from $a_{\rm max} = 0.25\,\rm\mu m$ to $a_{\rm max} = 1\,\rm\mu m$ and quantify their impact on the synthetic starlight polarization and the inferred inclination angles from the multi-wavelength polarization efficiency. Besides, the coagulation of small grains during the grain growth process could cause a deficit of smaller grain sizes (i.e., increasing $a_{\rm min}$), especially during the collapse of molecular cloud and the formation of stellar cores (see, e.g., \citealt{Ormel2011}; \citealt{Bate2022}). Nevertheless, these grains are much smaller than the aligned size $a_{\rm align}$ increasing with increasing gas density (i.e., $a_{\rm align} \varpropto n^{2/7}$, see Equation \ref{eq:align_ana}), and they are weakly aligned with B-fields with $f_{\rm align} < 0.1$ (see Figure \ref{fig:falign_a}). Hence, the increase in $a_{\rm min}$ less contributes to the total absorbed polarization and the calculation of inferred inclination angles, which is not considered in this numerical calculation.

\begin{table}[]
    \caption{Grain alignment models considered.}
    \begin{tabular}{l l l}
    \toprule
      Model  &  Aligned Sizes & Parameters \\
      \midrule
      Ideal RAT  &$[a_{\rm align},a_{\rm max}]$  & $R=1$ \cr
      MRAT   & $[a_{\rm align},a_{\rm max}]$ & PM, $f_{p}=0.1$ \cr
          & $[a_{\rm align},a_{\rm max}]$ &  SPM, $N_{\rm cl}=50, 10^{3}$ \cr
      \bottomrule
    \end{tabular}
    \label{tab:align_model}
\end{table}

\subsection{Modeling synthetic polarization from background stars}

To model the polarization of background stars, we first place a plane of unpolarized background stars with $N = 256 \times 256 $ behind the filamentary cloud. Each background source is considered to be a blackbody radiating at an effective temperature $T_{\rm eff} = 5000\,\rm K$. We assume that they are located at a long distance from the cloud so that their radiation does not affect the alignment of grains inside the cloud, which is determined by the interstellar radiation field only. We then model the synthetic polarization of background stars due to dichroic extinction by aligned grains at multi-bands in the Johnson-Cousin photometric system from optical (e.g., B-band: $\lambda_{\rm B} = 0.45\,\rm\mu m$; V-band: $\lambda_{\rm V} = 0.55\,\rm\mu m$; R-band: $\lambda_{\rm R} = 0.68\,\rm\mu m$) to NIR wavelengths (e.g., J-band: $\lambda_{\rm J} =1.22\,\rm\mu m$; H-band: $\lambda_{\rm H} = 1.63\,\rm\mu m$; K-band: $\lambda_{\rm K} = 2.19\,\rm\mu m$).

The RAT alignment physics was implemented in the POLARIS code by \cite{Reissl.2016}, whereas the MRAT mechanism was recently incorporated in the updated version by \cite{Giang.2023}. Here, we use the latest version of POLARIS code by \cite{Giang.2023} for our numerical studies. For a detailed description of the implementation of RAT alignment physics and polarized radiative transfer, please see \cite{Reissl.2016}. For a detailed description of the alignment model parameters, see \cite{Giang.2023}.


\section{Numerical Results}
\label{sec:results}

\subsection{Polarization Coefficient Fraction $f_{\rm pol}(\lambda)$}
\label{sec:fpol_result}

\begin{figure*}
    \centering
    \includegraphics[width = 0.48\textwidth]{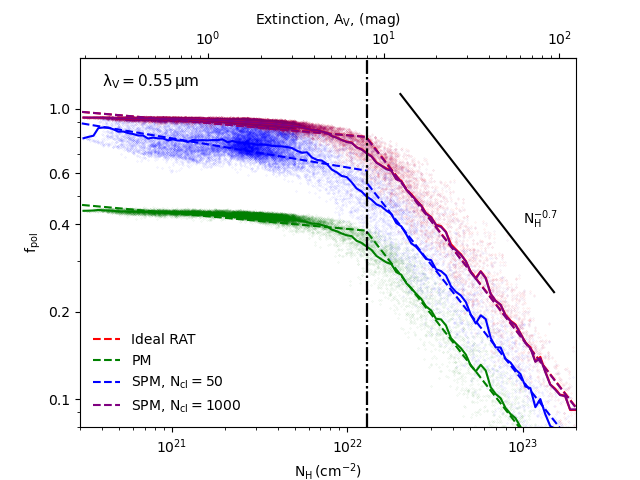}
    \includegraphics[width = 0.48\textwidth]{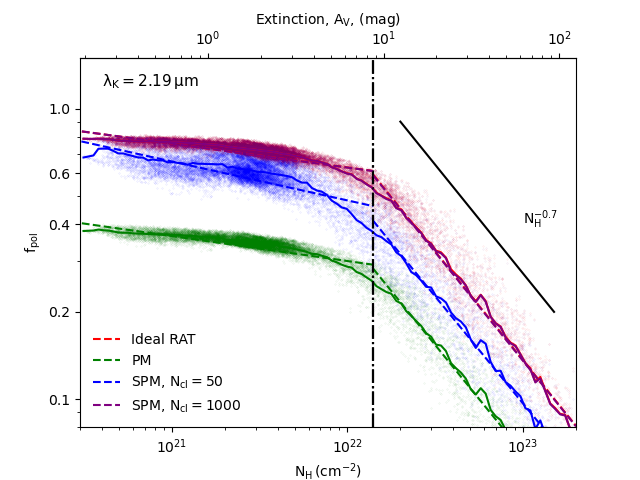}
    \caption{The variation of the polarization coefficient fraction $f_{\rm pol}$ calculated at V-band (i.e., $\lambda_{\rm V} = 0.55\,\rm\mu m$, left panel) and K-band (i.e., $\lambda_{\rm K} = 2.19\,\rm\mu m$, right panel) as a function of the gas column density $N_{\rm H}$. Different alignment models from Ideal RAT to MRAT alignment with PM and SPM grains with $N_{\rm cl} = 50, 1000$ and the maximum grain size $a_{\rm max} = 0.25\,\rm\mu m$ are considered. The optical extinction $A_{\rm V}$ is added at the top horizontal axis. The dashed-dotted black vertical line represents the transition column density where $f_{\rm pol}$ starts to drop due to grain alignment loss. The fraction $f_{\rm pol}$ is nearly constant in the outer cloud $N_{\rm H} < 1.3 \times 10^{22}\,\rm cm^{-2}$ (i.e., $A_{\rm V} < 8$) and decreases in the inner cloud with $f_{\rm pol} \varpropto N_{\rm H}^{-0.7}$ owing to increasing gas randomization with increasing density. The fraction $f_{\rm pol}$ is lower for PM grains, and can be enhanced if grains have high levels of iron inclusions $N_{\rm cl} > 50$. The results from MRAT alignment are the same as those in the Ideal RAT case for SPM grains having $N_{\rm cl} = 1000$.}
    \label{fig:fpol_Mag}
\end{figure*}

From the local properties of MRAT alignment (Section \ref{sec:method_grain_align}) and dust properties (Section \ref{sec:syn_dust_align}), we are able to determine the polarization coefficient fraction $f_{\rm pol}(\lambda)$ at a certain observed wavelength locally in the filamentary cloud (Equation \ref{eq:fpol}). Figure \ref{fig:fpol_Mag} shows the variation of $f_{\rm pol}$ at V- and K-bands with respect to the gas column density $N_{\rm H}$. The column density is calculated by integrating $n_{\rm H}$ from the MHD simulation along the sightline as $N_{\rm H} = \int_{\rm LOS} n_{\rm H}\,ds$. The visual extinction $A_{\rm V}$ is added at the top horizontal axis for comparison. The maximum grain size $a_{\rm max} = 0.25 \,\rm\mu m$ and different alignment models from Ideal RAT to MRAT for PM and SPM grains with iron inclusions are assumed. A power-law fit is performed to the running mean of $f_{\rm pol}$ to find the transition column density $N_{\rm trans}$ where $f_{\rm pol}$ starts to drop due to the loss of grain alignment and the slope of $f_{\rm pol}$ at $N_{\rm H} \gtrsim N_{\rm trans}$. For the Ideal RAT alignment, the fraction $f_{\rm pol}$ at V-band is around 0.9 in the outer region of the cloud with $N_{\rm H} < 1.3 \times 10^{22}\,\rm cm^{-2}$ (i.e., $A_{\rm V} < 8$), but cannot achieve to 1 since only grains larger than $a_{\rm align} \sim 0.01 - 0.02\,\rm\mu m$ can be aligned by RATs. Toward the inner region with $N_\H > 1.3 \times 10^{22} \,\cm^{-2}$ (i.e., $A_{\rm V} > 8$), only large grains from $a_{\rm align} \sim 0.1\,\rm\mu m$ to $a_{\rm max} = 0.25\,\rm\mu m$ can be aligned due to the increased gas randomization and the attenuated of ISRF (i.e., larger $a_{\rm align}$, see Equation \ref{eq:align_ana}). This results in the reduction of the alignment efficiency, and a significant decrease in $f_{\rm pol}$ with a slope of $\sim -0.7$. The patterns of $f_{\rm pol}$ are the same when observed at K-band, but the value is lower caused by the absorbed polarization of aligned grains $a < a_{\rm max} = 0.25\,\rm\mu m$ (see also Figure \ref{fig:fpol_wave} in Appendix \ref{sec:appendix_fpol_wave}).

The polarization coefficient fraction also depends on the grain magnetic properties, as shown in Figure \ref{fig:fpol_Mag}. For PM grains, they are aligned by RATs with a lower alignment degree $f_{\rm align}(a) \sim 0.475$ (see Section \ref{sec:method_align_degree}), then the fraction $f_{\rm pol}$ is considerably low in both V- and K-bands with $f_{\rm pol, max} \sim 0.4 - 0.5$ in the outer cloud. For SPM grains with high levels of embedded iron, they can achieve perfect magnetic alignment by MRAT with a higher alignment degree $f_{\rm align}(a) \sim 1$. Consequently, the total $f_{\rm pol}$ increases as the level of iron inclusions increases and can be close to the maximum value of 0.8 - 0.9 achieved in the Ideal alignment case for SPM grains having $N_{\rm cl} = 1000$.

\subsection{Synthetic Starlight Polarization}
\label{sec:pol_syn}
\begin{figure*}
    \centering
    \includegraphics[width = 0.48\textwidth]{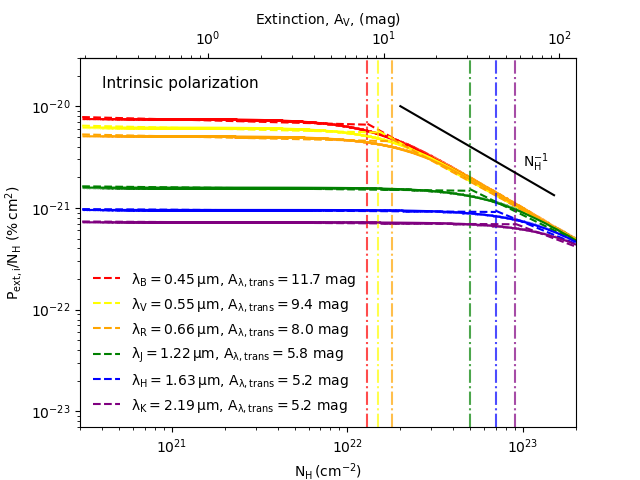}
    \includegraphics[width = 0.48\textwidth]{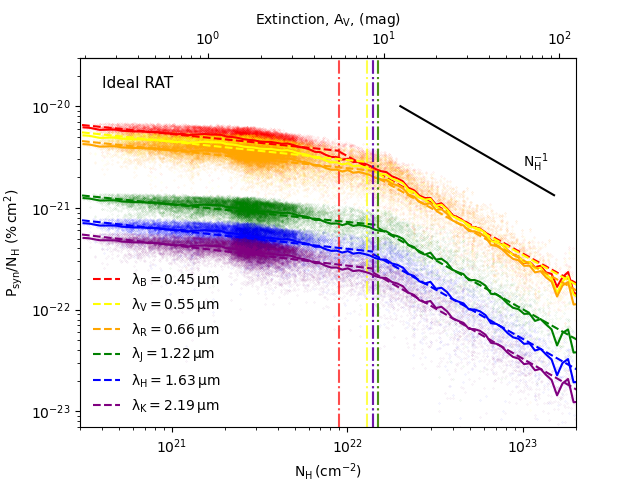}
    \caption{Left panel: The variation of the intrinsic polarization efficiency $P_{\rm ext,i}/N_{\rm H}$ (i.e., when all grain sizes have perfect alignment and the magnetic field is lying on the POS and well-ordered) at optical-NIR wavelengths with respect to the column density $N_{\rm H}$. The dashed-dotted color lines show the transition column density $N_{\rm H, trans}$ where the depolarization occurs. The slope of $-1$ is added for comparison. The intrinsic polarization is constant in the outer cloud, and drops in the inner dense cloud due to large polarized extinction ($\tau_{\rm pol} \gg 1$). Right panel: Similar but for the synthetic starlight polarization $P_{\rm syn}/N_{\rm H}$ when the Ideal RAT alignment and the realistic B-field morphology from MHD data (varying $\sin^2\gamma$ and $F_{\rm turb}$) are considered. The loss of grain alignment causes the reduction of $P_{\rm syn}/N_{\rm H} \varpropto N_{\rm H}^{-1}$ at high-density regions $N_{\rm H} > 1.3 \times 10^{22}\,\rm cm^{-2}$ ($A_{\rm V} > 8$).}
    \label{fig:P_NH_vsNH}
\end{figure*}

 The left panel of Figure \ref{fig:P_NH_vsNH} shows the change in the intrinsic polarization efficiency $P_{\rm ext, i}/N_{\H}$ in which all grain sizes are perfectly aligned within the cloud and the magnetic fields are lying on the POS (i.e., $\sin^2\gamma = 1$), as a function of the gas column density $N_{\rm H}$. The intrinsic polarization efficiency is calculated at multi-wavelength bands from optical (i.e., B-, V- and R-bands) to NIR (i.e., J-, H- and K-bands). The transition column density (i.e., where the polarization efficiency begins to drop with $P/N_{\rm H} \varpropto N_{\rm H}^{-1}$), denoted by $N_{\rm H, trans}$, and the slope at $N_{\rm H} \gtrsim N_{\rm H, trans}$ are estimated from the power-law fit to the running mean of the data. The extinction of each observed band at the transition column density, denoted by $A_{\rm\lambda, trans}$, is added for comparison. The intrinsic polarization efficiency is constant in the outer regions with $N_{\rm H} \lesssim N_{\rm H, trans}$, and decreases with increasing observed wavelength produced by the absorbed polarization of grains $a < 0.25\,\rm\mu m$. For high-density regions with $N_{\rm H} \gtrsim N_{\rm H, trans}$ and large polarized extinction ($\tau_{\rm pol} \gg 1$), most radiation from background stars is highly absorbed with the maximum starlight polarization degree to $P_{\rm ext,i} = 1$ (see Section \ref{sec:method_star_uniform}). Then, the polarization efficiency decreases with increasing $N_{\rm H}$. This effect is stronger for $P_{\rm ext, i}/N_{\H}$ at optical bands with high extinction $A_{\lambda} > 10$, corresponding to $N_{\rm H, trans} \approx 1.3 - 2 \times 10^{22}\,\cm^{-2}$ in the outer cloud. At NIR wavelengths, the polarized starlight radiation could be transparent to the dust with low extinction. The polarized extinction effect is then only significant at the densest part of the cloud with $N_{\rm H, trans}$ up to $\sim 10^{23}\,\cm^{-2}$ (i.e., $A_{\lambda} > 5$).




The right panel of Figure \ref{fig:P_NH_vsNH} shows the same for the synthetic starlight polarization efficiency $P_{\rm syn}/N_\H$, but when the Ideal RAT alignment model and the realistic B-field morphology from the MHD simulation (i.e., varying $\sin^2\gamma$ and $F_{\rm turb}$) are taken into account. Compared with the intrinsic ones, the values $P_{\rm syn}/N_\H$ are lower induced by a fraction of aligned grains $a > a_{\rm align} = 0.02 - 0.1\,\rm\mu m$. The loss of grain alignment is dominant in the denser region with $N_{\rm H, trans} \sim 1.3 \times 10^{22}\,\cm^{-2}$ (see also Figure \ref{fig:fpol_Mag}), resulting in a significant decrease in the polarization efficiency with a slope of $\sim -1$.

\begin{figure*}
    \centering
    \includegraphics[width = 0.48\textwidth]{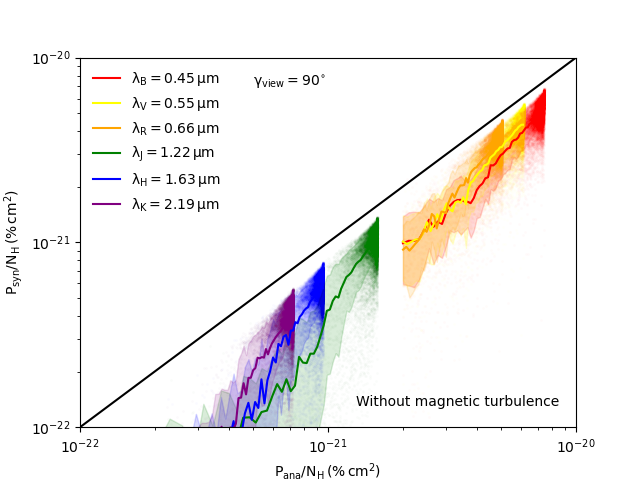}
     \includegraphics[width = 0.48\textwidth]{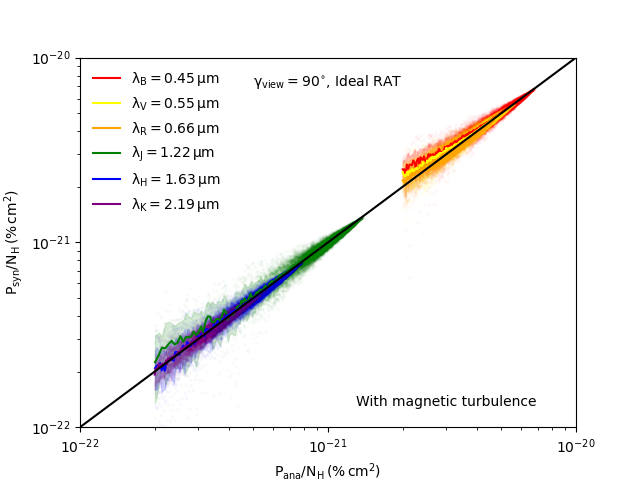}
    \caption{Synthetic polarization efficiency $P_{\rm syn}/N_\H$ from numerical calculations versus the analytical values $P_{\rm ana}/N_\H$ derived from Equation \ref{eq:p_NH_fpol}, considering two cases: without (i.e., $f_{\rm pol} = 1$ and $F_{\rm turb} = 1$, left panel) and with Ideal RAT alignment and magnetic turbulence (i.e., varying $f_{\rm pol}$ and $F_{\rm turb}$ from the MHD simulation, right panel). The analytical values are in agreement with the synthetic data at both optical and NIR bands when the effects of grain alignment and magnetic fluctuations are included.}
    \label{fig:P_NH_compare_wave}
\end{figure*}

To examine whether the inclination angles can be inferred from the synthetic data of the filamentary cloud when dust properties and magnetic fields are varying along the sightline, we calculate the polarization efficiency from the analytical model as shown in Equation \ref{eq:p_NH_fpol}, giving the multi-band intrinsic polarization efficiency in the left panel of Figure \ref{fig:P_NH_vsNH}, the mean fraction $f_{\rm pol}$ determined from our numerical calculations at optical-NIR wavelengths for the Ideal RAT alignment (Figure \ref{fig:fpol_Mag}), the mean inclination angles $\sin^2\gamma$ and the magnetic turbulence factor $F_{\rm turb}$ derived from the MHD simulation. We then compare with the synthetic polarization efficiency generated by the updated POLARIS, as illustrated in Figure \ref{fig:P_NH_compare_wave} at both optical and NIR bands. The effects of Ideal RAT alignment and magnetic turbulence across the cloud have been already incorporated during the synthetic modeling of $P_{\rm syn}/N_{\rm H}$ (\citealt{Giang.2023}, see the right panel of Figure \ref{fig:P_NH_vsNH}). For the analytical calculations of polarization efficiency $P_{\rm ana}/N_{\rm H}$, we consider two cases: without (i.e., $f_{\rm pol} = 1$ and $F_{\rm turb} = 1$, left panel) and with Ideal RAT alignment and magnetic turbulence from the MHD simulation (i.e., varying $f_{\rm pol}$ and $F_{\rm turb}$, right panel). For the first case, the analytical model mismatches with the synthetic observations, with $P_{\rm ana}/N_{\rm H}$ higher than $P_{\rm syn}/N_{\rm H}$. When the effects of grain alignment by RATs and magnetic fluctuations are included in the physical model of starlight polarization, the analytical values are well correlated with the synthetic ones at all optical-NIR wavelengths.


\subsection{Inferred inclination B-field angles from starlight polarization efficiency}
\label{sec:results_incl_PNH}
In Section \ref{sec:pol_syn}, we show the analytical model of polarization efficiency can well describe the synthetic polarimetric data once the impacts of grain alignment and magnetic turbulence are considered. We now apply the analytical model to calculate the inclination angles inferred from optical-NIR polarization efficiency $P/N_\H$ as presented in Equation \ref{eq:chi_ext}. We also discuss the major effects on the polarization efficiency and the inferred inclination angles as (1) the contribution of iron inclusions, (2) the viewing angles between mean fields and the LOS, and (3) the impact of grain growth.

\subsubsection{Using optical-NIR starlight polarization efficiency}
\begin{figure*}
    \includegraphics[width = 0.48\textwidth]{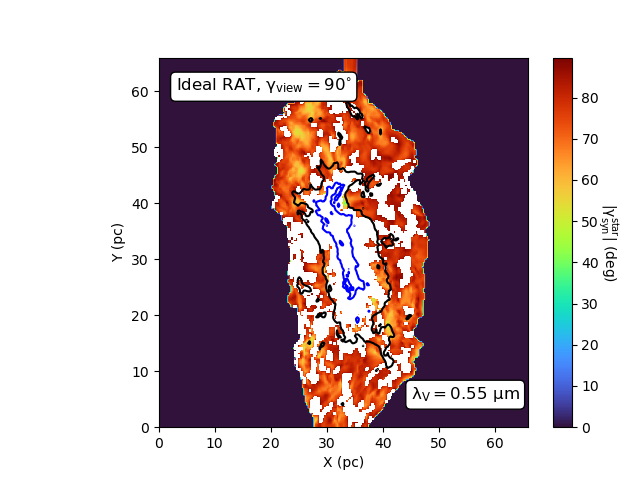}
    \includegraphics[width = 0.48\textwidth]{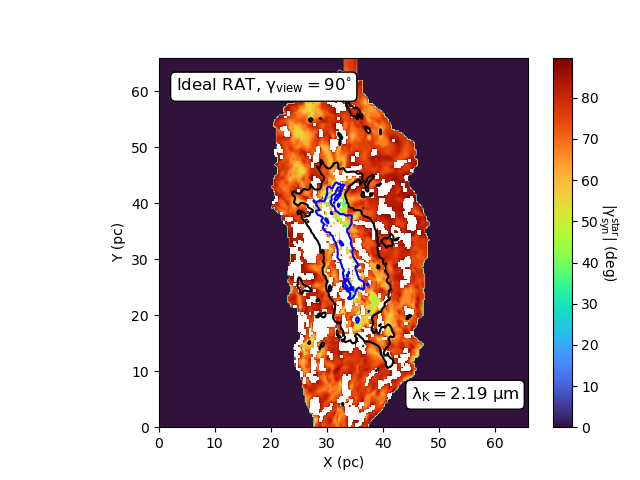}
    \caption{Maps of the inferred inclination angles $|\gamma^{\rm star}_{\rm syn}|$ from the polarization efficiency $P_{\rm syn}/N_\H$ at V-band (left panel) and K-band (right panel), considering the Ideal RAT alignment model and $\gamma_{\rm view} = 90^{\circ}$. At optical wavelengths, the inclination angles can be retrieved in the outer region of the cloud of $N_\H < 5 \times 10^{21}\,\cm^{-2}$ (i.e., $A_{\rm V} < 3$, black contour), and can be inferred in the high-density region up to $5 \times 10^{22}\,\cm^{-2}$ (i.e., $A_{\rm V} = 30$, blue contour) when being observed at NIR wavelengths.}
    \label{fig:Incl_Starpol_IdealRAT}
\end{figure*}

\begin{figure*}
    \centering
    \includegraphics[width = 1\textwidth]{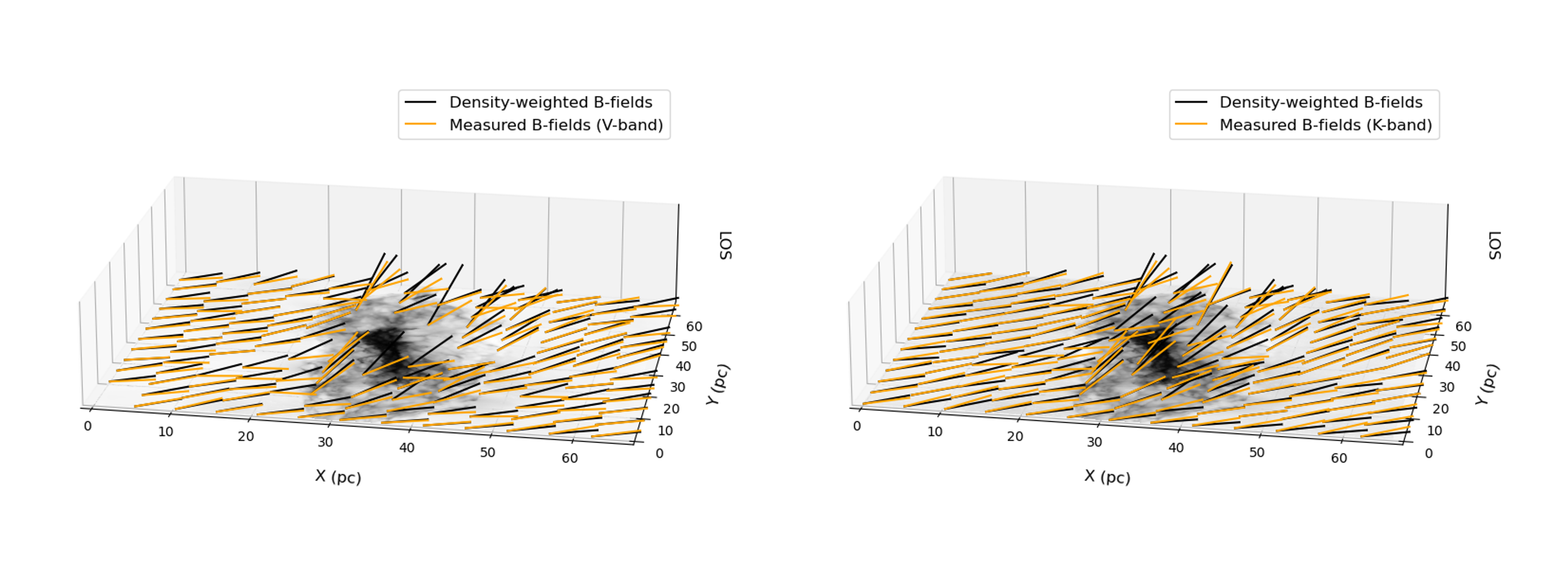}
    \caption{Illustrations of the orientation of mean B-fields with respect to the LOS retrieved from the inclination angles $|\gamma^{\rm star}_{\rm syn}|$ (orange lines) in the above Figure \ref{fig:Incl_Starpol_IdealRAT} at V-band (left panel) and K-band (right panel), in comparison with the true B-field orientations (black lines) derived from the density-weighted inclination angles $\gamma_{\rho}$. The starlight polarization efficiency maps $P/N_{\rm H}$ are placed in the POS (i.e., XY-plane), while the third axis represents the LOS with no distance information. The directions of mean fields are not provided owing to the unknown sign of $\gamma^{\rm star}_{\rm syn}$.}
    \label{fig:3DBfield_IdealRAT}
\end{figure*}

Figure \ref{fig:Incl_Starpol_IdealRAT} shows the maps of the inferred inclination angles $|\gamma^{\rm star}_{\rm syn}|$ from the V-band and K-band polarization efficiency $P_{\rm syn}/N_\H$ previously presented in Figure \ref{fig:P_NH_vsNH}. Figure \ref{fig:3DBfield_IdealRAT} demonstrates the orientation of mean B-fields with respect to the LOS derived from the above inferred angles $|\gamma^{\rm star}_{\rm syn}|$, in comparison with the true orientation from the MHD data. The colored contours illustrate the cloud regions where $A_{\rm V} = 3$ (black) and $A_{\rm V} = 30$ (blue). We determine the maximum column density $N_{\rm H, max}$ where the inclination angles can be retrieved from starlight polarization efficiency at each observed wavelength (i.e., $\sin^2\gamma \lesssim 1$). The NAN values (white pixels) in the inclination maps demonstrate the regions with $\sin^2\gamma > 1$ arising from the significant loss of grain alignment and the dominant magnetic turbulence, as predicted in Figure \ref{fig:sin_fpol_Fturb}. Nevertheless, the possibility of inferring inclination angles from starlight polarization efficiency depends on the observed wavelength. At optical bands (left panel), the polarized starlight is optically thin at low-density regions with $A_{\rm V} < 10$ (see the left panel of Figure \ref{fig:P_NH_vsNH}). Thus, the inclination angles of the mean B-fields can only be traced in the outer part of the cloud with $|\gamma_{\rm syn}^{\rm star}| \sim 75 - 80$ degrees, and the NAN pixels arise at $N_{\rm H} > N_{\rm H, max} \sim 5\times 10^{21}\,\cm^{-2}$ (i.e., $A_{\rm V} > 3$). The polarized starlight radiation at NIR wavelengths could be optically thin in high-density regions up to $A_{\rm V} \sim 50$. Subsequently, the inferred inclination angles can be traced further to the denser region with $N_\H$ up to $N_{\rm H, max} \sim 5 \times 10^{22}\,\cm^{-2}$ (i.e., $A_{\rm V} \sim 30$) (right panel). The values of $|\gamma^{\rm star}_{\rm syn}|$ tend to decrease to $\sim 40 - 60$ degrees caused by the back-and-forth bending of B-fields due to self-gravity, especially when clumps and dense cores are formed in this region (see \citealt{Ntormousi.2019}).

%

We then quantify our main method by comparing the inferred inclination angles from the synthetic starlight polarization efficiency with the true inclination angles of the mean B-fields derived from the MHD simulation. For each cell of the simulation box, the true inclination angles of B-fields are defined as
\bea
\sin{\gamma_{\rm B_{(i,j,k)}}}  = \sqrt{\frac{B_{x_{(i,j,k)}}^2 + B_{y_{(i,j,k)}}^2}{|B|_{(i,j,k)}^2}},\label{eq:gamma_B}
\ena
and the mean inclination angle is calculated by
\bea
\sin{\langle \gamma_{\rm B}} \rangle = \sqrt{\frac{\overline{B_x^2} + \overline{B_y^2}}{\overline{|B|^2}}}.\label{eq:gamma_meanB}
\ena

The inferred inclination angles can also be compared with the density-weighted of the mean inclination angle along the LOS, denoted by $\gamma_{\rho}$, which is determined by
\bea
\sin{\gamma_{\rho}}  = \sqrt{\frac{\int_{\rm LOS}\,n_\H \sin^2\gamma_{\rm B_{(i,j,k)}}\,\rm dz}{\int_{\rm LOS}\,n_\H\,\rm dz}},\label{eq:gamma_rho}
\ena
where $\gamma_{\rm B_{(i,j,k)}}$ are the local inclination angles derived from Equation \ref{eq:gamma_B}. 

%

\begin{figure*}
    \includegraphics[width = 0.48\textwidth]{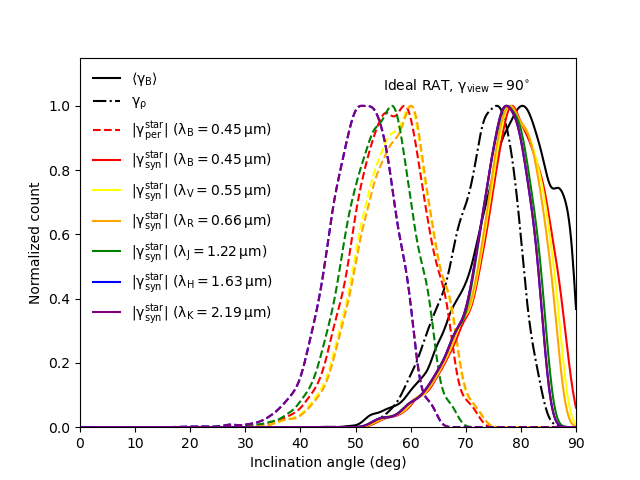}
    \includegraphics[width = 0.48\textwidth]{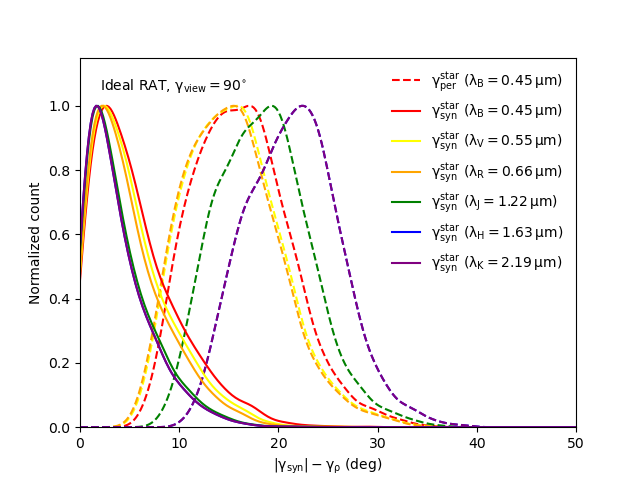}
    \caption{Left panel: Histogram of the inclination angles derived from our technique by using the polarization efficiency at optical-NIR wavelengths (solid color lines), compared with the true mean angles (solid black line) and the density-weighted angles (dashed black line) from the MHD data. The inclination angles in the perfect case (i.e., perfect alignment and without magnetic fluctuations) are added by dashed color lines, which differs from the realistic conditions in the MHD simulation (i.e., varying grain alignment and magnetic fluctuation across the cloud). Right panel: The distribution of the difference between the inferred inclination angles and the density-weighted angles $|\gamma_{\rm syn}| - \gamma_\rho$. The peak inclination angles derived from our technique incorporating grain alignment and magnetic fluctuations less deviate from the true value by $2^{\circ} - 6^{\circ}$}.
    \label{fig:Histogram_Starpol_wave}
\end{figure*}

Figure \ref{fig:Histogram_Starpol_wave} shows the distribution of the inferred inclination angles $|\gamma_{\rm syn}^{\rm star}|$ from our technique using the polarization efficiency $P/N_\H$ calculated at optical and NIR wavelengths (solid color lines), compared with the true angles (solid black line) and the density-weighted angles (dashed black line). The inferred angles $|\gamma_{\rm per}^{\rm star}|$ for the perfect case in which grain alignment is perfect and the magnetic fluctuations are minimal are added for comparison (dashed color lines). Table \ref{tab:incl_most_wave} summarizes the most probable values of the inclination angle from our technique and the true values from the MHD simulation. By considering the effects of grain alignment and magnetic fluctuations in the MHD simulation during model estimation (Figure \ref{fig:P_NH_compare_wave}), the peaks of inferred inclination angles retrieved from our technique are quite close to the mean inclination angles from the MHD data with a difference of only 2 - 6 degrees. Without these effects, the inferred angles are lower and more deviate by 10 - 30 degrees from the true value.


\begin{table*}
    \centering
    \caption{Summary of the most probable values (i.e., the peak inclination in the histograms) of the inclination angles derived from the starlight polarization efficiency $P/N_{\rm H}$ at optical and NIR bands in comparison with the true inclination angles in MHD simulation.}
    \begin{tabular}{c c c c c c c c}
    \toprule
      $\langle \gamma_{\rm B} \rangle^{\bar{\wedge}}$ & $ \gamma_{\rho}^{\bar{\wedge}}$ & $|\gamma_{\rm syn}^{\rm star, B^{\bar{\wedge}}}|$ & $|\gamma_{\rm syn}^{\rm star, V^{\bar{\wedge}}}|$ & $|\gamma_{\rm syn}^{\rm star, R^{\bar{\wedge}}}|$ & $|\gamma_{\rm syn}^{\rm star, J^{\bar{\wedge}}}|$ & $|\gamma_{\rm syn}^{\rm star, H^{\bar{\wedge}}}|$ & $|\gamma_{\rm syn}^{\rm star, K^{\bar{\wedge}}}|$\\
      \midrule
       $77.01^{\circ}$ & $73.04^{\circ}$ & $78.45^{\circ}$& $77.19^{\circ}$& $77.91^{\circ}$& $77.37^{\circ}$& $77.37^{\circ}$& $76.83^{\circ}$\\
      \bottomrule
    \end{tabular}
    \label{tab:incl_most_wave}
\end{table*}

\subsubsection{Effect of iron inclusions}
\label{sec:result_iron}

\begin{figure*}
    \centering
    \includegraphics[width = 0.48\textwidth]{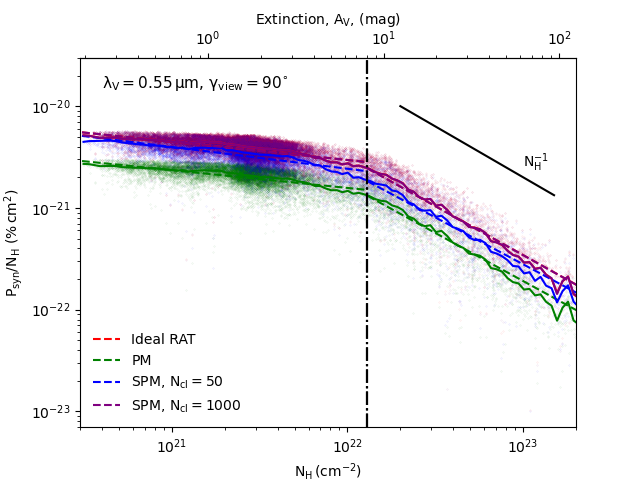}
    \includegraphics[width = 0.48\textwidth]{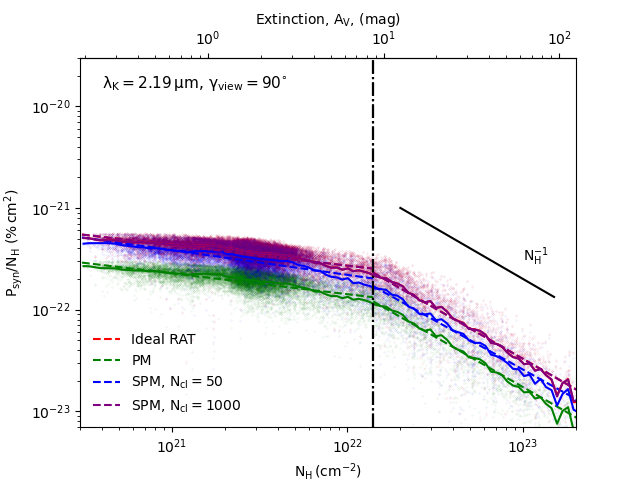}
    \caption{The variation of $P_{\rm syn}/N_{\rm H}$ at V-band (left) and K-band (right) with the gas column density $N_{\rm H}$ considering different alignment models and the viewing angle $\gamma_{\rm view} = 90^{\circ}$. The total polarization efficiency is lower for PM grains with lower alignment efficiency and can increase to the Ideal RAT case when grains are SPM and have high levels of iron inclusions.}
    \label{fig:P_NH_Mag}
    \includegraphics[width = 0.48\textwidth]{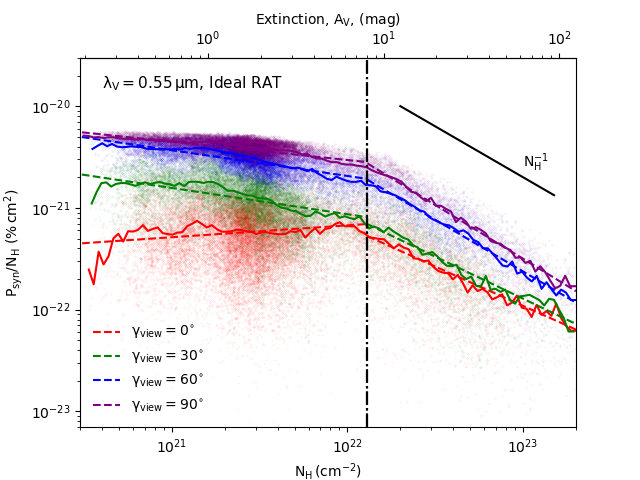}
    \includegraphics[width = 0.48\textwidth]{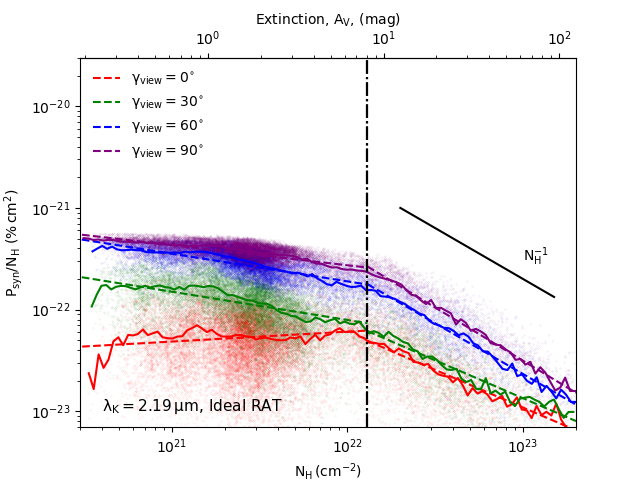}
    \caption{Similar to Figure \ref{fig:P_NH_Mag} but for various viewing angles $\gamma_{\rm view}$ and the Ideal RAT alignment. The effect of inclination angles is significant in the outer cloud with $N_{\rm H} < 1.3 \times 10^{22}\,\rm cm^{-2}$ ($A_{\rm V} < 8$), while the depolarization in the inner cloud is dominated by the grain alignment loss. The polarization efficiency increases with increasing $\gamma_{\rm view}$ (i.e., magnetic fields are perpendicular to the LOS).}
    \label{fig:P_NH_Incl}
\end{figure*}

The presence of iron inclusions inside grains enhances the magnetic alignment by MRAT (see \citealt{HoangLaz.2016}) and the synthetic results of polarization efficiency $P_{\rm syn}/N_\H$. Figure \ref{fig:P_NH_Mag} shows the variation of the polarization efficiency at V-band (left panel) and K-band (right panel) with respect to $N_{\rm H}$, considering Realistic alignment models as Ideal RAT alignment and MRAT alignment with PM and SPM grains with $N_{\rm cl} = 50$  and $N_{\rm cl} = 1000$. For the models of PM grains, the overall polarization efficiencies at optical and NIR bands are significantly low due to a lower alignment efficiency. For SPM grains with increasing levels of iron inclusions, the polarization efficiency is being enhanced for all wavelengths and can be recovered to the Ideal RAT case when $N_{\rm cl} \sim 1000$.

\begin{figure*}
    \centering
    \includegraphics[width = 0.48\textwidth]{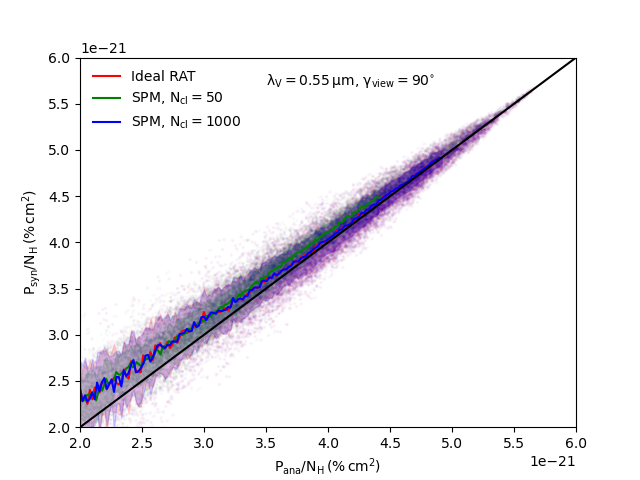}
    \includegraphics[width = 0.48\textwidth]{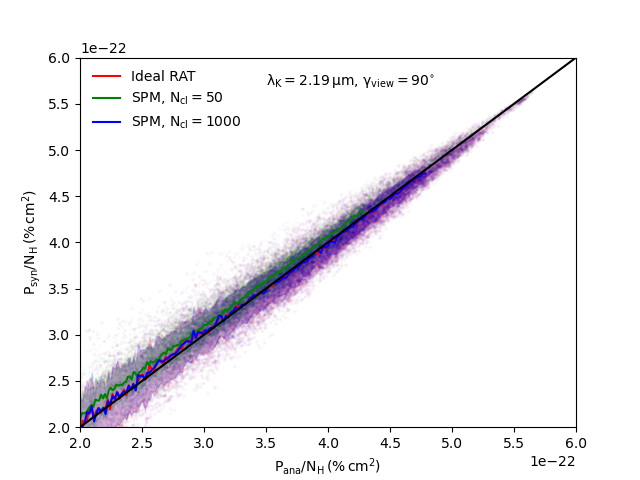}
    \caption{The comparison of the synthetic starlight polarization efficiency $P/N_\H$ at V-band (left panel) and K-band (right panel) with the analytical values for various models of grain alignment. The analytical calculations are well correlated with the synthetic data in all cases of grain alignment.}
    \label{fig:P_NH_compare_Mag}
    \includegraphics[width = 0.48\textwidth]{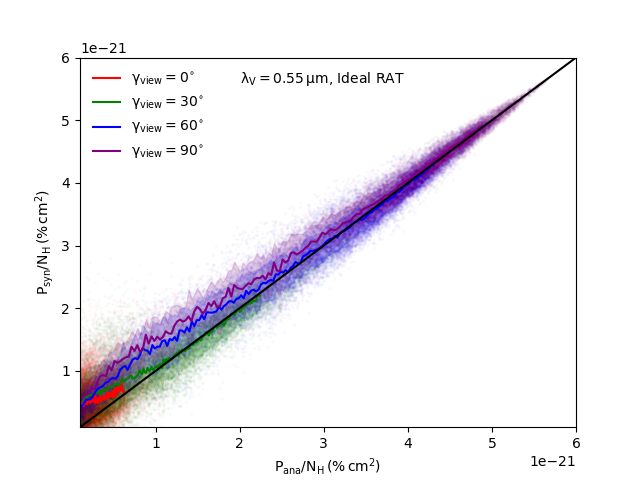}
    \includegraphics[width = 0.48\textwidth]{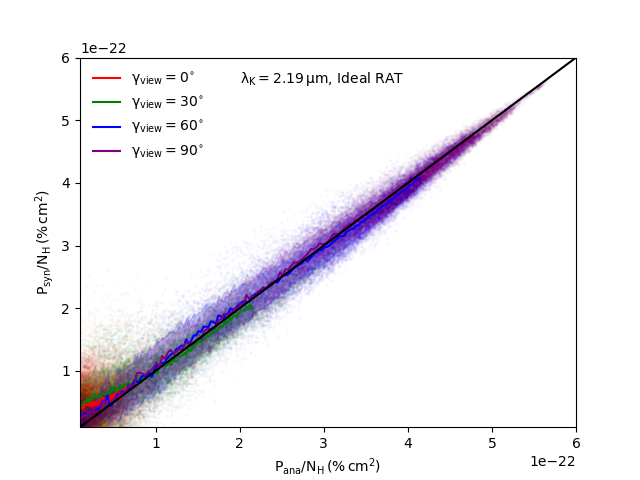}
    \caption{Similar to Figure \ref{fig:P_NH_compare_Mag}, but considering different viewing angles $\gamma_{\rm view} = 0^\circ - 90^\circ$ and the Ideal RAT alignment model.}
    \label{fig:P_NH_compare_Incl}
\end{figure*}

The comparison between the analytical polarization efficiency and the synthetic values for both Ideal RAT and Realistic alignment models are illustrated in the left panels of Figure \ref{fig:P_NH_compare_Mag} at V-band and K-band, assuming $\gamma_{\rm view} = 90^{\circ}$. When the magnetic properties of grains (i.e., PM or SPM grains) are characterized, the analytical calculations are in agreement with the synthetic polarimetric data for all alignment models.

\begin{figure*}
    \includegraphics[width = 1\textwidth]{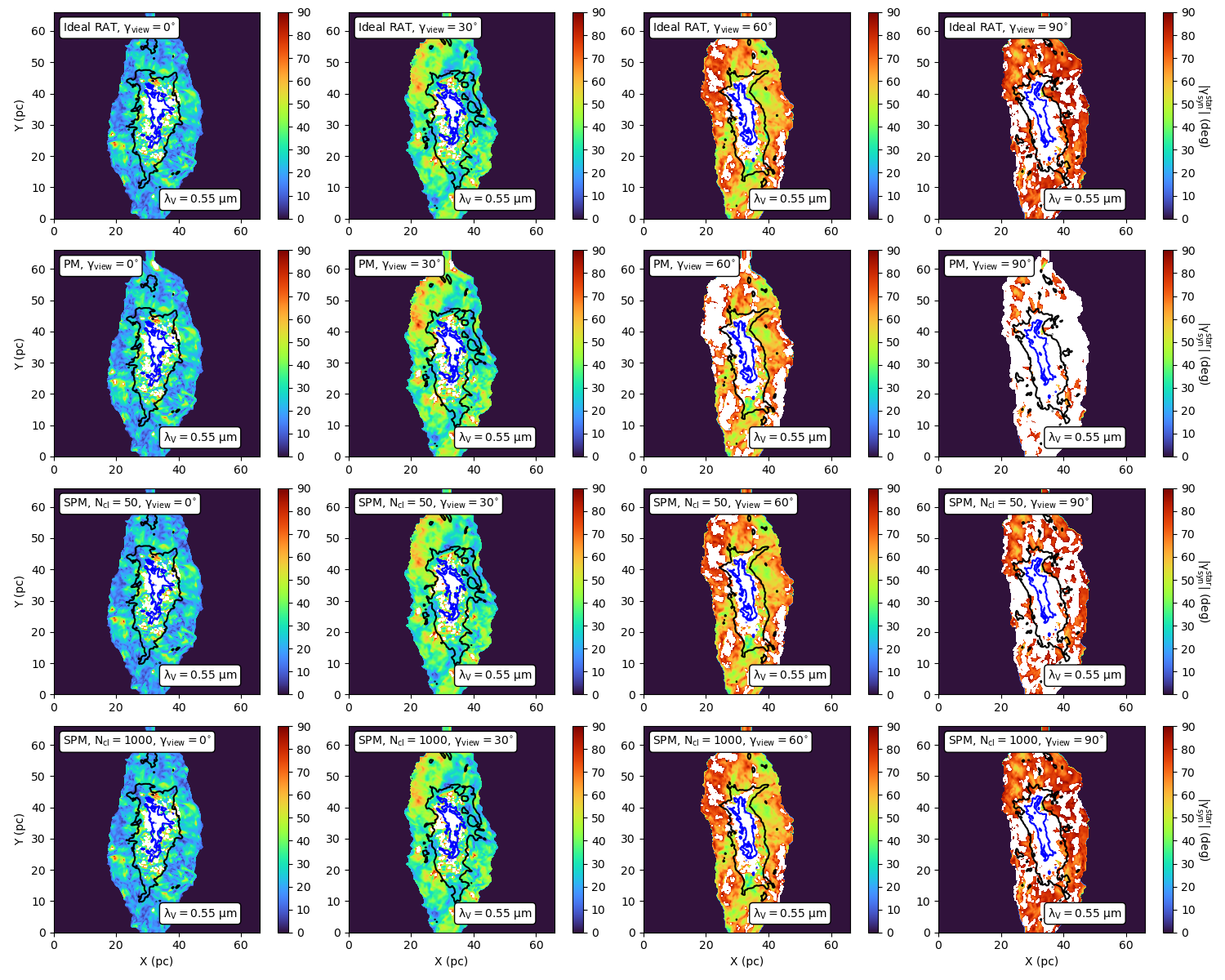}
    \caption{Resulting maps of the inclination angles inferred from the V-band polarization efficiency, assuming both Ideal and Realistic alignment models and the variation of the viewing angle $\gamma_{\rm view} = 0^\circ - 90^\circ$. The inferred values return fewer NAN values (white pixels) when grains are SPM with iron inclusions or smaller viewing angles $\gamma_{\rm view}$.}
    \label{fig:Incl_Starpol_Incl_Vband}
\end{figure*}

\begin{figure*}
    \includegraphics[width = 1\textwidth]{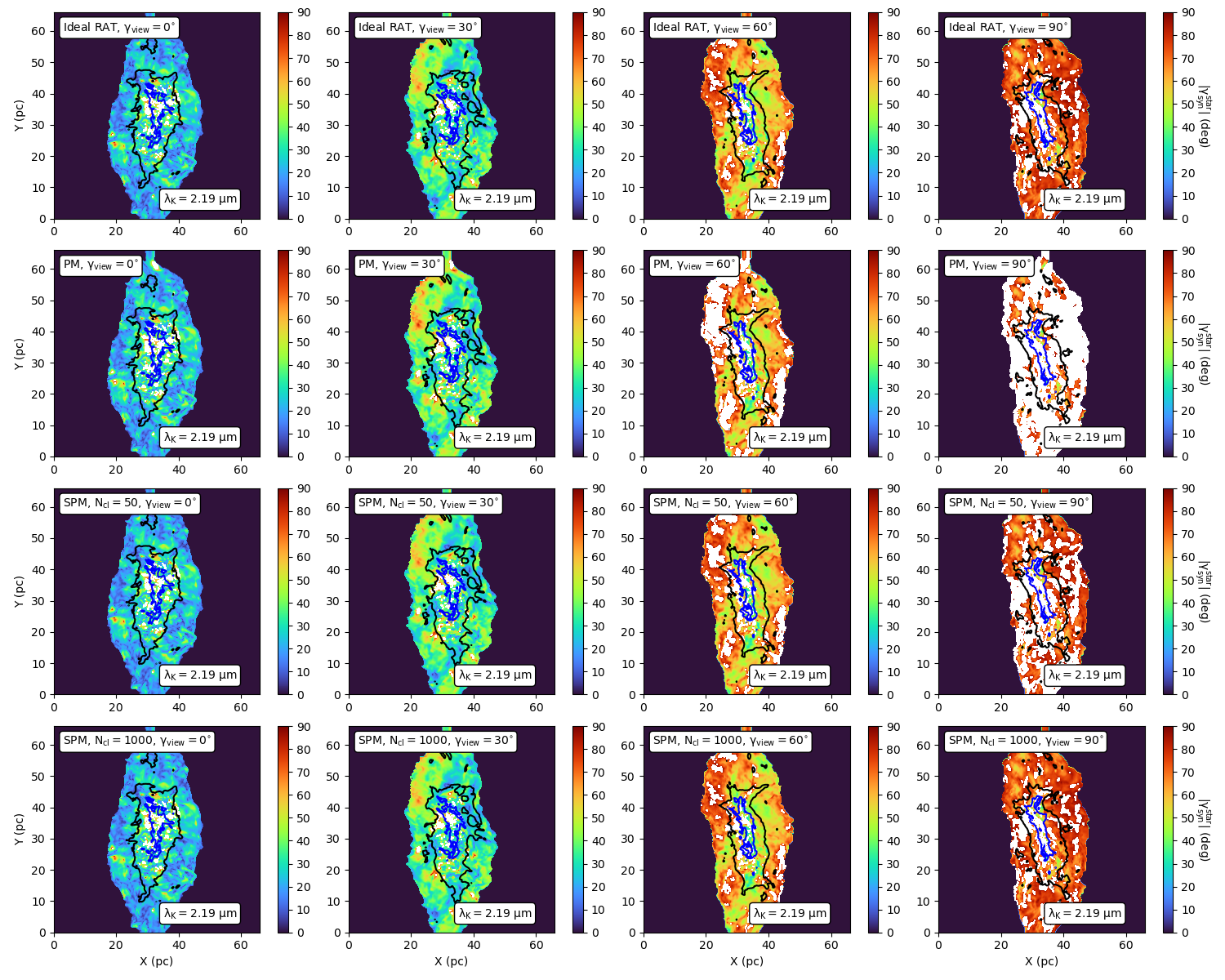}
    \caption{Same as Figure \ref{fig:Incl_Starpol_Incl_Vband} but tracing toward the denser region up to $A_{\rm V} = 30$ when being inferred from the K-band polarization efficiency.}
    \label{fig:Incl_Starpol_Incl_Kband}
\end{figure*}

Figure \ref{fig:Incl_Starpol_Incl_Vband} shows the inferred inclination angle maps extracted from the starlight polarization efficiency at V-band for different alignment models and viewing angles $\gamma_{\rm view}$. Figure \ref{fig:Incl_Starpol_Incl_Kband} shows the same, but tracing toward the denser region of the cloud $A_{\rm V} = 30$ when being observed at K-band. For PM grains, the inclination angle maps return more NAN values due to the reduced alignment efficiency with small $f_{\rm pol} < 0.4 - 0.5$ (see Figure \ref{fig:fpol_Mag}). The situation can be resolved when grains have high levels of iron inclusion with a higher fraction $f_{\rm pol} \sim 0.8 - 0.9$ due to the efficient alignment by MRAT, which results in fewer NAN values.

\begin{figure*}
    \centering
    \includegraphics[width = 0.48\textwidth]{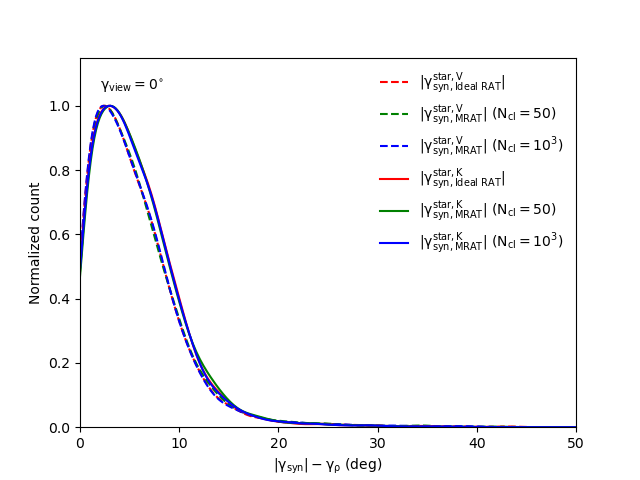}
    \includegraphics[width = 0.48\textwidth]{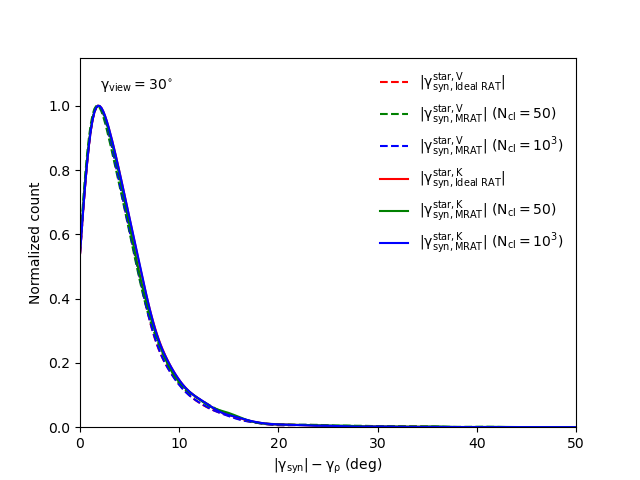}
    \includegraphics[width = 0.48\textwidth]{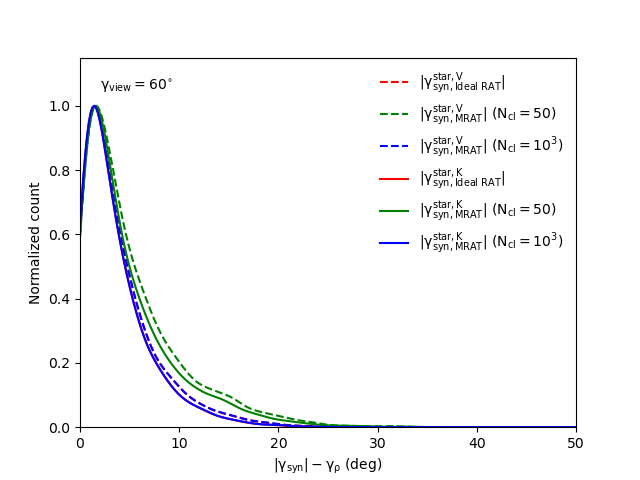}
    \includegraphics[width = 0.48\textwidth]{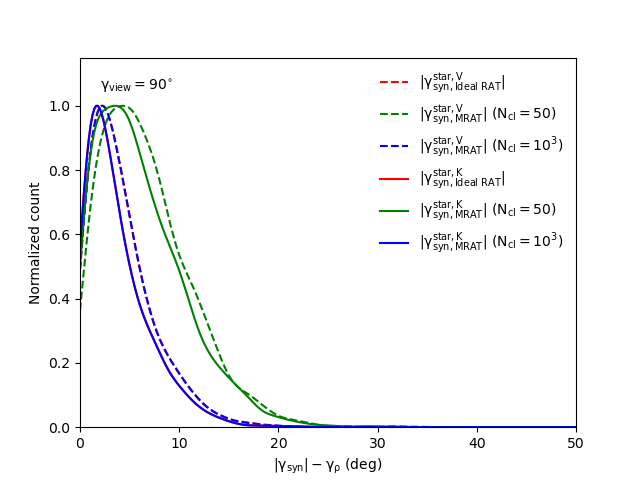}
    \caption{Histogram of the difference between the inferred inclination angles from the polarization efficiency at V-band (dashed color lines) and K-band (solid color lines) with the true mean inclination angles $\gamma_{\rho}$ for different alignment models and viewing angles. Regardless of grain magnetic properties and viewing angles, our method can accurately retrieve the inclination angles close to the true values from the MHD simulation with the difference of $2 - 6$ degrees.}
    \label{fig:Histogram_Starpol_Incl}
\end{figure*}

The histogram of the inferred inclination angles $|\gamma_{\rm syn}^{\rm star}|$ obtained from the starlight polarization at V- and K-bands for different grain alignment models and viewing angles in comparison with the mean inclination angles from the MHD data are illustrated in Figure \ref{fig:Histogram_Starpol_Incl}, while the peak values of the distribution are presented in Table \ref{tab:incl_most_incl}. Given known magnetic properties of grains (PM or SPM grains with iron inclusions), we can accurately determine the inferred inclination angles from the synthetic polarimetric observations, which are less different than the true values from the MHD simulation by $2^{\circ} - 6^{\circ}$.


\begin{table*}[]
\centering
    \caption{Same as Table \ref{tab:incl_most_wave} but for different grain alignment models and viewing angles $\gamma_{\rm view}$ between the LOS and the mean field.}
    \begin{tabular}{c c c c c c c c c}
    \toprule
      $\gamma_{\rm view}$ & $\langle \gamma_{\rm B} \rangle^{\bar{\wedge}}$  & $ \gamma_{\rho}^{\bar{\wedge}}$  & $|\gamma_{\rm syn, Ideal\,\,RAT}^{\rm star,V^{\bar{\wedge}}}|$ & $|\gamma_{\rm syn, MRAT}^{\rm star,V^{\bar{\wedge}}}|$ & $|\gamma_{\rm syn, MRAT}^{\rm star,V^{\bar{\wedge}}}|$ & $|\gamma_{\rm syn, Ideal\,\,RAT}^{\rm star,K^{\bar{\wedge}}}|$ & $|\gamma_{\rm syn, MRAT}^{\rm star,K^{\bar{\wedge}}}|$ & $|\gamma_{\rm syn, MRAT}^{\rm star,K^{\bar{\wedge}}}|$\\

      & & & & $(\rm N_{\rm cl} = 50)$& $(\rm N_{\rm cl} = 1000)$& & $(\rm N_{\rm cl} = 50)$& $(\rm N_{\rm cl} = 1000)$ \\
      \midrule
       $0^{\circ}$ & $11.36^{\circ}$ & $29.4^{\circ}$& $23.62^{\circ}$& $24.34^{\circ}$& $23.62^{\circ}$& $25.61^{\circ}$& $23.98^{\circ}$& $25.79^{\circ}$\\
       
       $30^{\circ}$ & $38.4^{\circ}$ & $42.38^{\circ}$& $42.56^{\circ}$& $42.38^{\circ}$& $42.56^{\circ}$& $42.38^{\circ}$& $42.02^{\circ}$& $42.38^{\circ}$\\
       
       $60^{\circ}$ & $67.81^{\circ}$ & $60.6^{\circ}$& $64.75^{\circ}$& $67.09^{\circ}$& $64.57^{\circ}$& $63.84^{\circ}$& $66.37^{\circ}$& $64.02^{\circ}$\\
       
       $90^{\circ}$ & $77.01^{\circ}$ & $73.04^{\circ}$& $77.19^{\circ}$& $79^{\circ}$& $77.19^{\circ}$& $76.83^{\circ}$& $81.34^{\circ}$& $76.65^{\circ}$\\
      \bottomrule
    \end{tabular}
    \label{tab:incl_most_incl}
\end{table*}

\subsubsection{Effect of viewing angles}

The viewing angle $\gamma_{\rm view}$ between the mean B-fields and the LOS can directly impact the calculation of the polarization efficiency $P_{\rm syn}/N_\H$ at V-band and K-band within the filamentary cloud, as illustrated in Figure \ref{fig:P_NH_Incl}. The effect of inclination angles on the polarization efficiency is dominant in the outer cloud with $N_{\rm H} < 1.3 \times 10^{22}\,\rm cm^{-2}$ (i.e., $A_{\rm V} < 8$), while the depolarization caused by grain alignment loss is dominant in the inner dense cloud (see Figure \ref{fig:fpol_Mag} and \ref{fig:P_NH_vsNH}). For large $\gamma_{\rm view}$, the polarization efficiency is higher as the mean B-fields are uniform and perpendicular to the LOS with $\sin^2\gamma \sim 1$. For small $\gamma_{\rm view}$, the projection effect by the inclined cloud-scale B-fields is significant, resulting in a much lower polarization efficiency. And the synthetic results are well correlated with the analytical calculations, as illustrated in Figure \ref{fig:P_NH_compare_Incl} for various viewing angles.


The impact of viewing angles $\gamma_{\rm view}$ on the calculation of the inferred inclination angles retrieved from $P_{\rm syn}/N_\H$ are also shown in Figure \ref{fig:Incl_Starpol_Incl_Vband} and \ref{fig:Incl_Starpol_Incl_Kband}. For large $\gamma_{\rm view} \sim 60^{\circ} - 90^{\circ}$, most of B-fields are perpendicular to the LOS. The depolarization is then dominated by the grain alignment loss and the effect of magnetic fluctuations, resulting in more NAN values in the inclination maps. For small $\gamma_{\rm view} \sim 0^{\circ} - 30^{\circ}$, the depolarization by cloud-scale inclined B-fields is more prominent, especially in dense regions where B-fields are bent due to the impact of self-gravity. As a result, it will return fewer NAN values when applying our new technique, as predicted in Figure \ref{fig:sin_fpol_Fturb}. In spite of these variations, the values of the inclination angle derived from our method are close to the true inclination angles from the MHD data in all cases of viewing angle, as shown in both Figure \ref{fig:Histogram_Starpol_Incl} and Table \ref{tab:incl_most_incl}.

\subsubsection{Effect of grain growth}
\label{sec:result_graingrowth}

\begin{figure*}
    \centering
    \includegraphics[width = 0.48\textwidth]{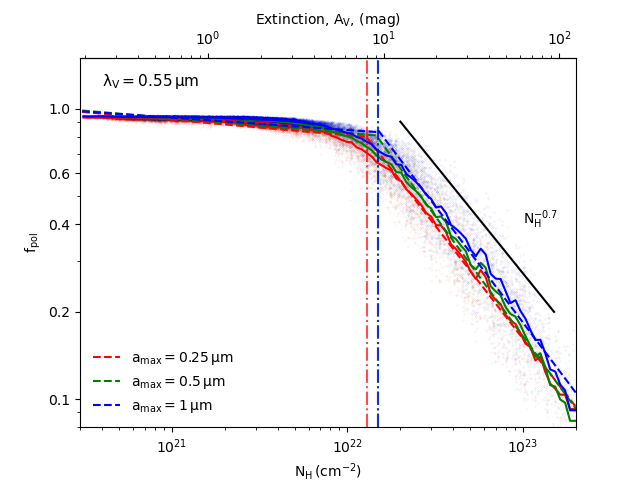}
    \includegraphics[width = 0.48\textwidth]{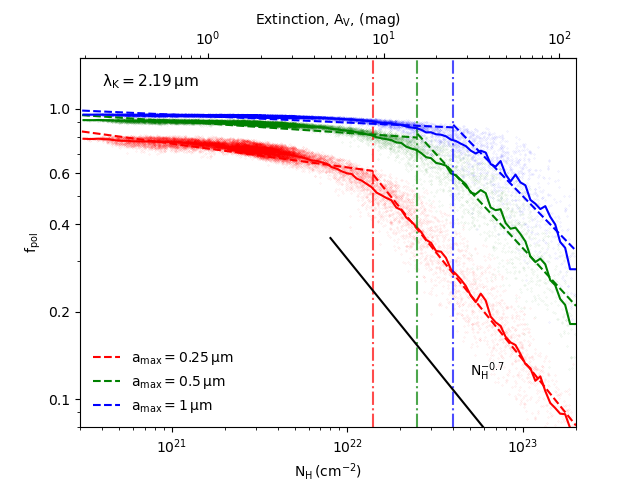}
    \caption{The impact of grain growth on the variation of $f_{\rm pol}$ at V-band (left panel) and K-band (right panel), considering $a_{\rm max} = 0.25 - 1\,\rm\mu m$. With increasing $a_{\rm max}$, larger grains $a > 0.25\,\rm\mu m$ can be aligned with B-fields by RATs, particularly in the inner cloud, resulting in a higher $f_{\rm pol}$ and an increase in the transition column density $N_{\rm H, trans}$. This effect is more prominent when measured at NIR wavelengths.}
    \label{fig:fpol_amax}
\end{figure*}

In this section, we examine the major impact of grain growth on the polarization coefficient fraction and the synthetic polarization efficiency. Figure \ref{fig:fpol_amax} shows the impact of grain growth on the variation of $f_{\rm pol}$ over gas column density $N_{\rm H}$, as the maximum grain size increases from $a_{\rm max} = 0.25\,\rm\mu m$ to $a_{\rm max} = 1\,\rm\mu m$. The presence of grain growth allows for the alignment of larger grains in the inner dense regions, which shows the increase in $N_{\rm trans}$ and a higher $f_{\rm pol} $ at all optical-NIR wavelengths. This effect is more prominent when observed at the K-band, with a significant increase in the transition column density from $N_{\rm H, trans} \sim 1.3 \times 10^{22}\,\cm^{-2}$ to $N_{\rm H, trans} \sim 4 \times 10^{22}\,\cm^{-2}$ when the maximum grain size increases to $1\,\rm\mu m$.

\begin{figure*}
    \centering
    \includegraphics[width = 0.48\textwidth]{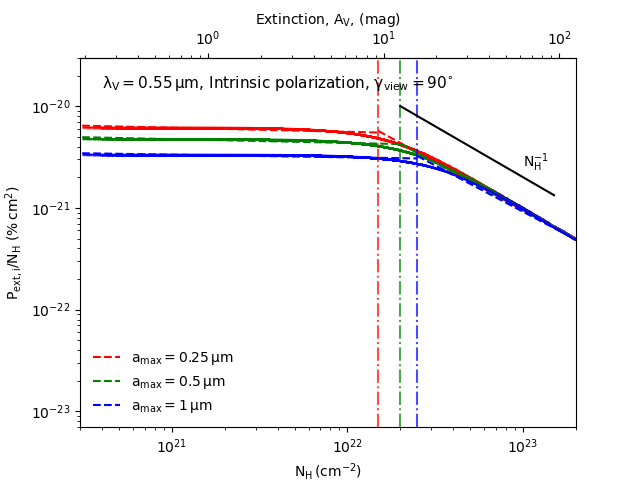}
    \includegraphics[width = 0.48\textwidth]{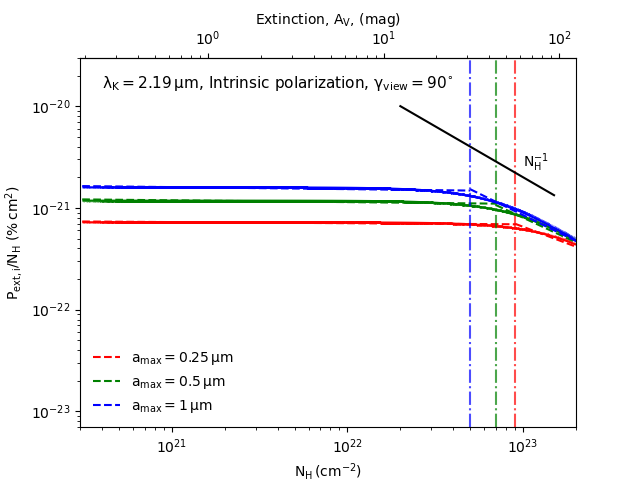}
    \caption{The effect of grain growth on the intrinsic polarization efficiency $P_{\rm ext,i}/N_{\rm H}$ at V-band (left panel) and K-band (right panel). The increasing grain sizes contribute to the decrease in the transition column density $N_{\rm H, trans}$ for the loss of K-band polarization efficiency by the polarized extinction effect, while it increases for V-band polarization efficiency.}
    \label{fig:P_NH_amax_Intrinsic}
    \includegraphics[width = 0.48\textwidth]{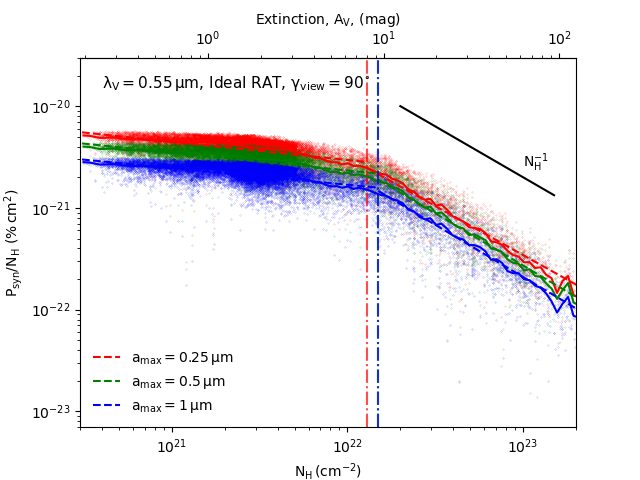}
    \includegraphics[width = 0.48\textwidth]{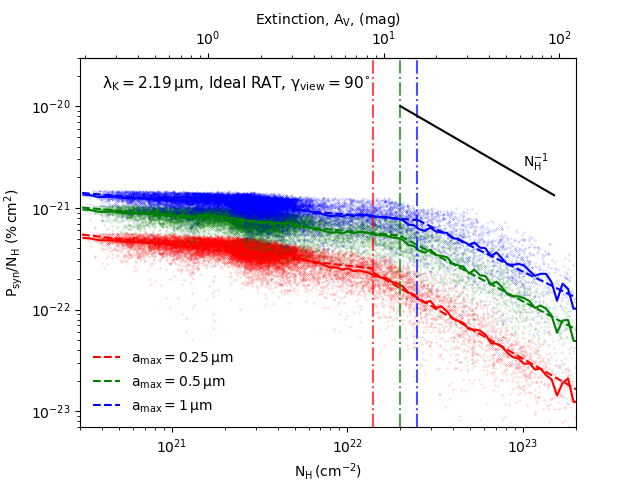}
    \caption{Same as Figure \ref{fig:P_NH_amax_Intrinsic} but for the synthetic polarization efficiency $P_{\rm syn}/N_{\rm H}$ considering the Ideal RAT alignment model. The alignment of larger grains $a > 0.25\,\rm\mu m$ is enhanced as $a_{\rm max}$ increases, leading to the increased transition column density for grain alignment loss to the inner dense cloud.}
    \label{fig:P_NH_amax_Ideal}
\end{figure*}

Figure \ref{fig:P_NH_amax_Intrinsic} shows the effect of grain growth on the intrinsic polarization efficiency $P_{\rm ext,i}/N_\H $ within the cloud at V-band and K-band, respectively. With the contribution of a higher population of larger grains, the absorbed polarization of background stars is more prominent at NIR than at optical wavelengths, with the shift of the peak wavelength of the starlight polarization toward NIR regimes (see also Figure \ref{fig:Pol_Spec_amax} in Appendix \ref{sec:appendix_spectrum}). This leads to a significant reduction of V-band intrinsic polarization efficiency and an increase in the K-band intrinsic polarization efficiency. Besides, the increasing grain sizes cause a decrease in optical extinction and an increase in NIR extinction. Therefore, the transition column density $N_{\rm H, trans}$ for the depolarization by extinction at V-band increases from $10^{22}$ to $2.5 \times 10^{22}\,\cm^{-2}$, whereas that transition value for K-band depolarization decreases from $\sim 10^{23}$ to $5 \times 10^{22}\,\cm^{-2}$.

Figure \ref{fig:P_NH_amax_Ideal} shows similar but for the synthetic V-band and K-band polarization efficiency $P_{\rm syn}/N_\H$ considering the Ideal RAT alignment. With increasing $a_{\rm max}$, the enhanced alignment of larger grains can produce higher polarization efficiency. This effect is more important in the inner dense cloud and causes the increase in $N_{\rm H, trans}$, especially at NIR wavelengths (see in Figure \ref{fig:fpol_amax}). For instance, for the K-band polarization efficiency, the contribution of large aligned grains causes the shift of the transition column density from $N_{\rm H, trans} \sim 1.3 \times 10^{22}\,\cm^{-2}$ to $N_{\rm H, trans} \sim 2.5 \times 10^{22}\,\cm^{-2}$.

\begin{figure*}
    \centering
    \includegraphics[width = 1\textwidth]{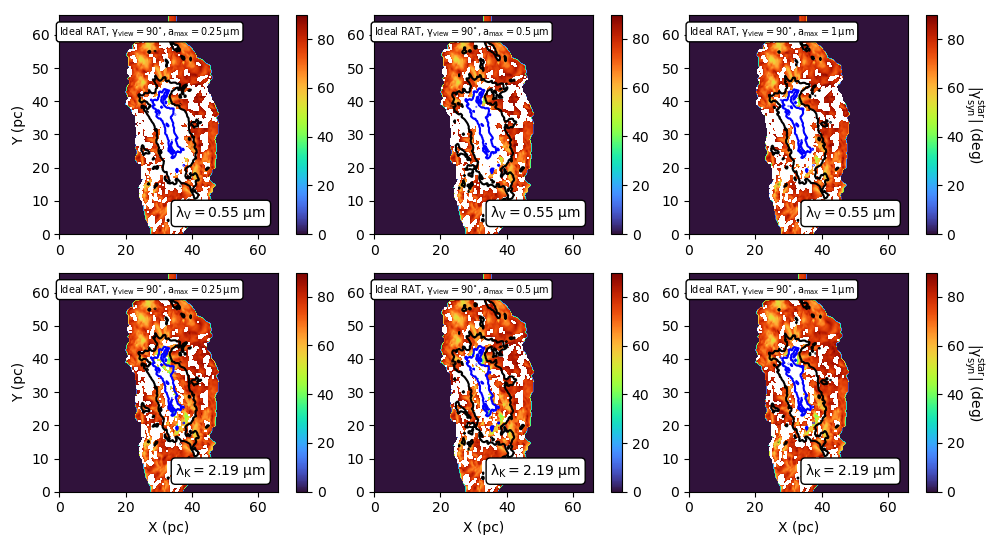}
    \caption{The resulting maps of inferred inclination angles from the V-band and K-band polarization efficiency for increasing $a_{\rm max} = 0.25 - 1\,\rm\mu m$. The values of $|\gamma^{\rm star}_{\rm syn}|$ seem to be unchanged once the grain growth effect on starlight polarization efficiency is considered.}
    \label{fig:Incl_Starpol_amax_Ideal}
\end{figure*}

Figure \ref{fig:Incl_Starpol_amax_Ideal} shows the results of the inclination angles retrieved from the V-band and K-band polarization efficiency under the effect of grain growth, assuming the case of Ideal RAT alignment and $\gamma_{\rm view} = 90^{\circ}$. The results of inferred inclination angles from our method seem to be unchanged when the impact of grain growth on grain alignment and the polarization efficiency within the filamentary cloud are all considered (Figure \ref{fig:P_NH_amax_Ideal}).


\subsection{Inferred Inclination Angles Using Starlight Polarization Integral}

As discussed in Section \ref{sec:method_polint}, the mean B-field inclination angles can potentially be retrieved from the observed starlight polarization integral $\Pi_{\rm obs}$. In this section, we present the synthetic results of the starlight polarization integral $\Pi_{\rm syn}$ and the derived inclination angles $\gamma^{\rm spec}_{\rm syn}$ as shown in Equation \ref{eq:incl_pi_obs}.

\subsubsection{Synthetic Starlight Polarization Integral}
\begin{figure}
    \includegraphics[width = 0.48\textwidth]{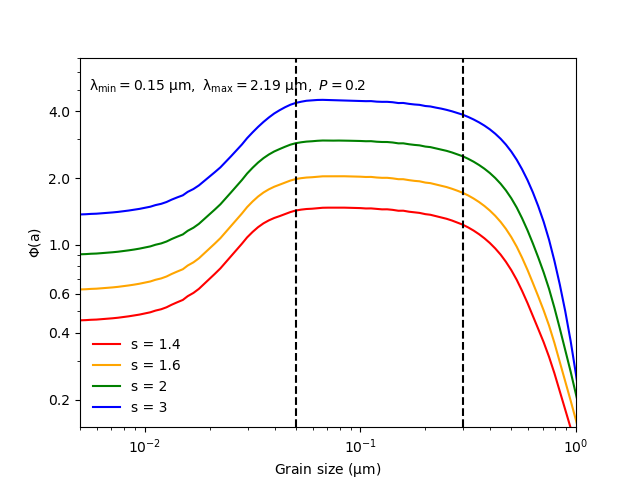}
    \caption{The starlight polarization efficiency integral $\Phi(a)$ over the wavelengths $\lambda = 0.15 - 2.19\,\rm\mu m$ as a function of grain sizes for the model of Astrodust by \cite{Draine.2021no}. Different axial ratios $s = 1.4 - 3$ for oblate grains, and the porosity of $20\%$ are assumed. The value of $\Phi(a)$ increases with increasing grain sizes, and becomes constant at the size range $a = 0.05 - 0.25\,\rm\mu m$ highlighted by black dashed lines. For large grains $a > 0.25\,\rm\mu m$, the polarization efficiency integral drops to $< 0.4$. The polarization efficiency integral is much higher when grains are highly elongated $s > 2$.}
    \label{fig:Pi_a}
\end{figure}

Figure \ref{fig:Pi_a} shows the starlight polarization efficiency integral $\Phi(a)$ from the Astrodust model calculated over the size range from $a_{\rm min} = 5\,\rm nm$ to $a_{\rm max} = 1\,\rm\mu m$ (\citealt{Draine.2021no}). The integral is taken over the range of wavelength from $\lambda_{\rm min} = 0.15\,\rm\mu m$ (UV) to $\lambda_{\rm max} = 2.19\,\rm\mu m$ (NIR) where the starlight polarization is observably strong. The oblate shape is considered with increasing axial ratio $s = 1.4 - 3$. One can see that the integral increases with increasing grain sizes and becomes saturated for grains with $a \sim 0.05 - 0.25\,\rm\mu m$. For large grains $a > 0.25 \,\rm\mu m$, they mostly polarize starlight radiation in longer wavelengths $\lambda > 1\,\rm\mu m$, then the value of $\Phi(a)$ significantly drop to $< 0.4$. And the polarization efficiency integral is higher to $\Phi(a) \sim 2-4$ when the grain shape is highly elongated with a large axial ratio $s > 2$. 

\begin{figure*}
    \centering
    \includegraphics[width = 0.48\textwidth]{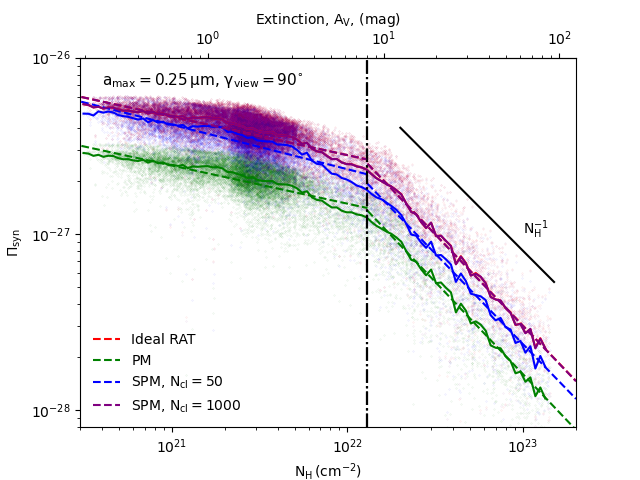}
    \includegraphics[width = 0.48\textwidth]{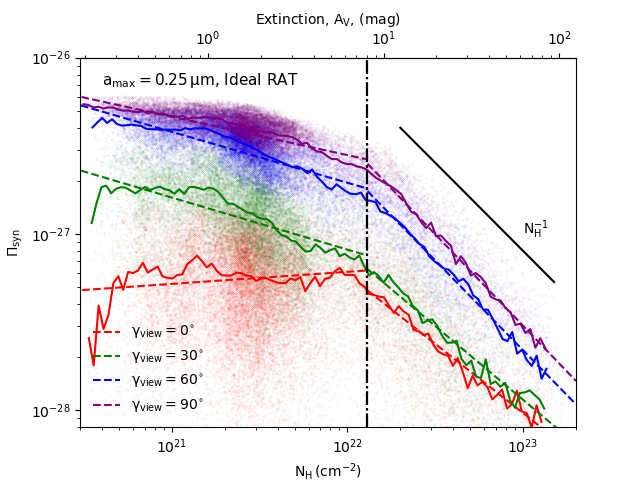}
    \caption{Left panel: The synthetic starlight polarization integral $\Pi_{\rm syn}$ vs. the gas column density $N_{\rm H}$. The starlight polarization integral is derived from the starlight polarization spectrum at $\lambda = 0.15 - 2.19\,\rm\mu m$. Both Ideal RAT alignment and MRAT alignment for PM and SPM grains are considered. The integral decreases toward the denser region of the cloud with $N_{\rm H} > 1.3 \times 10^{22}\,\rm cm^{-2}$ ($A_{\rm V} > 8$) caused by the grain alignment loss. The integral is significantly reduced for PM grains and becomes higher for SPM grains with high levels of iron inclusions. Right panel: Similar but for various viewing angles $\gamma_{\rm view} = 0 - 90^{\circ}$. The effect of inclined B-fields contributes to the decrease in starlight polarization integral for small $\gamma_{\rm view}$.}
    \label{fig:Pi_obs_Mag_Incl}
\end{figure*}

The left panel of Figure \ref{fig:Pi_obs_Mag_Incl} shows the change in the synthetic starlight polarization integral $\Pi_{\rm syn}$ calculated from the optical-NIR polarization spectrum of background stars (see Appendix \ref{sec:appendix_spectrum}) across the filamentary cloud (i.e., increasing $N_{\rm H}$), assuming both Ideal RAT alignment and Realistic models of PM and SPM grains. The right panel shows similar but for various viewing angles $\gamma_{\rm view}$. In the outer cloud with $N_{\rm H} < 1.3 \times 10^{22}\,\rm cm^{-2}$ ($A_{\rm V} < 8$), the effect of inclination angles on the starlight polarization integral is more prominent, while the depolarization by grain alignment loss dominates in the inner cloud with $\Pi_{\rm syn} \varpropto N_{\rm H}^{-1}$. The starlight polarization integral is much lower when grains are paramagnetic and increases when grains are superparamagnetic and have higher levels of iron inclusions. In addition, the mean B-field inclination angles impact the overall optical-NIR polarization spectrum and the observed starlight polarization integral, leading to a low $\Pi_{\rm syn}$ for small $\gamma_{\rm view}$, and becomes higher when the mean fields are perpendicular to the LOS (i.e., larger $\gamma_{\rm view}$).

\begin{figure*}
    \centering
    \includegraphics[width = 0.48\textwidth]{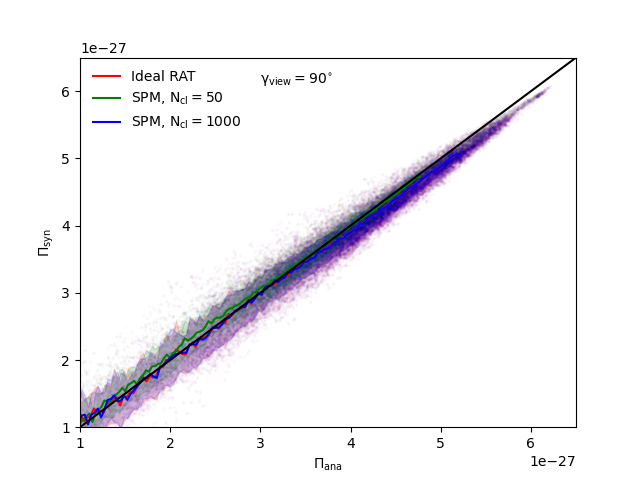}
    \includegraphics[width = 0.48\textwidth]{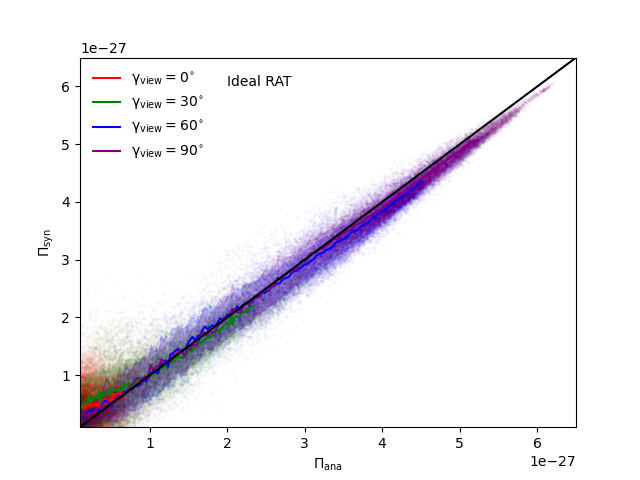}
    \caption{The analytical calculations of the starlight polarization integral $\Pi_{\rm ana}$ from Equation \ref{eq:Pi_obs} in comparison with the synthetic data when the variation of grain alignment models (left panel) and viewing angles (right panel) is taken into consideration. The analytical model describes well the synthetic integral taken from the starlight polarization spectrum.}
    \label{fig:Pi_compare}
\end{figure*}

Figure \ref{fig:Pi_compare} shows the comparison of the synthetic starlight polarization integral and the analytical one derived from Equation \ref{eq:Pi_obs} for different alignment models (left panel) and various viewing angles (right panel). As dust and magnetic properties are taken into account (Equation \ref{eq:Pi_obs_Fturb}), the analytical calculations of starlight polarization integral are in agreement with the synthetic data in all cases.

\begin{figure}
    \includegraphics[width = 0.48\textwidth]{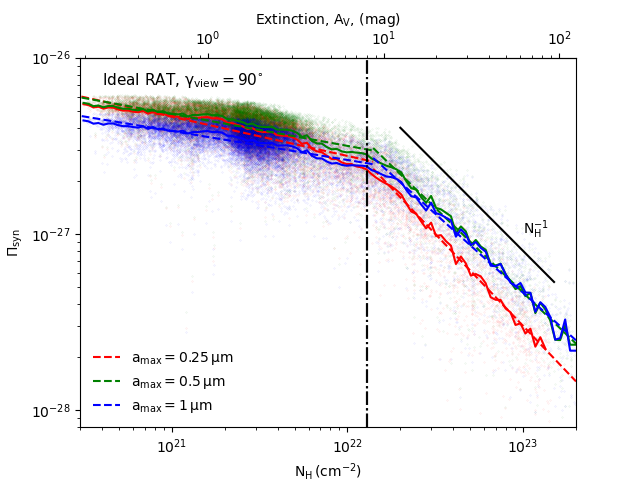}
    \caption{Same as Figure \ref{fig:Pi_obs_Mag_Incl} when the grain growth effect is taken into consideration. In the outer cloud with $N_{\rm H} < 1.3 \times 10^{22}\,\rm cm^{-2}$, the values of $\Pi_{\rm syn}$ decreases with increasing $a_{\rm max}$. In the inner cloud, the enhanced alignment of larger grains with increasing $a_{\rm max}$ causes the increase in $\Pi_{\rm syn}$ with a shallower slope of $< - 1$.}
    \label{fig:Pi_obs_amax}
\end{figure}

The effect of grain growth is taken into account in the numerical calculations of the starlight polarization integral, as presented in Figure \ref{fig:Pi_obs_amax}. In low-density regions with $N_{\rm H} < 1.3 \times 10^{22}\,\rm cm^{-2}$ and efficient alignment (Figure \ref{fig:fpol_amax}), the starlight polarization integral $\Pi_{\rm syn}$ decreases due to the decreasing polarization efficiency integral to $\Phi(a) < 0.4$ with increasing grain sizes $a > 0.25\,\rm\mu m$ (see Figure \ref{fig:Pi_a}). In high-density regions with $N_{\rm H} > 1.3 \times 10^{22}\,\rm cm^{-2}$, the impact of enhanced alignment for large grains with increasing $a_{\rm max}$ dominates. The values of $\Pi_{\rm syn}$ then become higher as $a_{\rm max}$ increases with a shallower slope of $< -1$ - the same effect as in the polarization efficiency presented in Figure \ref{fig:P_NH_amax_Ideal}.

\subsubsection{Inferred inclination angles from starlight polarization integral}
\label{sec:results_incl_Pi}
\begin{figure*}
    \includegraphics[width = 1\textwidth]{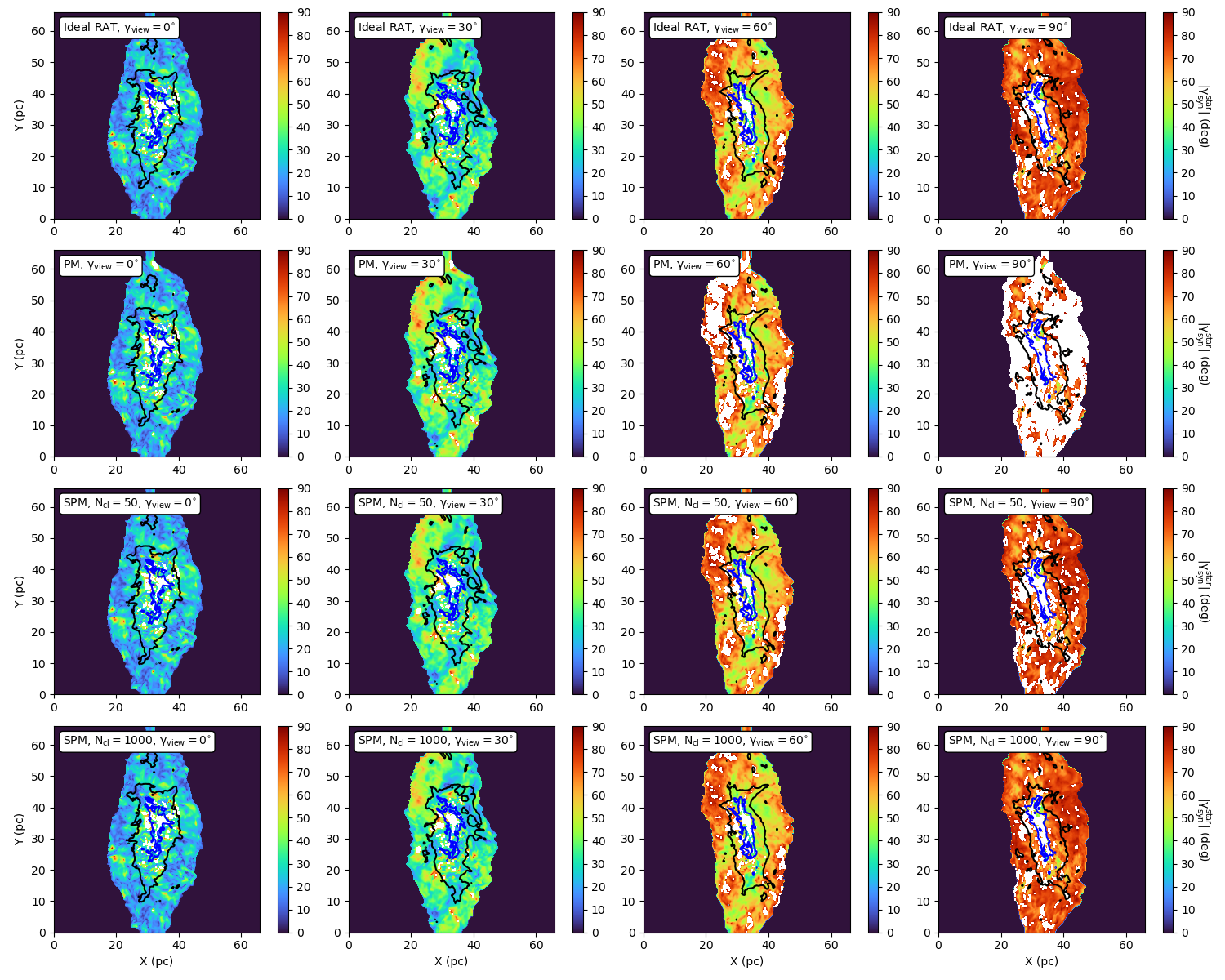}
    \caption{The maps of inferred inclination angles retrieved from the synthetic starlight polarization integral $\Pi_{\rm syn}$ for various alignment models and viewing angles. The results are insensitive to the wavelength and could be retrieved to the densest part of the cloud, in comparison with the previous ones using starlight polarization efficiency in Figure \ref{fig:Incl_Starpol_Incl_Vband} and \ref{fig:Incl_Starpol_Incl_Kband}.}
    \label{fig:Incl_spec_Incl}
\end{figure*}

Figure \ref{fig:Incl_spec_Incl} shows the maps of inferred inclination angles $|\gamma^{\rm spec}_{\rm syn}|$ from the synthetic starlight polarization integral for different alignment models and viewing angles $\gamma_{\rm view}$. In comparison with the previous technique using single-band polarization efficiency $P/N_\H$ in Figure \ref{fig:Incl_Starpol_Incl_Vband} and \ref{fig:Incl_Starpol_Incl_Kband}, the results are independent of the wavelength and can be retrieved up to the densest part of the cloud with $A_{\rm V} \approx 30$. Due to the dependence of starlight polarization integral on grain magnetic properties, the results show more NAN values when grains are paramagnetic, and becomes fewer when grains have embedded iron. Besides, the viewing angles and the fluctuations of magnetic fields affect the synthetic starlight polarization integral and the results of inferred inclination angles, which show fewer NAN values toward the denser region for small $\gamma_{\rm view}$. 

\begin{figure*}
    \centering
    \includegraphics[width = 0.48\textwidth]{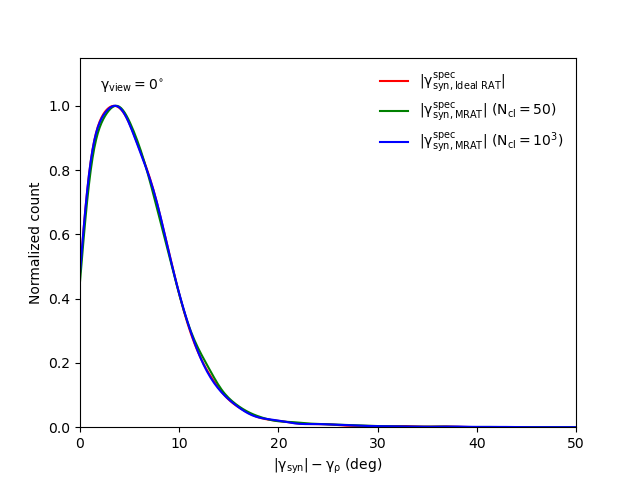}
    \includegraphics[width = 0.48\textwidth]{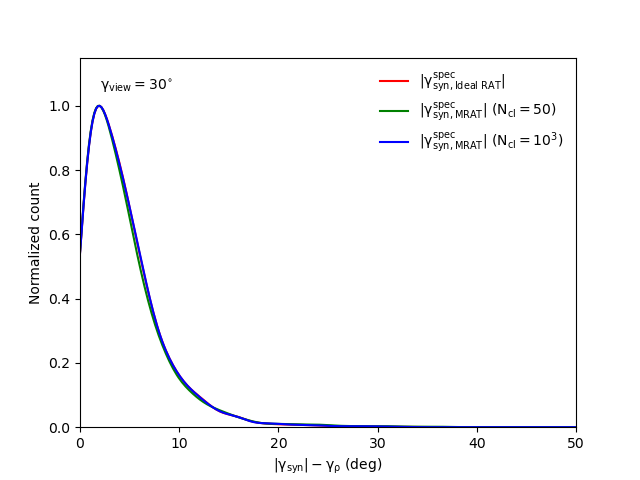}
    \includegraphics[width = 0.48\textwidth]{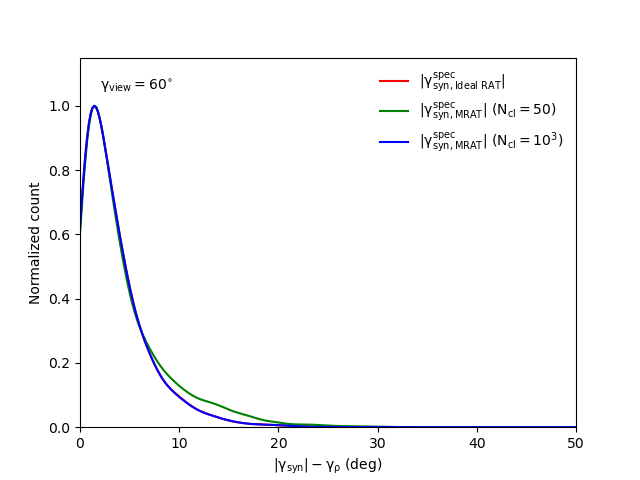}
    \includegraphics[width = 0.48\textwidth]{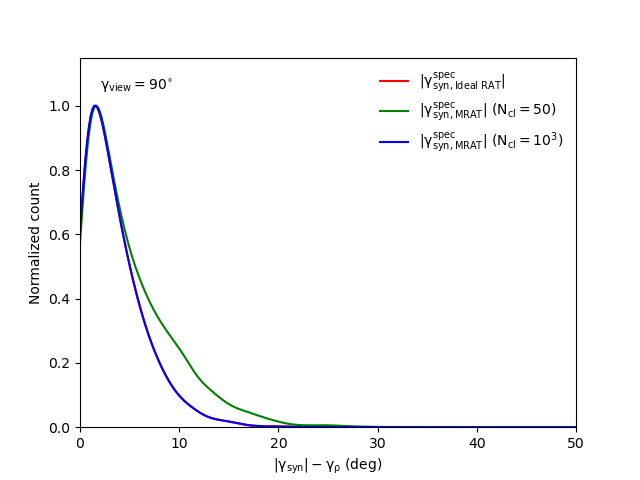}
    
    \caption{Similar to Figure \ref{fig:Histogram_Starpol_Incl} but for the inclination angles retrieved from the starlight polarization integral. As the effects of grain alignment, dust properties, and magnetic fluctuations are considered, the inferred results are in better agreement with the true inclination angle from MHD data for all grain alignment models and viewing angles.}
    \label{fig:Histogram_Spec_Incl}
\end{figure*}

The distribution of $|\gamma^{\rm spec}_{\rm syn}|$ and its probable values for various alignment models and viewing angles are summarized in Figure \ref{fig:Histogram_Spec_Incl} and Table \ref{tab:incl_most_spec}. Similar to the values of inferred inclination angle from the polarization efficiency obtained in Figure \ref{fig:Histogram_Starpol_Incl} and Table \ref{tab:incl_most_incl}, the results are less different than the mean values from the MHD simulation by 2 - 6 degrees as the effect of grain alignment and magnetic turbulence are both considered during the model estimation (Figure \ref{fig:Pi_compare}).

\begin{figure*}
    \includegraphics[width = 1\textwidth]{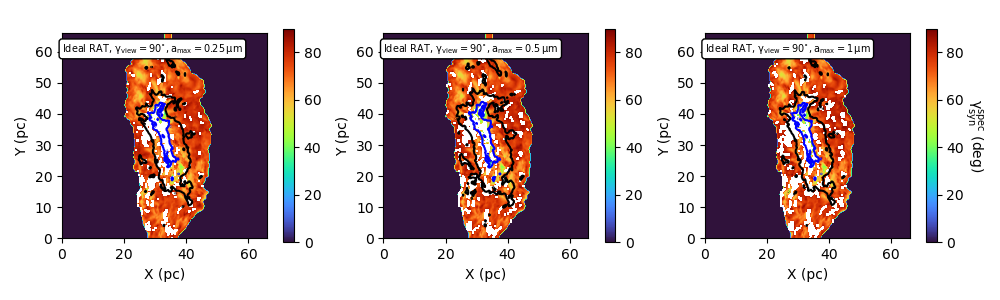}
    \caption{The local inferred inclination angles from the synthetic starlight polarization integral when the grain growth effect is considered. The values of the inclination angle do not change as $a_{\rm max}$ increases from $0.25\,\rm\mu m$ to $1\,\rm\mu m$.}
    \label{fig:Incl_spec_amax}
\end{figure*}

The grain growth effect is included in the calculation of inferred inclination angles from $\Pi_{\rm syn}$, which is illustrated in Figure \ref{fig:Incl_spec_amax}. Once the effect of grain growth on the starlight polarization integral is characterized (Figure \ref{fig:Pi_obs_amax}), the resulting inferred inclination angles do not change for different $a_{\rm max}$ and are the same as the synthetic results in Figure \ref{fig:Incl_spec_Incl}.

\begin{table*}[]
\centering
    \caption{Similar to Table \ref{tab:incl_most_incl} but for the inclination angles inferred from the starlight polarization integral $\Pi_{\rm syn}$.}
    \begin{tabular}{c c c c c c}
    \toprule
      $\gamma_{\rm view}$ & $\langle \gamma_{\rm B} \rangle^{\bar{\wedge}}$  & $ \gamma_{\rho}^{\bar{\wedge}}$  & $|\gamma_{\rm syn, Ideal\,\,RAT}^{\rm spec^{\bar{\wedge}}}|$ & $|\gamma_{\rm syn, MRAT}^{\rm spec^{\bar{\wedge}}}|$ & $|\gamma_{\rm syn, MRAT}^{\rm spec^{\bar{\wedge}}}|$\\

      & & & & $(\rm N_{\rm cl} = 50)$& $(\rm N_{\rm cl} = 1000)$ \\
      \midrule
       $0^{\circ}$ & $11.36^{\circ}$ & $29.4^{\circ}$& $25.07^{\circ}$& $23.86^{\circ}$& $25.43^{\circ}$\\
       
       $30^{\circ}$ & $38.4^{\circ}$ & $42.38^{\circ}$& $41.84^{\circ}$& $41.84^{\circ}$& $41.84^{\circ}$\\
       
       $60^{\circ}$ & $67.81^{\circ}$ & $60.6^{\circ}$& $62.94^{\circ}$& $64.74^{\circ}$& $62.76^{\circ}$\\
       
       $90^{\circ}$ & $77.01^{\circ}$ & $73.04^{\circ}$& $75.75^{\circ}$& $79.72^{\circ}$& $75.75^{\circ}$\\
      \bottomrule
    \end{tabular}
    \label{tab:incl_most_spec}
\end{table*}

\section{Discussion}\label{sec:discuss}

\subsection{Physical model of starlight polarization based on grain alignment theory}

A detailed physical model of absorbed polarization from background stars incorporating modern grain alignment theories and the properties of B-fields' morphology is necessary to investigate its potential of constraining 3D B-fields and dust properties from polarimetric data at optical-NIR wavelengths. In this paper, we introduced the general model of the wavelength-dependent starlight polarization efficiency $P_{\rm ext}(\lambda)/N_{\rm H}$ as shown in Equation \ref{eq:p_NH_fpol} based on (1) the intrinsic polarization efficiency $P_{\rm ext,i}(\lambda)/N_{\rm H}$, (2) the fraction of polarization efficient $f_{\rm pol}(\lambda) = \sigma_{\rm pol, alig}/\sigma_{\rm pol}$, (3) the factor $F_{\rm turb}$ describing the levels of magnetic turbulence and (4) the mean inclination angles $\sin^2\gamma$. The intrinsic polarization efficiency $P_{\rm ext,i}/N_{\rm H}$ describes the starlight polarization per H by all grains at wavelength $\lambda$ in an ideal condition when B-fields are well-ordered and completely in the POS, which then depends only the grain shape. The intrinsic polarization efficiency could be constant, but it is reduced in very high-density regions due to the large polarized extinction effect with respect to the observed wavelength (i.e., $\tau_{\rm pol} \gg 1$, see the left panel of Figure \ref{fig:P_NH_compare_wave}). The fraction $f_{\rm pol}(\lambda)$ describes the starlight polarization by aligned grains only and strongly depends on the alignment properties $f_{\rm align}(a)$ calculated from the MRAT theory for a given environmental condition (e.g., local gas density, radiation intensity, and B-field strength) and grain magnetic properties (e.g., PM or SPM grains with embedded iron) as shown in Equation \ref{eq:align_ana} - \ref{eq:falign}. The fraction $f_{\rm pol}$ could be around $0.8 - 0.9$ in low-density regions $N_{\rm H} < 1.3 \times 10^{22}\,\rm cm^{-2}$ ($A_{\rm V} < 8$); however, it decreases significantly with $f_{\rm pol} \varpropto N_{\rm H}^{-0.7}$ caused by the loss of grain alignment by increasing gas randomization in high-density regions. The values of $f_{\rm pol}$ are lower when grains are PM materials, and can be enhanced if they have iron inclusions (Figure \ref{fig:fpol_Mag}). The fraction $f_{\rm pol}$ could be further enhanced once the enhanced alignment of large grains $a > 0.25\,\rm\mu m$ by grain growth takes place, especially when measured at NIR wavelengths with constant $f_{\rm pol} \sim 0.8 - 0.9$ up to $N_{\rm H} \sim 4 \times 10^{22}\,\rm cm^{-2}$ (i.e., $A_{\rm V} \sim 11.5$, see Figure \ref{fig:fpol_amax}). These variations of $f_{\rm pol}$ with the observed wavelength, dust properties, and local environments are important and should be taken into account for precisely determining inferred inclination angles using starlight polarization efficiency.

Our physical model of starlight polarization was quantified and compared with the synthetic starlight polarization efficiency from the background stars behind the filamentary cloud generated by the updated POLARIS code. The effects of dust properties (e.g., intrinsic dust properties and grain alignment by MRAT) and B-field properties (e.g., B-field inclination angles and magnetic tangling) were included during the synthetic modeling (\citealt{Giang.2023}). Our analytical calculation of polarization efficiency (Equation \ref{eq:p_NH_fpol}) well describes the synthetic polarimetric data once these effects are incorporated in the physical model of starlight polarization (see Figures \ref{fig:P_NH_compare_wave}, \ref{fig:P_NH_compare_Mag} and \ref{fig:P_NH_compare_Incl}). This broadens the full capability of starlight polarization in constraining 3D B-fields, grain alignments, and dust properties from single-band polarimetry at optical and NIR wavelengths. 

Based on the physical model of starlight polarization, we improved the analytical model of starlight polarization integral $\Pi_{\rm obs}$ introduced by \cite{Draine.2021no}. This parameter is determined by taking the integral over the observed starlight polarization spectrum at optical-NIR wavelengths and depends on both B-fields' properties (e.g., the level of magnetic turbulence $F_{\rm turb}$ and the mean inclination angles $\sin^2\gamma$) and grain properties (e.g., the starlight polarization efficiency integral $\Phi(a)$, the size distribution $n_d(a)$ and the alignment degree $f_{\rm align}(a)$, see in Equation \ref{eq:Pi_obs}). Here, the dimensionless starlight polarization efficiency $\Phi(a)$ is used for constraining the intrinsic properties of grains responsible for the starlight polarization, e.g., sizes, shapes, and porosities. Our study considers the dependence of $\Phi(a)$ on both grain elongations and their sizes from nano-sized grains $a = 5\,\rm nm$ to micron-sized grains $a = 1\,\rm \mu m$, differing from the constant $\Phi(a)$ for grains in the diffuse ISM $a = 0.05 - 0.25\,\rm\mu m$ in \cite{Draine.2021no} (see the calculation in Figure \ref{fig:Pi_a}). Consequently, the applications of our new model can be extended in various astrophysical environments, from MCs to dense cores, in which grain growth and grain destruction are commonly taken into consideration. The grain alignment can be characterized from the minimum aligned size $a_{\rm align}$ and the alignment degree $f_{\rm align}(a)$ locally (Section \ref{sec:method_grain_align}) and is independent of the observed wavelength. Once these properties are considered, our analytical calculation of starlight polarization integral is consistent with the one derived from the synthetic starlight polarization spectrum (Figure \ref{fig:Pi_compare}), implying the potential of inferring 3D B-fields and constraining dust properties from background stars having multi-wavelength polarimetric observations. 

\subsection{Tracing 3D B-fields in star-forming regions}

The studies of 3D B-fields are essential for fully understanding their roles in astrophysical processes in star-forming regions, e.g., the giant filamentary cloud formation (\citealt{HennebelleInutsuka.2019}; \citealt{Tahani.2022}) or the hourglass-shape B-field in magnetized dense cores (\citealt{Padovani.2012}; \citealt{Basu.2024}). The effect of inclined 3D B-fields on the degree of starlight polarization was presented in optical-NIR polarimetric observations of MCs (\citealt{Lee.1985}; \citealt{Angarita.2023}) and to smaller scales as dense cores (\citealt{Kandori.2018}; \citealt{Kandori:2020gr}); however, accurate determination of 3D B-fields from starlight polarization has not been proposed and still challenging.

In this study, we introduced a new technique for tracing 3D B-fields by inferring the inclination angles from the starlight polarization efficiency $P/N_{\rm H}$ incorporating modern grain alignment theories. We tested with the synthetic polarization from background stars and found that the new technique can retrieve the inclination angles from optical polarization only in the low-density region $N_{\rm H} < 5 \times 10^{21}\,\rm \cm^{-2}$ corresponding to $A_{\rm V} < 3$ (the left panel of Figure \ref{fig:Incl_Starpol_IdealRAT} and Figure \ref{fig:Incl_Starpol_Incl_Vband}) . This is applicable in inferring 3D B-fields in the outer region of giant filamentary clouds with efficient alignment by MRAT with $f_{\rm pol} \sim 0.8 - 0.9$ (Figure \ref{fig:fpol_Mag}). Meanwhile, by using NIR polarization, the local inclination angles can be inferred toward high-density regions up to $N_{\rm H} \sim 5 \times 10^{22}\,\cm^{-2}$ (i.e., $A_{\rm V} \approx 30$) where dense filaments/cores are formed and evolved (the right panel of Figure \ref{fig:Incl_Starpol_IdealRAT} and Figure \ref{fig:Incl_Starpol_Incl_Kband}). In these dense regions, the grain alignment loss by gas randomization is significant with $f_{\rm pol} \varpropto N_{\rm H}^{-0.7}$ at $A_{\rm V} \sim 8 - 30$ (Figures \ref{fig:fpol_Mag} and \ref{fig:fpol_amax}) and should be considered when interpreting inclination angles from NIR polarization. Once dust and magnetic properties are well determined across the observed wavelength, our new technique provides the local inclination angles close to the true angles derived from MHD simulations by 2-6 degrees (Figures \ref{fig:Histogram_Starpol_wave} and \ref{fig:Histogram_Starpol_Incl}) - similar to what was obtained in the previous technique using thermal dust polarization degree (Paper I, \citealt{HoangBao.2024}). Therefore, it is possible to apply the new technique in constraining local 3D B-fields using multi-wavelength polarization on multiple scales of star-forming regions from MCs by optical polarimeters (\citealt{Panopoulou.2019a}; \citealt{Doi.2021}; \citealt{Angarita.2023}) to filaments (\citealt{Chapman.2011}; \citealt{Sugitani.2019}; \citealt{Wang.2020}; \citealt{Chen.2022}; \citealt{Chen.2023}) and even in dense cores (\citealt{Kandori.2018}; \citealt{Kandori.2020}) by NIR polarimeters with high sensitivity.

\subsection{Tracing 3D B-fields in the diffuse ISM}


On a large scale of $>$ 1kpc as the diffuse ISM, the presence of 3D B-fields impacts the formation and evolution of multi-phase ISM structures (see the review of \citealt{Haverkorn.2015}). As discussed in the above section, by using our new technique, we can trace the inclination angles from optical starlight polarization efficiency produced by background stars with low-extinction $A_{\rm V} < 3$ and efficient MRAT alignment $f_{\rm pol} \sim 0.8 - 0.9$ in the magnetized ISM (Figure \ref{fig:Incl_Starpol_IdealRAT}). This can be achievable through modern optical polarization surveys from the Interstellar Polarization Survey in the General ISM (IPS-GI; \citealt{Versteeg.2023}; \citealt{Angarita.2023}) at low- and intermediate-latitude $|b| < 10^{\circ}$ to the Polar-Areas Stellar Imaging Polarization High Accuracy Experiment (PASIPHAE; \citealt{Panopoulou.2019a}) at high-latitude $|b| > 50^{\circ}$, providing the morphology of Galactic B-fields in the three-dimensional space.

Note that the 3D ISM B-fields can be inferred by using the starlight polarization integral $\Pi_{\rm obs}$ (see Equation \ref{eq:incl_pi_obs}) for a few samples of background stars with $A_{\rm V} < 4$ having spectropolarimetric observations from optical to NIR wavelengths (see \citealt{Chapman.2011}; \citealt{Vaillancourt:2020ch}). Same as the above technique using starlight polarization efficiency, the new method using starlight polarization integral can accurately determine the mean inclination angles, given known dust properties and grain alignment in local environments (Figure \ref{fig:Incl_spec_Incl} and \ref{fig:Histogram_Spec_Incl}). In the case of background stars in the diffuse ISM, the determination of inclination angles from $\Pi_{\rm obs}$ can be simplified by using $\Phi(a) \approx \Phi(a_{\rm align})$ and the local mass-weighted alignment efficiency $\langle f_{\rm align} \rangle$ (see Equation \ref{eq:incl_pi_ism}), considering the constraint on grain elongation $s \gtrsim 1.4$ (see \citealt{Hensley.2023}). This unlocks a new possibility of constraining both 3D Galactic B-fields, dust properties, and grain alignment physics in the diffuse medium from spectropolarimetric observations.




\subsection{Applications for 3D B-field tomography}
The advantage of starlight polarization is that it can reveal the 3D B-fields arising from multiple cloud components along the LOS when the information on the background star's distances is available, whereas polarized thermal emission tells only the average 3D B-field due to the integration effect (see \citealt{Panopoulou.2019a}; \citealt{Pelgrims.2023}; \citealt{Doi.2024}). The measurement of B-fields' components along the LOS has been extensively studied with the combination of starlight polarization observations and accurate determination of stellar distances up to a few kpc by the Gaia astrometric mission. The latest study of \cite{Doi.2024} separated 2D magnetic field components along the LOS in the Sagittarius Arm using R-band polarization by Hiroshima Optical and Near-InfraRed (HONIR) camera combined with distance measurement in the Gaia Data Release 3 catalog (DR3, \citealt{Gaia.2023}). They found the variation of polarization efficiency $P/A_{\rm V} < 1.4\%\,\rm mag^{-1}$ in each isolated cloud component, which could be explained by the effect of Galactic B-fields' inclination angles with respect to the observed POS.

Throughout our study, we have emphasized the applications of our new technique using starlight polarization efficiency to infer the 3D B-fields for multiple clouds along the sightline. Given the polarization efficiency and polarization angle dispersion for each LOS cloud component separated by stellar distances combined with the constrained dust properties and grain alignment physics, we can derive the inclination angles of each LOS magnetic field component (see Appendix \ref{sec:discuss_application}). Due to the sensitive dependence of starlight polarization on the sightline of background stars for different observed wavelengths (\citealt{Doi.2021}; \citealt{Chen.2022}; \citealt{Chen.2023}), the application to multi-wavelength will provide the 3D B-fields of the different layers/clouds along the LOS once the grain alignment and dust properties are well constrained across the wavelengths (see Sections \ref{sec:fpol_result} and \ref{sec:pol_syn}). This contributes to reconstructing the 3D structures of B-fields from MCs to the diffuse ISM, which is essential for investigating their influences on star formation processes. The complete applications of 3D tomographic imaging of B-fields combined with our new method will be presented in the forthcoming studies. 

\subsection{Applications for constraining dust properties}

The polarization of starlight from aligned grains is a powerful tool to constrain dust properties and grain alignment physics in local environments (see the review of \citealt{Andersson:2015bq}). For instance, the slope $\alpha$ of $P/A_{\rm V} \varpropto A^{-\alpha}_{\rm V}$ at $A_{\rm V} > 10$ is commonly used for characterizing grain alignment physics in MCs and dense cores with and without embedded protostars (\citealt{AnderssonPotter.2007}; \citealt{Whittet.2008}; \citealt{Hoang.2021}). Meanwhile, the maximum grain size can be constrained by observing the maximum visual extinction $A_{\rm V, max}$ where grain alignment still exists, providing evidence of grain growth in dense environments (\citealt{Hoang.2021}). The signature of grain growth can also be detected from the starlight polarization spectrum with large peak wavelengths $\lambda_{\rm max} > 0.55\,\rm\mu m$ toward high-extinction regions $A_{\rm V} > 4$ (\citealt{Whittet.2008}; \citealt{Vaillancourt:2020ch}). The observations of the starlight polarization spectrum can also allow the constraint on intrinsic grain properties, e.g., shapes (elongation) and porosities, through the starlight polarization integral $\Pi_{\rm obs}$ (see \citealt{Draine.2021no}; \citealt{Hensley.2023}). Additionally, \cite{HoangLaz.2016} have demonstrated the major role of iron inclusions in enhancing dust polarization degree by the MRAT mechanism, especially in dense environments as dense cores and protostars, providing the capability to constrain the levels of embedded iron through both thermal dust polarization and starlight polarization degree.

Our study has demonstrated the importance of dust properties in accurate calculations of inferred inclination angles from starlight polarization efficiency $P/N_\H$ and the starlight polarization integral $\Pi_{\rm obs}$. Given the constrained dust properties, e.g., intrinsic and magnetic properties (Section \ref{sec:result_iron},  \ref{sec:result_graingrowth} and Appendix \ref{sec:appendix_ratio}), we can determine the local inclination angles of the mean fields with high accuracy. Therefore, the potential of constraining 3D B-fields and dust properties using starlight polarization are associated with each other. And vice versa, our physical model of starlight polarization can help constrain the dust properties and grain alignment physics when the information on 3D B-fields is available. This possibility can be done by comparing the inferred inclination angles from our techniques with the available inclination angles derived from other methods, e.g., the combination of dust polarization and Faraday rotation (\citealt{Tahani.2018}; \citealt{Tahani.2022b2h}) or the combination of dust polarization and Zeeman splitting (\citealt{Reissl.2021}), which will be investigated further in the future studies.

\subsection{Uncertainties of estimating $F_{\rm turb}$ in starlight polarization observations}
\label{sec:Fturb_estimation}
Up to now, we have demonstrated the importance of considering grain alignment and dust properties when accurately retrieving the inclination angles from starlight polarization. We note that the effect of B-field fluctuations has a significant impact on the decrease in starlight polarization degree (Section \ref{sec:method_star_fluc}). By including the magnetic turbulence factor $F_{\rm turb}$ derived from the deviation between local and mean B-fields in the MHD simulations, we can infer B-field inclination angles from starlight polarization efficiency $P/N_{\rm H}$ (Section \ref{sec:results_incl_PNH}) and starlight polarization integral $\Pi$ (Section \ref{sec:results_incl_Pi}) with high accuracy.

Practically, the magnetic turbulence contributes to the depolarization in observational data (\citealt{AnderssonPotter2005}; \citealt{Chapman.2011}; \citealt{Chen.2023}). If the turbulence is isotropic (i.e., $\delta B_{\rm POS, \perp} = \sqrt{2}\delta B_{\rm LOS}$), the factor $F_{\rm turb}$ can be characterized through the polarization angle dispersion $\delta \phi$ in a 2D POS map. If the turbulence is anisotropic (i.e., $\delta B_{\rm POS, \perp} \neq \sqrt{2}\delta B_{\rm LOS}$) as for the filamentary cloud used in this paper, the estimation of $F_{\rm turb}$ from $\delta \phi$ could be affected by the viewing angle $\gamma_{\rm view}$, which is realistic in the diffuse ISM and star-forming regions (e.g., \citealt{Ostriker.2001}; \citealt{Falceta.2008}; \citealt{Skalidis.Tassis2021}). In Paper I, \cite{HoangBao.2024} demonstrated that the magnetic turbulence factor $F_{\rm turb}$ is correlated with the POS polarization angle dispersion with $F_{\rm turb} \varpropto \delta \phi^{-0.3}$ when $\delta \phi > 5^{\circ}$ and $\gamma_{\rm view} = 90^{\circ}$; nevertheless, the slope is shallower with $|\eta| < 0.3$ for smaller $\gamma_{\rm view}$ due to the dominance of LOS magnetic fluctuations (see Figure 26 in Paper I, \citealt{HoangBao.2024}). In addition, the integration effect along the sightline could lower the POS polarization angle dispersion $\delta \phi$ and underestimate the factor $F_{\rm turb}$ (e.g., \citealt{Cho.2016}; \citealt{Skalidis.Tassis2021}).

In real observations of starlight polarization, the polarization angle dispersion $\delta \phi$ is determined over a small area sampled from the observed field-of-view (see \citealt{AnderssonPotter2005}; \citealt{Chapman.2011}; \citealt{Sugitani.2019}; \citealt{Chen.2023}). The density of polarized background stars within a sampling cell size should be sufficiently high (i.e., $N_{\rm star} > 3$, see \citealt{AnderssonPotter2005}; \citealt{Chen.2023}), and the spatial sampling cell size should be shorter than the magnetic field scale length (\citealt{Houde.2009}) in order to resolve the turbulent components of B-fields and estimate $\delta \phi$ and $F_{\rm turb}$. Besides, the depolarization caused by B-field tangling within the sampling cell is weaker for smaller sampling sizes, and the inferred inclination angles can be inferred with higher accuracy - a similar effect when using smaller beam sizes in thermal dust polarization (see Section 6.5 in Paper I, \citealt{HoangBao.2024}). These uncertainties of the estimation of $F_{\rm turb}$ and the calculation of inferred inclination angles in real observational data will be quantified further in follow-up studies.




\section{Summary}\label{sec:summ}
We have extended our new technique for probing 3D magnetic fields using thermal dust polarization at far-IR/sub-mm and grain alignment theory from Paper I (\citealt{HoangBao.2024}) to starlight polarization at optical/NIR. Our results are summarized as follows:
\begin{enumerate}
\item We introduced a physical model of starlight polarization efficiency using the grain alignment theory based on MRAT combined with the effect of magnetic turbulence. At optical/NIR wavelengths, the degree of starlight polarization is a function of the intrinsic polarization, the fraction of polarization coefficient from aligned grains induced by MRAT, the magnetic fluctuation effect, and the B-field inclination angles.

\item The polarization coefficient fraction $f_{\rm pol}$ is introduced to describe the absorbed polarization from aligned grains by MRAT at optical-NIR wavelengths. We examined the variation of $f_{\rm pol}$ within a filamentary cloud from the MHD simulations. The fraction $f_{\rm pol}$ is 0.8 - 0.9 in the low-density regions having efficient MRAT alignment, while this fraction can decrease considerably at high-density regions $N_{\rm H} > 1.3 \times 10^{22}\,\cm^{-2}$ (i.e., $A_{\rm V} > 8$) due to inefficient alignment caused by increasing gas randomization.

\item We tested our physical polarization model incorporating modern MRAT alignment theory and the magnetic fluctuation effect by performing synthetic starlight polarization observations of MHD simulations for a filamentary cloud using the upgraded POLARIS code. We found that the physical polarization model can accurately describe the synthetic absorbed polarization from background stars.

\item Using our tested physical polarization model and synthetic polarization data, we inferred the inclination angles of the mean 3D B-fields. Using optical starlight polarization, the inclination angles can only be inferred in low-density regions $A_{\rm V} < 3$ with efficient MRAT alignment (i.e., $f_{\rm pol} \sim 0.8 - 0.9$), while they can be inferred toward high-density regions with the loss of grain alignment (i.e., $f_{\rm pol} \varpropto N_{\rm H}^{-0.7}$) at $A_{\rm V} \sim 8 - 30$ when using NIR polarization. As the effects of grain alignment, dust properties, and B-field fluctuation on starlight polarization are determined across the observed wavelength, our new method provides an accurate inference of inclination angle close to the actual values from MHD simulations by $2 - 6$ degrees.

\item Our new method can be applied to real observation data to constrain 3D B-fields using multi-wavelength starlight polarization data in multi-scale star-forming regions, from large scales as the diffuse ISM/molecular clouds with $A_{\rm V} < 3$ at optical wavelengths to small scales as dense filaments/cores with $A_{\rm V} \sim 8 - 30$ at NIR wavelengths. 

\item Our new method can be applied to determine the B-field inclination angles from multiple cloud components along the line-of-sight, providing tomographic imaging of true Galactic 3D B-fields using multi-wavelength polarization data combined with the accurate determination of stellar distances.

\item We also introduced a new technique of inferring inclination angles using starlight polarization integral $\Pi_{\rm obs}$ derived from the starlight polarization spectrum incorporating grain intrinsic properties (i.e., sizes and shapes) and modern MRAT alignment theory. The new technique can provide an accurate measurement of the inclination angles and the 3D magnetic fields. Our new method can potentially be applied in constraining both dust properties and 3D B-fields using available spectropolarimetric data of low-extinction background stars $A_{\rm V} < 4$ in the diffuse ISM.

\item When the information of 3D B-fields is provided, our physical model of starlight polarization can help constrain the dust properties, such as the grain elongation and magnetic properties, and grain alignment physics.

\end{enumerate}

\acknowledgments
We thank the anonymous referee for helpful comments that improves our paper. T.H. acknowledges the support from the main research project (No. 2025186902) from Korea Astronomy and Space Science (KASI). This work was partly supported by a grant from the Simons Foundation to IFIRSE, ICISE (916424, N.H.). We thank the ICISE staff for excellent support and hospitality.

\bibliographystyle{apj}
\bibliography{ms}

\appendix
\section{The dependence of polarization coefficient fraction $f_{\rm pol}(\lambda)$ on the observed wavelength}
\label{sec:appendix_fpol_wave}

\begin{figure*}
    \centering
    \includegraphics[width=0.48\linewidth]{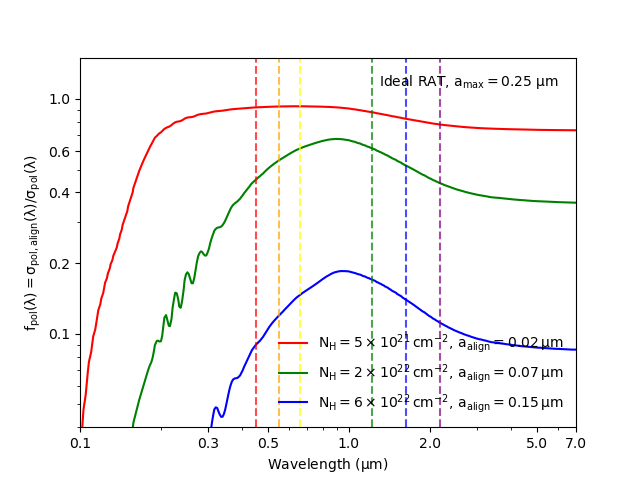}
    \includegraphics[width=0.48\linewidth]{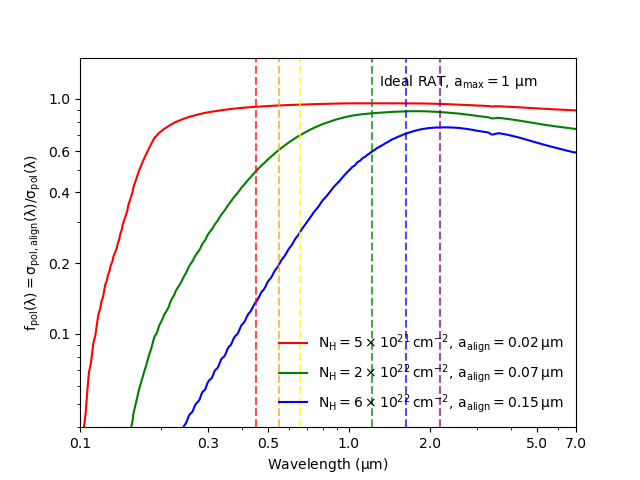}
    \caption{The wavelength-dependent fraction of linear polarization coefficient $f_{\rm pol}(\lambda)$ calculated from the outer to the inner cloud with increasing column density $N_\H$, considering the grain growth effect with increased maximum grain size from $a_{\rm max} = 0.25\,\rm\mu m$ to $a_{\rm max} = 1\,\rm\mu m$. The dashed color vertical lines correspond to the fraction $f_{\rm pol}$ measured at optical-NIR polarimetric bands. The Ideal RAT alignment model is being assumed. The fraction $f_{\rm pol}$ is lower in the inner dense cloud under the effect of grain alignment loss by gas randomization, and the peak of $f_{\rm pol}(\lambda)$ tends to be shifted to longer wavelengths.}
    \label{fig:fpol_wave}
\end{figure*}

The left panel of Figure \ref{fig:fpol_wave} shows the fraction of linear polarization coefficient $f_{\rm pol}(\lambda)$ as a function of wavelength measured from the outer to the inner regions of the cloud with increasing $N_\H$, considering the Ideal RAT alignment model and $a_{\rm max} = 0.25\,\rm\mu m$. The fraction $f_{\rm pol}(\lambda)$ is around 0.8 - 0.9 in the outer cloud, and $f_{\rm pol}(\lambda)$ at optical bands is higher than at NIR bands produced by aligned grains $a < 0.25\,\rm\mu m$. This fraction decreases significantly to $< 0.2$ in the inner dense cloud due to the reduced RAT alignment efficiency caused by the increased gas randomization (i.e., increasing $a_{\rm align}$, see Equation \ref{eq:align_ana}). The peak $f_{\rm pol, peak}$ tends to be shifted to longer wavelengths $\lambda \sim 0.5 - 1\,\rm\mu m$ in this region.

The presence of grain growth strongly impacts the wavelength-dependent $f_{\rm pol}(\lambda)$, as illustrated in the right panel of Figure \ref{fig:fpol_wave} when increasing maximum grain sizes to $a_{\rm max} = 1\,\rm\mu m$. A higher proportion of large grains $a > 0.25 \,\rm\mu m$ can be aligned with RATs, producing stronger absorbed polarization at NIR bands rather than optical ones. The peak $f_{\rm pol, peak}$ is then shifted further to $\lambda > 1\,\rm\mu m$. This effect is significant in the dense inner cloud with $N_{\rm H} > 1.3 \times 10^{22}\,\rm cm^{-2}$, leading to an increase $f_{\rm pol}(\lambda)$ to $\sim 0.8 - 0.9$ when being observed at NIR wavelengths.

\section{Effect of grain elongation on the calculations of inferred inclination angles from starlight polarization}
\label{sec:appendix_ratio}

In this section, we present the impact of grain elongation on the calculations of inferred inclination angles from starlight polarization. We take the Astrodust model of oblate grains by \cite{Draine.2021no} for different axial ratios from $s = 1.4$ (less elongated) to $s = 2$ (highly elongated). We quantify their main effect on the synthetic starlight polarization efficiency $P/N_\H$ and the inferred inclination angles $|\gamma^{\rm star}_{\rm syn}|$.

\begin{figure*}
    \centering
    \includegraphics[width = 0.48\textwidth]{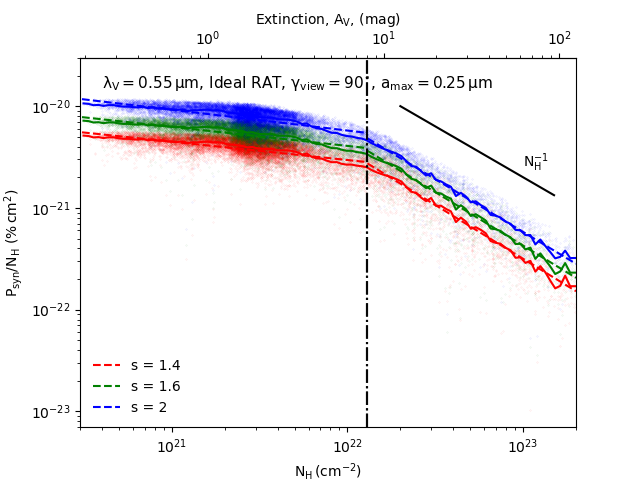}
    \includegraphics[width = 0.48\textwidth]{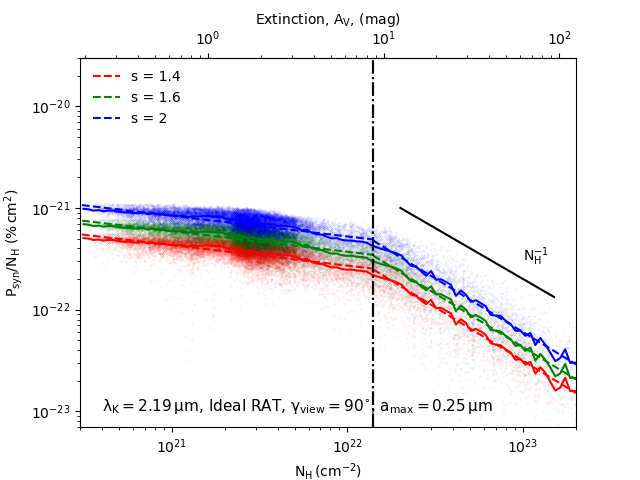}
    \caption{The resulting V-band (left panel) and K-band (right panel) starlight polarization efficiency $P_{\rm syn}/N_\H$ for different axial ratios $s = 1.4 - 2$. As grains become highly elongated to $s = 2$, the polarization efficiency in the entire filamentary cloud increases by a factor of $\sim 2$.}
    \label{fig:Pol_ratio_Ideal}
\end{figure*}

\begin{figure*}
    \centering
    \includegraphics[width = 1\textwidth]{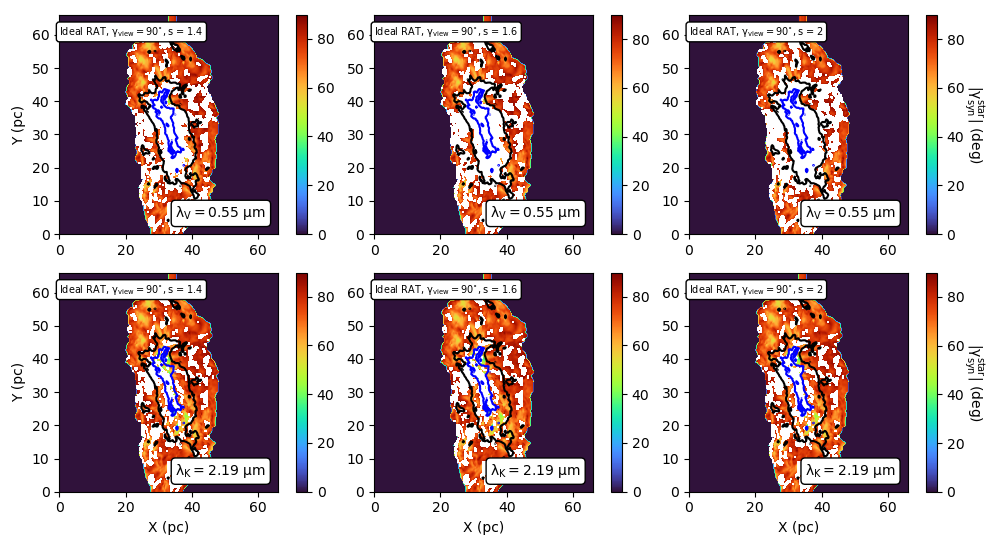}
    \caption{The inferred inclination angles from starlight polarization efficiency at V-band (top panels) and K-band (bottom panels), but considering the variation of grain elongation. The values of $\gamma^{\rm star}_{\rm syn}$ are unchanged in comparison to the results in Figure \ref{fig:Incl_Starpol_Incl_Vband} and \ref{fig:Incl_Starpol_Incl_Kband}.}
    \label{fig:Incl_obs_ratio_Ideal}
\end{figure*}

Figure \ref{fig:Pol_ratio_Ideal} shows the V-band (top) and K-band (bottom) starlight polarization efficiency $P_{\rm syn}/N_\H$ for increasing axial ratios $s$, while Figure \ref{fig:Incl_obs_ratio_Ideal} illustrates the inclination angles inferred from the above polarization efficiency. The polarization efficiency at all observed wavelengths increases by a factor of 2 as grains become highly elongated (i.e., $s \approx 2$). The inferred inclination angles do not change and are comparable to the results in Figure \ref{fig:Incl_Starpol_Incl_Vband} and \ref{fig:Incl_Starpol_Incl_Kband} as the grain elongation are determined from starlight polarization efficiency. This emphasizes the applications of our new technique in inferring 3D B-fields using starlight polarization to arbitrary grain elongation. Note that in addition to oblate grains, ISM grains could have other shape geometries (\citealt{Draine2024a}) and porosities (\citealt{Draine2024b}) affecting the results of starlight polarization. The determination of 3D B-fields using starlight polarization is accurate when these intrinsic properties are constrained.

\section{Synthetic Starlight Polarization Spectrum}
\label{sec:appendix_spectrum}

\begin{figure*}
    \includegraphics[width = 0.48\textwidth]{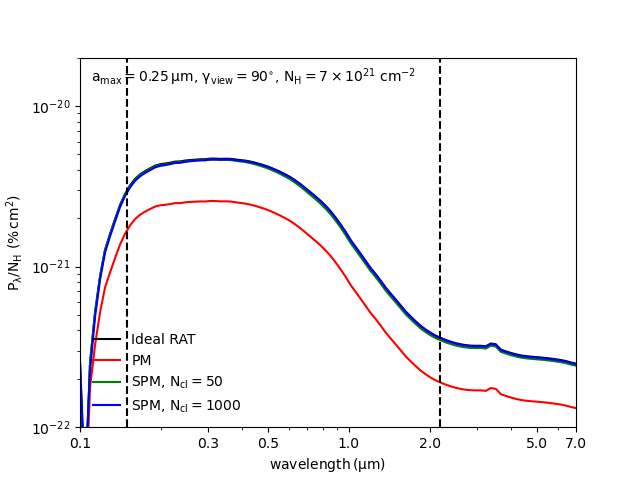}
    \includegraphics[width = 0.48\textwidth]{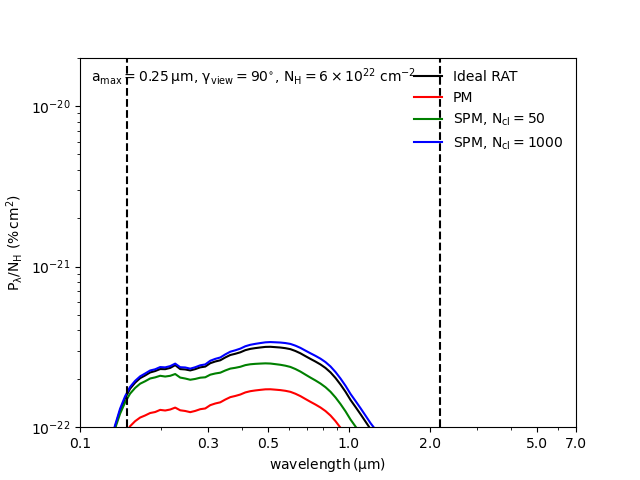}
    \caption{The synthetic optical-NIR polarization spectrum from background stars located at the outer part (i.e., $N_{\rm H} = 7 \times 10^{21}\,\cm^{-2}$, left panel) and the inner part of the cloud (i.e., $N_{\rm H} = 6 \times 10^{22}\,\cm^{-2}$, right panel). Black dashed lines show $\lambda_{\rm min} = 0.15\,\rm\mu m$ and $\lambda_{\rm max} = 2.19\,\rm\mu m$ for the calculation of $\Pi_{\rm syn}$. Both Ideal and Realistic alignment models for PM and SPM grains with $N_{\rm cl} = 50 - 1000$ are being assumed. The peak of the polarization spectrum shifts to longer wavelengths $\lambda \sim 0.5\,\rm\mu m$ produced by aligned large grains in the denser cloud. The polarization efficiency is higher in all wavelengths as grains have high levels of embedded iron.}
    \label{fig:Pol_Spec_Mag}

    \includegraphics[width = 0.48\textwidth]{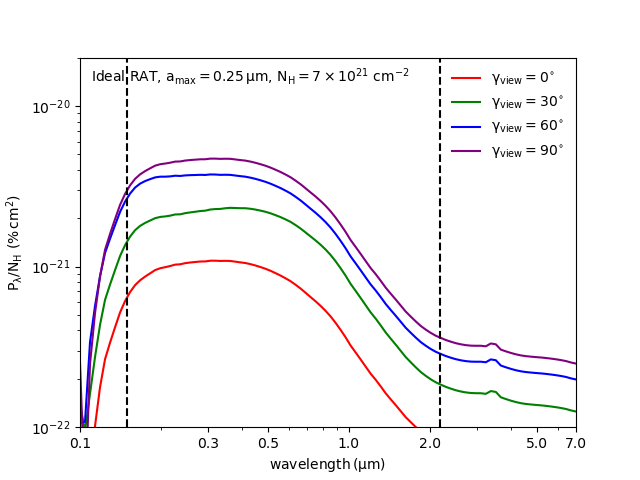}
    \includegraphics[width = 0.48\textwidth]{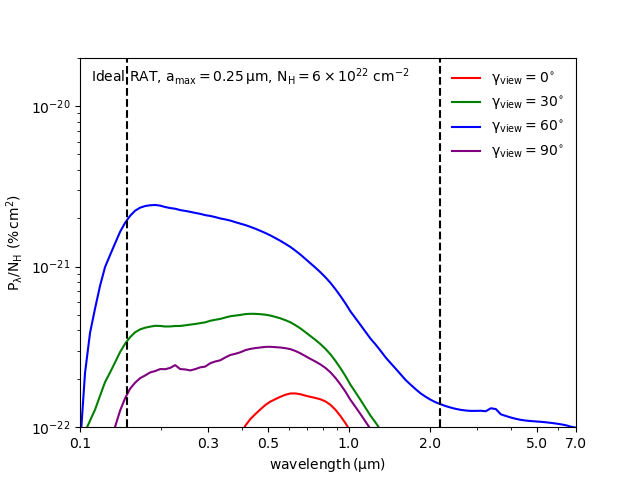}
    \caption{Similar to Figure \ref{fig:Pol_Spec_Mag} but for various viewing angles and the Ideal RAT alignment model. The polarization efficiency is reduced in all optical-NIR wavelengths when $\gamma_{\rm view}$ is smaller.}
    \label{fig:Pol_Spec_Incl}

    \includegraphics[width = 0.48\textwidth]{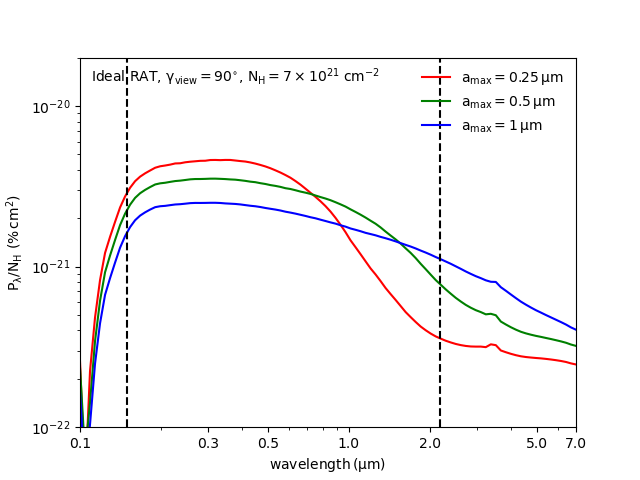}
    \includegraphics[width = 0.48\textwidth]{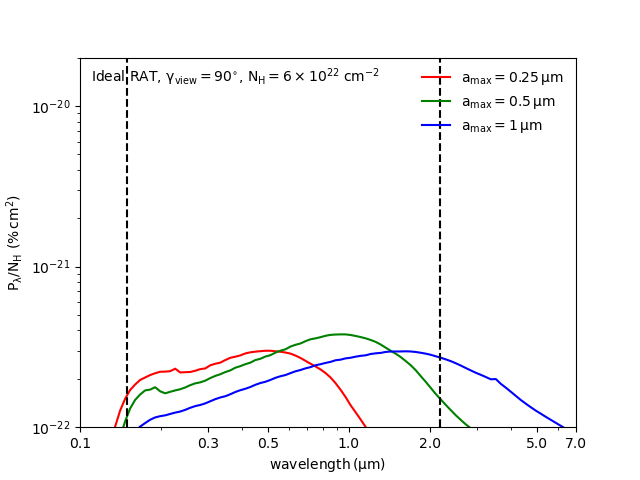}
    \caption{Same as Figure \ref{fig:Pol_Spec_Mag} but considering increasing maximum grain sizes from $0.25\rm\mu m$ to $1\,\rm\mu m$. The polarization efficiency increases significantly at NIR wavelengths with the spectrum peak at $\lambda \sim 1 - 2\,\rm\mu m$ in the denser part of the cloud as $a_{\rm max}$ increases. }
    \label{fig:Pol_Spec_amax}
\end{figure*}

Figure \ref{fig:Pol_Spec_Mag} presents the synthetic polarization spectrum at $\lambda = 0.1 - 7\,\rm\mu m$ for background stars located in the outer region (i.e., $N_\H = 7 \times 10^{21}\,\cm^{-2}$, left panel) and inner region (i.e., $N_\H = 6 \times 10^{22}\,\cm^{-2}$, right panel) of the filamentary cloud. Both Ideal and Realistic alignment models, $a_{\rm max} = 0.25\,\rm\mu m$ and the viewing angle $\gamma_{\rm view} = 90^{\circ}$ are considered. The wavelength-dependent polarization efficiency is high in the outer cloud with efficient MRAT alignment with $(P_{\lambda}/N_{\rm H})_{\rm max} \sim 4 \times 10^{-21}\,\rm\%\,cm^{2}$. The polarization spectrum is lower with $(P_{\lambda}/N_{\rm H})_{\rm max} \sim 3 \times 10^{-22}\,\rm\%\,cm^{2}$ in the denser region $N_{\rm H} > 1.3 \times 10^{22}\,\rm cm^{-2}$ with the dominant alignment loss (see Figure \ref{fig:fpol_Mag}). The peak spectrum is shifted to longer wavelengths $\lambda \sim 0.4 - 0.5\,\rm\mu m$ owing to the increase in $a_{\rm align}$ in this region. The overall polarization spectrum can be much lower by a factor of 1.5 for PM grains, and increase significantly to the Ideal case as grains are SPM materials and have iron inclusions $N_{\rm cl} > 50$.

Figure \ref{fig:Pol_Spec_Incl} shows the effect of viewing angles $\gamma_{\rm view}$ on the starlight polarization spectrum observed in the outer and inner regions. The polarization spectrum is significantly low due to the projection effect of the inclined mean B-fields on the POS for small $\gamma_{\rm view}$, and becomes higher with increasing $\gamma_{\rm view}$ (i.e., $\sin^2\gamma \sim 1$).

The impact of grain growth on the starlight polarization spectrum is presented in Figure \ref{fig:Pol_Spec_amax} with increasing maximum grain sizes $a_{\rm max} = 0.25 - 1\,\rm\mu m$. The wavelength-dependent polarization efficiency gets decreased at optical wavelengths $\lambda = 0.1 - 0.8\,\rm\mu m$, while it gets increased at NIR wavelengths  $\lambda > 1 \,\rm\mu m$. In the inner dense region $N_\H = 6 \times 10^{22}\,\cm^{-2}$, the spectrum is shifted toward NIR regimes with the peak wavelength of $\lambda_{\rm max} \sim 1 - 2 \,\rm\mu m$ produced by the alignment of larger grains $a > 0.25\,\rm\mu m$ enhanced by grain growth.

\section{Applications to observational data}
\label{sec:discuss_application}
So far, we have applied our new technique to synthetic data to infer 3D B-fields using starlight polarization when the effects of grain alignment, intrinsic dust properties, and B-field tangling are all included. In real observations, grain alignment properties by MRAT could be constrained locally once the local environmental conditions (e.g., gas density, radiation field strength, dust and gas temperature) are given from multiple observations (see \citealt{TramHoang.2022} and references therein), while intrinsic dust properties could vary with local environments due to grain evolution (i.e., varying $a_{\rm max}$, see \citealt{Hirashita.2012}; \citealt{Bate2022}; and varying grain elongation, see \citealt{Hoang.2022}. In this section, we outline the following steps for applying the new technique to real polarimetric data from observed background stars at optical-NIR bands, considering the constraint on grain alignment, dust properties, and B-field fluctuation in local environments.

\begin{enumerate}

    \item We first constrain the grain properties (e.g., size distribution and elongation) to be used for calculations of the intrinsic polarization efficiency $P_{\rm ext,i}/N_{\rm H}$ within the cloud of interest. These properties can be constrained by fitting the observed maximum starlight polarization efficiency at optical/NIR with the modeled one in which the magnetic field is lying in the POS (i.e., $\sin^2\gamma = 1$), well-ordered (i.e., $F_{\rm turb} = 1$) and Ideal RAT alignment (i.e., $R = 1$) as $(P_{\rm ext}/N_\H)_{\rm max} \approx (P_{\rm mod}/N_\H)_{\rm max}$ (see \citealt{Hensley.2023}). The grain size distribution can be constrained independently by the dust extinction curve at optical/NIR (see \citealt{Mathis.1977}; \citealt{1989ApJ...345..245C}; \citealt{Weingartner:2001p3480}) or by the dust opacity spectral index $\beta$ of thermal dust emission at far-IR/sub-mm (see \citealt{Draine.2006}; \citealt{Kwon.2009}).

    \item Derive the volume number gas density $n_{\rm H}$ from the column density $N_{\rm H}$ along the LOS as $N_\H = \int_{\rm LOS} n_\H \rm ds$ and the radiation field from dust temperature in the local environments. 

    \item Derive the polarization angular dispersion ($\delta \phi$) and the turbulence velocity dispersion from the velocity gas profile ($\delta v$) to apply the DCF method to calculate the B-field strength, $B_{\rm POS}$ (see in \citealt{Chapman.2011}).

    \item Calculate the minimum grain alignment size $a_{\rm align}$ from Equation \ref{eq:align_ana}, the magnetic relaxation $\delta_m$ from Equation \ref{eq:delta_m} and the alignment degree $f_{\rm align}(a)$ for given magnetic properties of grains (e.g., PM or SPM grains with $N_{\rm cl}$) from Equation \ref{eq:falign} (see also Figure \ref{fig:falign_a}). Combined with the polarization cross-section $C_{\rm pol}(a, \lambda)$ from the constrained size distribution and shape, the fraction of polarization coefficient $f_{\rm pol}(\lambda)$ can be computed as presented in Equation \ref{eq:fpol}.

    \item Calculate $F_{\rm turb}$ using the polarization angle dispersion of $\delta \phi$ (see the discussion in Section \ref{sec:Fturb_estimation}).

    \item With $P_{\rm ext,i}/N_{\rm H}$, $f_{\rm pol}$, $F_{\rm turb}$, and $P_{\rm ext}/N_{\rm H}$ available at a certain polarimetric band, the local B-field inclination angles are inferred using Equation \ref{eq:chi_ext}.
    
\end{enumerate}

Note that this technique can be applicable to trace 3D B-fields in the diffuse ISM and nearby star-forming regions from the ratio of observed 
polarization to visual extinction $P_{\lambda}/A_{\rm V}$ as $N_\H/A_{\rm V} \approx 1.87 \times 10^{21}\,\H\cm^{-2}\rm mag^{-1}$, or from the ratio of observed polarization to reddening $P_{\lambda}/E(B - V)$ as $N_\H/E(B - V) \approx 5.8 \times 10^{21}\,\cm^{-2}\rm mag^{-1}$ in the averaged Galactic line-of-sight (\citealt{Bohlin.1978}). For multiple layers/clouds along the LOS, the 3D B-fields can be retrieved from the $P_{\lambda}/A_{\rm V}$ at each cloud component, where $A_{\rm V}$ is determined over stellar distances from Gaia astrometric mission (see \citealt{Zucker.2021}; \citealt{Angarita.2023}). We will apply this technique to interpret the 3D B-fields morphologies in multi-scale star-forming regions, or in multiple layers/clouds along the sightline in the ISM using multi-wavelength polarimetric data in the follow-up studies.

\end{document}